\documentclass[longauth]{aa}
\usepackage{pdflscape}
\usepackage{graphicx}
\usepackage{natbib}
\usepackage{scalerel}
\usepackage{tabularx} 
\usepackage{pdflscape}
\usepackage{rotating}
\usepackage{subcaption} 
\usepackage[font=small]{caption}
\usepackage[labelfont=bf]{caption} 
\usepackage[singlelinecheck=false]{caption}
\usepackage{csquotes}
\usepackage[table]{xcolor}
\usepackage[nolist,nohyperlinks]{acronym}
\usepackage{txfonts}
\usepackage[]{hyperref}
\hypersetup{
    colorlinks=true,
    linkcolor=blue,
    filecolor=magenta,      
    urlcolor=blue,
    citecolor=blue}
\usepackage{orcidlink}
\usepackage[nameinlink,capitalise]{cleveref}
\crefname{section}{Sect.}{Sects.}
\Crefname{section}{Section}{Sections}
\crefname{figure}{Fig.}{Figs.}
\Crefname{figure}{Figure}{Figures}
\crefname{equation}{Eq.}{Eqs.}
\Crefname{equation}{Equation}{Equations}

\usepackage{lastpage}
\usepackage[utf8]{inputenc}
\usepackage[switch, modulo]{lineno}
\usepackage{float}
\usepackage{euclid}

\definecolor{bostonuniversityred}{rgb}{0.8, 0.0, 0.0}

\begin{document}

\acrodef{AGN}{active galactic nucleus}
\acrodef{ASIC}{application specific integrated circuit}
\acrodef{BFE}{brighter-fatter effect}
\acrodef{BCDs}{blue compact dwarfs}
\acrodef{BCGs}{blue compact galaxies}
\acrodef{BCG}{brightest cluster galaxy}
\acrodef{CaLA}{camera-lens assembly}
\acrodef{CCD}{charge-coupled device}
\acrodef{CEA}{Comité Energie Atomique}
\acrodef{CoLA}{corrector-lens assembly}
\acrodef{CDS}{Correlated Double Sampling}
\acrodef{CFC}{cryo-flex cable}
\acrodef{CFHT}{Canada-France-Hawaii Telescope}
\acrodef{CGH}{computer-generated hologram}
\acrodef{CNES}{Centre National d'Etude Spacial}
\acrodef{CPPM}{Centre de Physique des Particules de Marseille}
\acrodef{CPU}{central processing unit}
\acrodef{CTE}{coefficient of thermal expansion}
\acrodef{DCU}{Detector Control Unit}
\acrodef{DES}{Dark Energy Survey}
\acrodef{DGL}{diffuse galactic light}
\acrodef{DPU}{Data Processing Unit}
\acrodef{DS}{Detector System}
\acrodef{EDS}{Euclid Deep Survey}
\acrodef{EE}{encircled energy}
\acrodef{ERO}{Early Release Observations}
\acrodef{ESA}{European Space Agency}
\acrodef{EWS}{Euclid Wide Survey}
\acrodef{FDIR}{Fault Detection, Isolation and Recovery}
\acrodef{FGS}{fine guidance sensor}
\acrodef{FOM}[FoM]{figure of merit}
\acrodef{FOV}[FoV]{field of view}
\acrodef{FPA}{focal plane array}
\acrodef{FWA}{filter-wheel assembly}
\acrodef{FWC}{full-well capacity}
\acrodef{FWHM}{full width at half maximum}
\acrodef{GC}{globular cluster}
\acrodef{GWA}{grism-wheel assembly}
\acrodef{H2RG}{HAWAII-2RG}
\acrodef{HST}{{\it Hubble} space telescope}
\acrodef{HSC}{Subaru-Hyper Suprime-Cam}
\acrodef{ISM}{interstellar medium}
\acrodef{IP2I}{Institut de Physique des 2 Infinis de Lyon}
\acrodef{JWST}{{\em James Webb} Space Telescope}
\acrodef{IAD}{ion-assisted deposition}
\acrodef{ICGC}{intracluster globular cluster}
\acrodef{ICU}{instrument control unit}
\acrodef{ICL}{intra-cluster light}
\acrodef{ICM}{intra-cluster medium}
\acrodef{IGM}{intragalactic medium}
\acrodef{IMF}{initial mass function}
\acrodef{IPC}{inter-pixel capacitance}
\acrodef{LAM}{Laboratoire d'Astrophysique de Marseille}
\acrodef{LED}{light-emitting diode}
\acrodef{LF}{Luminosity function}
\acrodef{LSB}{low surface brightness}
\acrodef{LSST}{Legacy Survey of Space and Time}
\acrodef{MACC}{Multiple Accumulated}
\acrodef{MLI}{multi-layer insulation}
\acrodef{MMU}{Mass Memory Unit}
\acrodef{MPE}{Max-Planck-Institut für extraterrestrische Physik}
\acrodef{MPIA}{Max-Planck-Institut für Astronomie}
\acrodef{MW}{Milky Way}
\acrodef{NA}{numerical aperture}
\acrodef{NASA}{National Aeronautic and Space Administration}
\acrodef{JPL}{NASA Jet Propulsion Laboratory}
\acrodef{MZ-CGH}{multi-zonal computer-generated hologram}
\acrodef{NGVS}{Next Generation Virgo Survey}
\acrodef{NI-CU}{NISP calibration unit}
\acrodef{NI-OA}{near-infrared optical assembly}
\acrodef{NI-GWA}{NISP Grism Wheel Assembly}
\acrodef{NIR}{near-infrared}
\acrodef{NISP}{near-infrared spectrometer and photometer}
\acrodef{NSCs}{Nuclear star clusters}
\acrodef{PA}{position angle}
\acrodef{PARMS}{plasma-assisted reactive magnetron sputtering}
\acrodef{PLM}{payload module}
\acrodef{PTFE}{polytetrafluoroethylene}
\acrodef{PV}{performance verification}
\acrodef{PWM}{pulse-width modulation}
\acrodef{PSF}{point spread function}
\acrodef{QE}{quantum efficiency}
\acrodef{RGB}{red-green-blue}
\acrodef{RMS}{root mean square}
\acrodef{ROI}[RoI]{region of interest}
\acrodef{ROIC}{readout-integrated chip}
\acrodef{ROS}{reference observing sequence}
\acrodef{SBF}{surface brightness fluctuation}
\acrodef{SCA}{sensor chip array}
\acrodef{SCE}{sensor chip electronic}
\acrodef{SCS}{sensor chip system}
\acrodef{SED}{spectral energy distribution}
\acrodef{SDSS}{Sloan Digital Sky Survey}
\acrodef{SGS}{science ground segment}
\acrodef{SHS}{Shack-Hartmann sensor}
\acrodef{SMF}{stellar mass function}
\acrodef{SNR}[SNR]{signal-to-noise ratio}
\acrodef{SED}{spectral energy distribution}
\acrodef{SiC}{silicon carbide}
\acrodef{SVM}{service module}
\acrodef{UCDs}{ultra compact dwarfs}
\acrodef{UDGs}{ultra diffuse galaxies}
\acrodef{UNIONS}{Ultraviolet Near Infrared Optical Northern Survey}
\acrodef{VGC}{Virgo cluster catalogue}
\acrodef{VIS}{visible imager}
\acrodef{WD}{white dwarf}
\acrodef{WCS}{world coordinate system}
\acrodef{WFE}{wavefront error}
\acrodef{ZP}{zero point}

\title{\Euclid: Early Release Observations -- The surface brightness and colour profiles of the far outskirts of galaxies in the Perseus cluster\thanks{This paper is published on behalf of the Euclid Consortium}}

\newcommand{\orcid}[1]{} 	   
\author{M.~Mondelin\orcid{0009-0004-5954-0930}\thanks{\email{maelie.mondelin@cea.fr}}\inst{\ref{aff1}}
\and F.~Bournaud\inst{\ref{aff1}}
\and J.-C.~Cuillandre\orcid{0000-0002-3263-8645}\inst{\ref{aff1}}
\and S.~Codis\inst{\ref{aff1}}
\and C.~Stone\orcid{0000-0002-9086-6398}\inst{\ref{aff2},\ref{aff3},\ref{aff4}}
\and M.~Bolzonella\orcid{0000-0003-3278-4607}\inst{\ref{aff5}}
\and J.~G.~Sorce\orcid{0000-0002-2307-2432}\inst{\ref{aff6},\ref{aff7}}
\and M.~Kluge\orcid{0000-0002-9618-2552}\inst{\ref{aff8}}
\and N.~A.~Hatch\orcid{0000-0001-5600-0534}\inst{\ref{aff9}}
\and F.~R.~Marleau\orcid{0000-0002-1442-2947}\inst{\ref{aff10}}
\and M.~Schirmer\orcid{0000-0003-2568-9994}\inst{\ref{aff11}}
\and H.~Bouy\orcid{0000-0002-7084-487X}\inst{\ref{aff12},\ref{aff13}}
\and F.~Buitrago\orcid{0000-0002-2861-9812}\inst{\ref{aff14},\ref{aff15}}
\and C.~Tortora\orcid{0000-0001-7958-6531}\inst{\ref{aff16}}
\and L.~Quilley\orcid{0009-0008-8375-8605}\inst{\ref{aff17}}
\and K.~George\orcid{0000-0002-1734-8455}\inst{\ref{aff18}}
\and M.~Baes\orcid{0000-0002-3930-2757}\inst{\ref{aff19}}
\and T.~Saifollahi\orcid{0000-0002-9554-7660}\inst{\ref{aff20}}
\and P.~M.~Sanchez-Alarcon\orcid{0000-0002-6278-9233}\inst{\ref{aff21},\ref{aff22}}
\and J.~H.~Knapen\orcid{0000-0003-1643-0024}\inst{\ref{aff21},\ref{aff22}}
\and N.~Aghanim\orcid{0000-0002-6688-8992}\inst{\ref{aff7}}
\and A.~Amara\inst{\ref{aff23}}
\and S.~Andreon\orcid{0000-0002-2041-8784}\inst{\ref{aff24}}
\and C.~Baccigalupi\orcid{0000-0002-8211-1630}\inst{\ref{aff25},\ref{aff26},\ref{aff27},\ref{aff28}}
\and A.~Balestra\orcid{0000-0002-6967-261X}\inst{\ref{aff29}}
\and S.~Bardelli\orcid{0000-0002-8900-0298}\inst{\ref{aff5}}
\and P.~Battaglia\orcid{0000-0002-7337-5909}\inst{\ref{aff5}}
\and A.~Biviano\orcid{0000-0002-0857-0732}\inst{\ref{aff26},\ref{aff25}}
\and E.~Branchini\orcid{0000-0002-0808-6908}\inst{\ref{aff30},\ref{aff31},\ref{aff24}}
\and M.~Brescia\orcid{0000-0001-9506-5680}\inst{\ref{aff32},\ref{aff16}}
\and J.~Brinchmann\orcid{0000-0003-4359-8797}\inst{\ref{aff33},\ref{aff34}}
\and V.~Capobianco\orcid{0000-0002-3309-7692}\inst{\ref{aff35}}
\and C.~Carbone\orcid{0000-0003-0125-3563}\inst{\ref{aff36}}
\and M.~Castellano\orcid{0000-0001-9875-8263}\inst{\ref{aff37}}
\and G.~Castignani\orcid{0000-0001-6831-0687}\inst{\ref{aff5}}
\and S.~Cavuoti\orcid{0000-0002-3787-4196}\inst{\ref{aff16},\ref{aff38}}
\and A.~Cimatti\inst{\ref{aff39}}
\and G.~Congedo\orcid{0000-0003-2508-0046}\inst{\ref{aff40}}
\and C.~J.~Conselice\orcid{0000-0003-1949-7638}\inst{\ref{aff41}}
\and L.~Conversi\orcid{0000-0002-6710-8476}\inst{\ref{aff42},\ref{aff43}}
\and Y.~Copin\orcid{0000-0002-5317-7518}\inst{\ref{aff44}}
\and F.~Courbin\orcid{0000-0003-0758-6510}\inst{\ref{aff45},\ref{aff46}}
\and H.~M.~Courtois\orcid{0000-0003-0509-1776}\inst{\ref{aff47}}
\and M.~Cropper\orcid{0000-0003-4571-9468}\inst{\ref{aff48}}
\and G.~De~Lucia\orcid{0000-0002-6220-9104}\inst{\ref{aff26}}
\and X.~Dupac\inst{\ref{aff43}}
\and M.~Fabricius\orcid{0000-0002-7025-6058}\inst{\ref{aff8},\ref{aff18}}
\and M.~Farina\orcid{0000-0002-3089-7846}\inst{\ref{aff49}}
\and F.~Faustini\orcid{0000-0001-6274-5145}\inst{\ref{aff37},\ref{aff50}}
\and S.~Ferriol\inst{\ref{aff44}}
\and S.~Fotopoulou\orcid{0000-0002-9686-254X}\inst{\ref{aff51}}
\and B.~Gillis\orcid{0000-0002-4478-1270}\inst{\ref{aff40}}
\and C.~Giocoli\orcid{0000-0002-9590-7961}\inst{\ref{aff5},\ref{aff52}}
\and F.~Grupp\inst{\ref{aff8},\ref{aff18}}
\and S.~V.~H.~Haugan\orcid{0000-0001-9648-7260}\inst{\ref{aff53}}
\and W.~Holmes\inst{\ref{aff54}}
\and F.~Hormuth\inst{\ref{aff55}}
\and A.~Hornstrup\orcid{0000-0002-3363-0936}\inst{\ref{aff56},\ref{aff57}}
\and K.~Jahnke\orcid{0000-0003-3804-2137}\inst{\ref{aff11}}
\and M.~Jhabvala\inst{\ref{aff58}}
\and E.~Keih\"anen\orcid{0000-0003-1804-7715}\inst{\ref{aff59}}
\and S.~Kermiche\orcid{0000-0002-0302-5735}\inst{\ref{aff60}}
\and M.~Kilbinger\orcid{0000-0001-9513-7138}\inst{\ref{aff1}}
\and B.~Kubik\orcid{0009-0006-5823-4880}\inst{\ref{aff44}}
\and M.~K\"ummel\orcid{0000-0003-2791-2117}\inst{\ref{aff18}}
\and H.~Kurki-Suonio\orcid{0000-0002-4618-3063}\inst{\ref{aff61},\ref{aff62}}
\and A.~M.~C.~Le~Brun\orcid{0000-0002-0936-4594}\inst{\ref{aff63}}
\and S.~Ligori\orcid{0000-0003-4172-4606}\inst{\ref{aff35}}
\and P.~B.~Lilje\orcid{0000-0003-4324-7794}\inst{\ref{aff53}}
\and V.~Lindholm\orcid{0000-0003-2317-5471}\inst{\ref{aff61},\ref{aff62}}
\and I.~Lloro\orcid{0000-0001-5966-1434}\inst{\ref{aff64}}
\and D.~Maino\inst{\ref{aff65},\ref{aff36},\ref{aff66}}
\and E.~Maiorano\orcid{0000-0003-2593-4355}\inst{\ref{aff5}}
\and O.~Mansutti\orcid{0000-0001-5758-4658}\inst{\ref{aff26}}
\and S.~Marcin\inst{\ref{aff67}}
\and O.~Marggraf\orcid{0000-0001-7242-3852}\inst{\ref{aff68}}
\and M.~Martinelli\orcid{0000-0002-6943-7732}\inst{\ref{aff37},\ref{aff69}}
\and E.~Medinaceli\orcid{0000-0002-4040-7783}\inst{\ref{aff5}}
\and Y.~Mellier\inst{\ref{aff70},\ref{aff71}}
\and E.~Merlin\orcid{0000-0001-6870-8900}\inst{\ref{aff37}}
\and G.~Meylan\inst{\ref{aff72}}
\and L.~Moscardini\orcid{0000-0002-3473-6716}\inst{\ref{aff73},\ref{aff5},\ref{aff52}}
\and S.-M.~Niemi\orcid{0009-0005-0247-0086}\inst{\ref{aff74}}
\and C.~Padilla\orcid{0000-0001-7951-0166}\inst{\ref{aff75}}
\and F.~Pasian\orcid{0000-0002-4869-3227}\inst{\ref{aff26}}
\and K.~Pedersen\inst{\ref{aff76}}
\and W.~J.~Percival\orcid{0000-0002-0644-5727}\inst{\ref{aff77},\ref{aff78},\ref{aff79}}
\and V.~Pettorino\inst{\ref{aff74}}
\and S.~Pires\orcid{0000-0002-0249-2104}\inst{\ref{aff1}}
\and M.~Poncet\inst{\ref{aff80}}
\and L.~A.~Popa\inst{\ref{aff81}}
\and L.~Pozzetti\orcid{0000-0001-7085-0412}\inst{\ref{aff5}}
\and A.~Renzi\orcid{0000-0001-9856-1970}\inst{\ref{aff82},\ref{aff83}}
\and G.~Riccio\inst{\ref{aff16}}
\and E.~Romelli\orcid{0000-0003-3069-9222}\inst{\ref{aff26}}
\and R.~Saglia\orcid{0000-0003-0378-7032}\inst{\ref{aff18},\ref{aff8}}
\and P.~Schneider\orcid{0000-0001-8561-2679}\inst{\ref{aff68}}
\and A.~Secroun\orcid{0000-0003-0505-3710}\inst{\ref{aff60}}
\and S.~Serrano\orcid{0000-0002-0211-2861}\inst{\ref{aff84},\ref{aff85},\ref{aff86}}
\and C.~Sirignano\orcid{0000-0002-0995-7146}\inst{\ref{aff82},\ref{aff83}}
\and J.~Steinwagner\orcid{0000-0001-7443-1047}\inst{\ref{aff8}}
\and I.~Tereno\orcid{0000-0002-4537-6218}\inst{\ref{aff87},\ref{aff15}}
\and R.~Toledo-Moreo\orcid{0000-0002-2997-4859}\inst{\ref{aff88}}
\and F.~Torradeflot\orcid{0000-0003-1160-1517}\inst{\ref{aff89},\ref{aff90}}
\and I.~Tutusaus\orcid{0000-0002-3199-0399}\inst{\ref{aff91}}
\and L.~Valenziano\orcid{0000-0002-1170-0104}\inst{\ref{aff5},\ref{aff92}}
\and T.~Vassallo\orcid{0000-0001-6512-6358}\inst{\ref{aff18},\ref{aff26}}
\and Y.~Wang\orcid{0000-0002-4749-2984}\inst{\ref{aff93}}
\and J.~Weller\orcid{0000-0002-8282-2010}\inst{\ref{aff18},\ref{aff8}}
\and F.~M.~Zerbi\inst{\ref{aff24}}
\and E.~Zucca\orcid{0000-0002-5845-8132}\inst{\ref{aff5}}
\and C.~Burigana\orcid{0000-0002-3005-5796}\inst{\ref{aff94},\ref{aff92}}
\and V.~Scottez\inst{\ref{aff70},\ref{aff95}}}
	
\institute{Universit\'e Paris-Saclay, Universit\'e Paris Cit\'e, CEA, CNRS, AIM, 91191, Gif-sur-Yvette, France\label{aff1}
\and
Department of Physics, Universit\'{e} de Montr\'{e}al, 2900 Edouard Montpetit Blvd, Montr\'{e}al, Qu\'{e}bec H3T 1J4, Canada\label{aff2}
\and
Ciela Institute - Montr{\'e}al Institute for Astrophysical Data Analysis and Machine Learning, Montr{\'e}al, Qu{\'e}bec, Canada\label{aff3}
\and
Mila - Qu{\'e}bec Artificial Intelligence Institute, Montr{\'e}al, Qu{\'e}bec, Canada\label{aff4}
\and
INAF-Osservatorio di Astrofisica e Scienza dello Spazio di Bologna, Via Piero Gobetti 93/3, 40129 Bologna, Italy\label{aff5}
\and
Univ. Lille, CNRS, Centrale Lille, UMR 9189 CRIStAL, 59000 Lille, France\label{aff6}
\and
Universit\'e Paris-Saclay, CNRS, Institut d'astrophysique spatiale, 91405, Orsay, France\label{aff7}
\and
Max Planck Institute for Extraterrestrial Physics, Giessenbachstr. 1, 85748 Garching, Germany\label{aff8}
\and
School of Physics and Astronomy, University of Nottingham, University Park, Nottingham NG7 2RD, UK\label{aff9}
\and
Universit\"at Innsbruck, Institut f\"ur Astro- und Teilchenphysik, Technikerstr. 25/8, 6020 Innsbruck, Austria\label{aff10}
\and
Max-Planck-Institut f\"ur Astronomie, K\"onigstuhl 17, 69117 Heidelberg, Germany\label{aff11}
\and
Laboratoire d'Astrophysique de Bordeaux, CNRS and Universit\'e de Bordeaux, All\'ee Geoffroy St. Hilaire, 33165 Pessac, France\label{aff12}
\and
Institut universitaire de France (IUF), 1 rue Descartes, 75231 PARIS CEDEX 05, France\label{aff13}
\and
Departamento de F\'{i}sica Te\'{o}rica, At\'{o}mica y \'{O}ptica, Universidad de Valladolid, 47011 Valladolid, Spain\label{aff14}
\and
Instituto de Astrof\'isica e Ci\^encias do Espa\c{c}o, Faculdade de Ci\^encias, Universidade de Lisboa, Tapada da Ajuda, 1349-018 Lisboa, Portugal\label{aff15}
\and
INAF-Osservatorio Astronomico di Capodimonte, Via Moiariello 16, 80131 Napoli, Italy\label{aff16}
\and
Centre de Recherche Astrophysique de Lyon, UMR5574, CNRS, Universit\'e Claude Bernard Lyon 1, ENS de Lyon, 69230, Saint-Genis-Laval, France\label{aff17}
\and
Universit\"ats-Sternwarte M\"unchen, Fakult\"at f\"ur Physik, Ludwig-Maximilians-Universit\"at M\"unchen, Scheinerstrasse 1, 81679 M\"unchen, Germany\label{aff18}
\and
Sterrenkundig Observatorium, Universiteit Gent, Krijgslaan 281 S9, 9000 Gent, Belgium\label{aff19}
\and
Universit\'e de Strasbourg, CNRS, Observatoire astronomique de Strasbourg, UMR 7550, 67000 Strasbourg, France\label{aff20}
\and
Instituto de Astrof\'{\i}sica de Canarias, V\'{\i}a L\'actea, 38205 La Laguna, Tenerife, Spain\label{aff21}
\and
Universidad de La Laguna, Departamento de Astrof\'{\i}sica, 38206 La Laguna, Tenerife, Spain\label{aff22}
\and
School of Mathematics and Physics, University of Surrey, Guildford, Surrey, GU2 7XH, UK\label{aff23}
\and
INAF-Osservatorio Astronomico di Brera, Via Brera 28, 20122 Milano, Italy\label{aff24}
\and
IFPU, Institute for Fundamental Physics of the Universe, via Beirut 2, 34151 Trieste, Italy\label{aff25}
\and
INAF-Osservatorio Astronomico di Trieste, Via G. B. Tiepolo 11, 34143 Trieste, Italy\label{aff26}
\and
INFN, Sezione di Trieste, Via Valerio 2, 34127 Trieste TS, Italy\label{aff27}
\and
SISSA, International School for Advanced Studies, Via Bonomea 265, 34136 Trieste TS, Italy\label{aff28}
\and
INAF-Osservatorio Astronomico di Padova, Via dell'Osservatorio 5, 35122 Padova, Italy\label{aff29}
\and
Dipartimento di Fisica, Universit\`a di Genova, Via Dodecaneso 33, 16146, Genova, Italy\label{aff30}
\and
INFN-Sezione di Genova, Via Dodecaneso 33, 16146, Genova, Italy\label{aff31}
\and
Department of Physics "E. Pancini", University Federico II, Via Cinthia 6, 80126, Napoli, Italy\label{aff32}
\and
Instituto de Astrof\'isica e Ci\^encias do Espa\c{c}o, Universidade do Porto, CAUP, Rua das Estrelas, PT4150-762 Porto, Portugal\label{aff33}
\and
Faculdade de Ci\^encias da Universidade do Porto, Rua do Campo de Alegre, 4150-007 Porto, Portugal\label{aff34}
\and
INAF-Osservatorio Astrofisico di Torino, Via Osservatorio 20, 10025 Pino Torinese (TO), Italy\label{aff35}
\and
INAF-IASF Milano, Via Alfonso Corti 12, 20133 Milano, Italy\label{aff36}
\and
INAF-Osservatorio Astronomico di Roma, Via Frascati 33, 00078 Monteporzio Catone, Italy\label{aff37}
\and
INFN section of Naples, Via Cinthia 6, 80126, Napoli, Italy\label{aff38}
\and
Dipartimento di Fisica e Astronomia "Augusto Righi" - Alma Mater Studiorum Universit\`a di Bologna, Viale Berti Pichat 6/2, 40127 Bologna, Italy\label{aff39}
\and
Institute for Astronomy, University of Edinburgh, Royal Observatory, Blackford Hill, Edinburgh EH9 3HJ, UK\label{aff40}
\and
Jodrell Bank Centre for Astrophysics, Department of Physics and Astronomy, University of Manchester, Oxford Road, Manchester M13 9PL, UK\label{aff41}
\and
European Space Agency/ESRIN, Largo Galileo Galilei 1, 00044 Frascati, Roma, Italy\label{aff42}
\and
ESAC/ESA, Camino Bajo del Castillo, s/n., Urb. Villafranca del Castillo, 28692 Villanueva de la Ca\~nada, Madrid, Spain\label{aff43}
\and
Universit\'e Claude Bernard Lyon 1, CNRS/IN2P3, IP2I Lyon, UMR 5822, Villeurbanne, F-69100, France\label{aff44}
\and
Institut de Ci\`{e}ncies del Cosmos (ICCUB), Universitat de Barcelona (IEEC-UB), Mart\'{i} i Franqu\`{e}s 1, 08028 Barcelona, Spain\label{aff45}
\and
Instituci\'o Catalana de Recerca i Estudis Avan\c{c}ats (ICREA), Passeig de Llu\'{\i}s Companys 23, 08010 Barcelona, Spain\label{aff46}
\and
UCB Lyon 1, CNRS/IN2P3, IUF, IP2I Lyon, 4 rue Enrico Fermi, 69622 Villeurbanne, France\label{aff47}
\and
Mullard Space Science Laboratory, University College London, Holmbury St Mary, Dorking, Surrey RH5 6NT, UK\label{aff48}
\and
INAF-Istituto di Astrofisica e Planetologia Spaziali, via del Fosso del Cavaliere, 100, 00100 Roma, Italy\label{aff49}
\and
Space Science Data Center, Italian Space Agency, via del Politecnico snc, 00133 Roma, Italy\label{aff50}
\and
School of Physics, HH Wills Physics Laboratory, University of Bristol, Tyndall Avenue, Bristol, BS8 1TL, UK\label{aff51}
\and
INFN-Sezione di Bologna, Viale Berti Pichat 6/2, 40127 Bologna, Italy\label{aff52}
\and
Institute of Theoretical Astrophysics, University of Oslo, P.O. Box 1029 Blindern, 0315 Oslo, Norway\label{aff53}
\and
Jet Propulsion Laboratory, California Institute of Technology, 4800 Oak Grove Drive, Pasadena, CA, 91109, USA\label{aff54}
\and
Felix Hormuth Engineering, Goethestr. 17, 69181 Leimen, Germany\label{aff55}
\and
Technical University of Denmark, Elektrovej 327, 2800 Kgs. Lyngby, Denmark\label{aff56}
\and
Cosmic Dawn Center (DAWN), Denmark\label{aff57}
\and
NASA Goddard Space Flight Center, Greenbelt, MD 20771, USA\label{aff58}
\and
Department of Physics and Helsinki Institute of Physics, Gustaf H\"allstr\"omin katu 2, 00014 University of Helsinki, Finland\label{aff59}
\and
Aix-Marseille Universit\'e, CNRS/IN2P3, CPPM, Marseille, France\label{aff60}
\and
Department of Physics, P.O. Box 64, 00014 University of Helsinki, Finland\label{aff61}
\and
Helsinki Institute of Physics, Gustaf H{\"a}llstr{\"o}min katu 2, University of Helsinki, Helsinki, Finland\label{aff62}
\and
Laboratoire d'etude de l'Univers et des phenomenes eXtremes, Observatoire de Paris, Universit\'e PSL, Sorbonne Universit\'e, CNRS, 92190 Meudon, France\label{aff63}
\and
SKA Observatory, Jodrell Bank, Lower Withington, Macclesfield, Cheshire SK11 9FT, UK\label{aff64}
\and
Dipartimento di Fisica "Aldo Pontremoli", Universit\`a degli Studi di Milano, Via Celoria 16, 20133 Milano, Italy\label{aff65}
\and
INFN-Sezione di Milano, Via Celoria 16, 20133 Milano, Italy\label{aff66}
\and
University of Applied Sciences and Arts of Northwestern Switzerland, School of Computer Science, 5210 Windisch, Switzerland\label{aff67}
\and
Universit\"at Bonn, Argelander-Institut f\"ur Astronomie, Auf dem H\"ugel 71, 53121 Bonn, Germany\label{aff68}
\and
INFN-Sezione di Roma, Piazzale Aldo Moro, 2 - c/o Dipartimento di Fisica, Edificio G. Marconi, 00185 Roma, Italy\label{aff69}
\and
Institut d'Astrophysique de Paris, 98bis Boulevard Arago, 75014, Paris, France\label{aff70}
\and
Institut d'Astrophysique de Paris, UMR 7095, CNRS, and Sorbonne Universit\'e, 98 bis boulevard Arago, 75014 Paris, France\label{aff71}
\and
Institute of Physics, Laboratory of Astrophysics, Ecole Polytechnique F\'ed\'erale de Lausanne (EPFL), Observatoire de Sauverny, 1290 Versoix, Switzerland\label{aff72}
\and
Dipartimento di Fisica e Astronomia "Augusto Righi" - Alma Mater Studiorum Universit\`a di Bologna, via Piero Gobetti 93/2, 40129 Bologna, Italy\label{aff73}
\and
European Space Agency/ESTEC, Keplerlaan 1, 2201 AZ Noordwijk, The Netherlands\label{aff74}
\and
Institut de F\'{i}sica d'Altes Energies (IFAE), The Barcelona Institute of Science and Technology, Campus UAB, 08193 Bellaterra (Barcelona), Spain\label{aff75}
\and
DARK, Niels Bohr Institute, University of Copenhagen, Jagtvej 155, 2200 Copenhagen, Denmark\label{aff76}
\and
Waterloo Centre for Astrophysics, University of Waterloo, Waterloo, Ontario N2L 3G1, Canada\label{aff77}
\and
Department of Physics and Astronomy, University of Waterloo, Waterloo, Ontario N2L 3G1, Canada\label{aff78}
\and
Perimeter Institute for Theoretical Physics, Waterloo, Ontario N2L 2Y5, Canada\label{aff79}
\and
Centre National d'Etudes Spatiales -- Centre spatial de Toulouse, 18 avenue Edouard Belin, 31401 Toulouse Cedex 9, France\label{aff80}
\and
Institute of Space Science, Str. Atomistilor, nr. 409 M\u{a}gurele, Ilfov, 077125, Romania\label{aff81}
\and
Dipartimento di Fisica e Astronomia "G. Galilei", Universit\`a di Padova, Via Marzolo 8, 35131 Padova, Italy\label{aff82}
\and
INFN-Padova, Via Marzolo 8, 35131 Padova, Italy\label{aff83}
\and
Institut d'Estudis Espacials de Catalunya (IEEC),  Edifici RDIT, Campus UPC, 08860 Castelldefels, Barcelona, Spain\label{aff84}
\and
Satlantis, University Science Park, Sede Bld 48940, Leioa-Bilbao, Spain\label{aff85}
\and
Institute of Space Sciences (ICE, CSIC), Campus UAB, Carrer de Can Magrans, s/n, 08193 Barcelona, Spain\label{aff86}
\and
Departamento de F\'isica, Faculdade de Ci\^encias, Universidade de Lisboa, Edif\'icio C8, Campo Grande, PT1749-016 Lisboa, Portugal\label{aff87}
\and
Universidad Polit\'ecnica de Cartagena, Departamento de Electr\'onica y Tecnolog\'ia de Computadoras,  Plaza del Hospital 1, 30202 Cartagena, Spain\label{aff88}
\and
Port d'Informaci\'{o} Cient\'{i}fica, Campus UAB, C. Albareda s/n, 08193 Bellaterra (Barcelona), Spain\label{aff89}
\and
Centro de Investigaciones Energ\'eticas, Medioambientales y Tecnol\'ogicas (CIEMAT), Avenida Complutense 40, 28040 Madrid, Spain\label{aff90}
\and
Institut de Recherche en Astrophysique et Plan\'etologie (IRAP), Universit\'e de Toulouse, CNRS, UPS, CNES, 14 Av. Edouard Belin, 31400 Toulouse, France\label{aff91}
\and
INFN-Bologna, Via Irnerio 46, 40126 Bologna, Italy\label{aff92}
\and
Infrared Processing and Analysis Center, California Institute of Technology, Pasadena, CA 91125, USA\label{aff93}
\and
INAF, Istituto di Radioastronomia, Via Piero Gobetti 101, 40129 Bologna, Italy\label{aff94}
\and
ICL, Junia, Universit\'e Catholique de Lille, LITL, 59000 Lille, France\label{aff95}}    

\abstract{
The Perseus field captured by \Euclid as part of its Early Release Observations provides a unique opportunity to study cluster environment ranging from outskirts to dense regions. Leveraging unprecedented optical and near-infrared depths, we investigate the stellar structure of massive disc galaxies in this field.

This study focuses on outer disc profiles, including simple exponential (Type\,I), down-bending break (Type\,II) and up-bending break (Type\,III) profiles, and their associated colour gradients, to trace late assembly processes across various environments. Type\,II profiles, though relatively rare in high dense environments, appear stabilised by internal mechanisms like bars and resonances, even within dense cluster cores. Simulations suggest that in dense environments, Type\,II profiles tend to evolve into Type\,I profiles over time. Type\,III profiles often exhibit small colour gradients beyond the break, hinting at older stellar populations, potentially due to radial migration or accretion events.

We analyse correlations between galaxy mass, morphology, and profile types. Mass distributions show weak trends of decreasing mass from the centre to the outskirts of the Perseus cluster. Type\,III profiles become more prevalent, while Type\,I profiles decrease in lower-mass galaxies with cluster centric distance. Type\,I profiles dominate in spiral galaxies, while Type\,III profiles are more common in S0 galaxies. Type\,II profiles are consistently observed across all morphological types. While the limited sample size restricts statistical power, our findings shed light on the mechanisms shaping galaxy profiles in cluster environments. Future work should extend observations to the cluster outskirts to enhance statistical significance and explore looser environments. Additionally, 3D velocity maps are needed to achieve a non-projected view of galaxy positions, offering deeper insights into spatial distribution and dynamics.
}


\titlerunning{\Euclid: ERO -- The density and colour profiles of the far outskirts of galaxies in the Perseus cluster}
\authorrunning{Mondelin et al.}
   
\maketitle

\section{\label{sc:Intro}Introduction}

The study of the radial surface brightness distribution of galaxies enables decoding intricate patterns of light and the variations in brightness across different regions of galaxies, providing in turn crucial insights into their structure, formation, and evolution. By analysing these distributions, one can infer the presence of various components such as bulges, discs, and bars, and gain a deeper understanding of the underlying physical processes \citep{deVaucouleurs1948, Freeman1970, Sersic1973}.
\cite{deVaucouleurs1948} specifically describes galaxy discs as having declining exponential profiles. However, later observational results \citep{vanderKruit1979} have shown that these simple exponential profiles do not necessarily hold for some galaxies beyond a certain distance from the core. The development of deeper surveys, beyond a surface brightness of 25 $\mathrm{mag~arcsec^{-2}}$, has revealed that disc profiles can be classified into three categories \citep{PohlenTrujillo2006, ErwinPohlenBeckman2008}: Type\,I corresponds to the traditional decreasing exponential profile down to the twenty-fifth isophote or even beyond; Type\,II galaxies show a truncation and can be fitted by down-bending double exponentials; and Type\,III are up-bending break profiles, also known as up-bending double exponentials.

Studies on several hundreds of galaxies \citep{ErwinBeckmanPohlen2005, ErwinPohlenBeckman2008} have revealed that down- and up-bending breaks occur near a surface brightness of 25 $\mathrm{mag~arcsec^{-2}}$ in the visible spectrum, and are also present in the infrared profiles of galaxies \citep{MunozMateos2013}. It is important to clarify the terminology used in this context: some authors, such as \cite{BuitragoTrujillo2024}, refer to the outermost break in a galaxy’s surface brightness or mass profile (which coincides with the disc termination) as down-bending breaks or galaxy edges. However, in this paper, we use the term 'down-bending break' specifically to describe a bending in the galaxy surface brightness profile, regardless of its location relative to the disc edge.

Additionally, \cite{Gutierrez2011} showed that the proportion of each Type\,I in the optical $R$-band varies according to the morphology of galaxies. More than 80\% of late-type spiral galaxies have Type\,II profiles.. The fraction of these types changes gradually with the evolution of galaxies. For early-type galaxies, including lenticulars and early-type spirals, Type\,III profiles dominate at 50\%, with the remaining galaxies having profiles split between Type\,I and Type\,II.

Given these observations, the origin of down- and up-bending disc breaks has been questioned since their discovery. Down-bending disc breaks, meaning Type\,II profiles, can be explained by two main scenarios. The first scenario encompasses several explanations related to dynamical phenomena. The role of bars and their outer Lindblad resonance has been questioned. \cite{ElmegreenHunter2006} propose that the outer Lindblad resonance of bars may cause angular momentum redistribution, contributing to the formation of down-bending breaks.

Observations show that the link between the break radius and outer rings or outer lenses is undeniable, especially for early-type spirals or lenticulars \citep{Laine2016}. The formation of clumps at higher redshifts could also explain these down-bending disc breaks \citep{Si-YueYu2024}. The second scenario primarily applies to young spiral galaxies and proposes that down-bending disc breaks occur because the distribution of cold gas in the discs falls below a critical density threshold for star formation beyond the truncation radius \citep{Kennicutt1989, Schaye2004, ElmegreenHunter2006}.  The phenomenon of stellar migration also reveals that stars formed in the inner disc can migrate to the outermost regions of the disc, populating them with old red stars \citep{SellwoodBinney2002, Debattista2006, Roskar2008}. These down-bending disc breaks in the profiles are also observed in simulations \citep{Bournaud2007, Elmegreen2013, Struck2016}.

Regarding Type\,III profiles, meaning up-bending break profiles, the hypotheses are related to internal disc phenomena as well as external environmental factors. Observations of young star emission in the outer regions of discs suggest ongoing star formation, indicating the presence of cold gas in these regions \citep{GildePaz2005, Thilker2005, Tsvetkov2024}. In contrast, in the context of hierarchical evolution, spiral galaxies undergo mergers and interactions with neighbouring galaxies over their evolutionary history. This can lead to an excess of red stars in the outer regions. Simulations and observations, particularly in our own galaxy \citep{Toomre1972}, show that gravitational interactions, major and minor mergers can lead to the formation of tidal tails, shell, loops and disc perturbations. For major mergers, works such as \cite{Borlaff2014} indicate that an up-bending break profile could be produced after an accretion event, especially with a break at higher surface brightness. In many cases, the minor mergers have also an important role in the evolution of disc galaxies and result in episodes of dwarf galaxies accretion within a spiral galaxy. These satellites are gradually assimilated into the galactic halo and eventually into the outer disc \citep{Helmi1999, Helmi2018, Belokurov2018,Koppelman2019b,Myeong2019,Horta2021}. This process also disrupts the dynamical equilibrium of the disc, kinetically heating the disc stars and moving them to outer disc orbits, and even into the stellar halo \citep{NissenSchuster2010, Haywood2018}. Therefore, stars originating from both the remnants of satellite galaxies and the initial disc stars could be responsible for these up-bending disc breaks \citep{Younger2007, Eliche-Moral2011}, in other words, an excess of light beyond a certain radius. \textit{Gaia} observations of our Galaxy also confirm the presence of two disc components in the Galaxy, thin and thick, with different scale lengths. 

The role of the environment is therefore debated for Type\,III galaxies and for Type\,II, as the phenomenon of disc warping could be a source of truncation \citep{Roskar2008a, Sanchez2009}. Early studies on clusters reached ambiguous conclusions. A study of S0 galaxies in the Virgo cluster \citep{Erwin2012} showed that the cluster galaxies can be compared to field galaxies. First, the proportions of up-bending disc breaks within the cluster and outside are equal. Additionally, Type\,II down-bending break profiles are not present in this cluster. Thus, as galaxies approach the cluster, due to various effects such as ram-pressure stripping, strangulation and harassment, they may lose their truncation, becoming either Type\,I galaxies or, in some cases, initially Type\,I galaxies becoming up-bending break. However, this result is questioned by \cite{Laine2016} which suggests that Type\,II profiles are not absent from the Virgo cluster because of three down-bending break profiles on 24 Virgo cluster members in the sample. In addition, the Coma cluster as studied for instance by \cite{Head2015} challenges these previous results on S0 galaxies in clusters. According to their results, disc down-bending breaks are present in the cluster up to the core. This study also highlighted the role of bars in stabilising down- and up-bending disc breaks. Overall, the above-mentioned findings on two clusters, Coma and Virgo, reach different conclusions relative to the impact of environmental factors on disc profiles. Recent findings have found more perturbed galaxies in Type\,II and III than in Type\,I \citep{Sanchez2023}. This supports the formation scenario in which Type\,III discs are formed via interactions such as major mergers, and Type\,II discs stem from a star formation threshold \citep{Laine2016, Watkins2019, Pranger2017}. As of today, this question is consequently left unanswered but could be explored further by improving statistics, in the number of clusters observed, and the quality of observations.

The Perseus cluster, located at 72 Mpc from us, was previously covered by the {\it Hubble} Space Telescope at 30\% sky coverage through multiple stacks. The new \Euclid Release Observation -- Perseus programme \citep{pipelineERO,LF} now offers an unprecedented view of a 0.7 ${\rm deg}^{2}$ (1 Mpc$^2$) field at Perseus cluster distance, providing a groundbreaking perspective of the cluster. The combination of the high resolution of the visible imager \citep{EuclidSkyVIS,Mellier2024}, called VIS, and the infrared information from the \ac{NISP} is unprecedented over such a large field. This wide field allows for a comprehensive and controlled analysis with good statistical reliability. Note that observing down- and up-bending disc breaks in the cores of clusters has historically been challenging due to the dominance of intracluster light \citep{ICL}. 

Thanks to its remarkable depth and capacity to capture diffuse stellar halos \citep{pipelineERO}, \Euclid \citep{Mellier2024} reveals the outermost regions of galaxies, specifically areas beyond the location of the break for Type\,II or III. The measured luminosity profiles extend beyond 28 $\mathrm{mag~arcsec^{-2}}$ for most galaxies in the cluster in both the optical and the \ac{NIR}, enabling us to study the various components of the disc and the fainter parts of the galaxy in great detail. Additionally, the unique capabilities of the \ac{NISP} instrument \citep{EuclidSkyNISP} across three photometric bands provide valuable near-infrared information, which allows us to examine the distribution of older stars within the discs. This combination of deep field optical and near-infrared imaging offers a comprehensive view of both the young and old stellar populations, contributing significantly to our understanding of galaxy formation and evolution in cluster environments.

Within the framework of these \cite{EROcite}, we study the disc profiles of galaxies observed within the Perseus cluster, focusing mainly on S0 galaxies but also including some early-type spirals. This paper is organised as follows. In \cref{sc:method}, we will describe precise photometric extraction, followed by fitting different models to detect morphological characteristics. Subsequently, in \cref{sc:charectisation}, we will study the characteristics of down- and up-bending disc breaks, such as surface brightness and break radius, as well as the influence of mass and colour. In \cref{sc:Conclusions}, we will discuss the implications of our observations and conclude on possible processes of galaxy evolution in the Perseus cluster.

\section{Method}\label{sc:method}
\subsection{Data and sample selection}

\begin{figure}[htbp!] 
\centering
\includegraphics[width=0.49\textwidth]{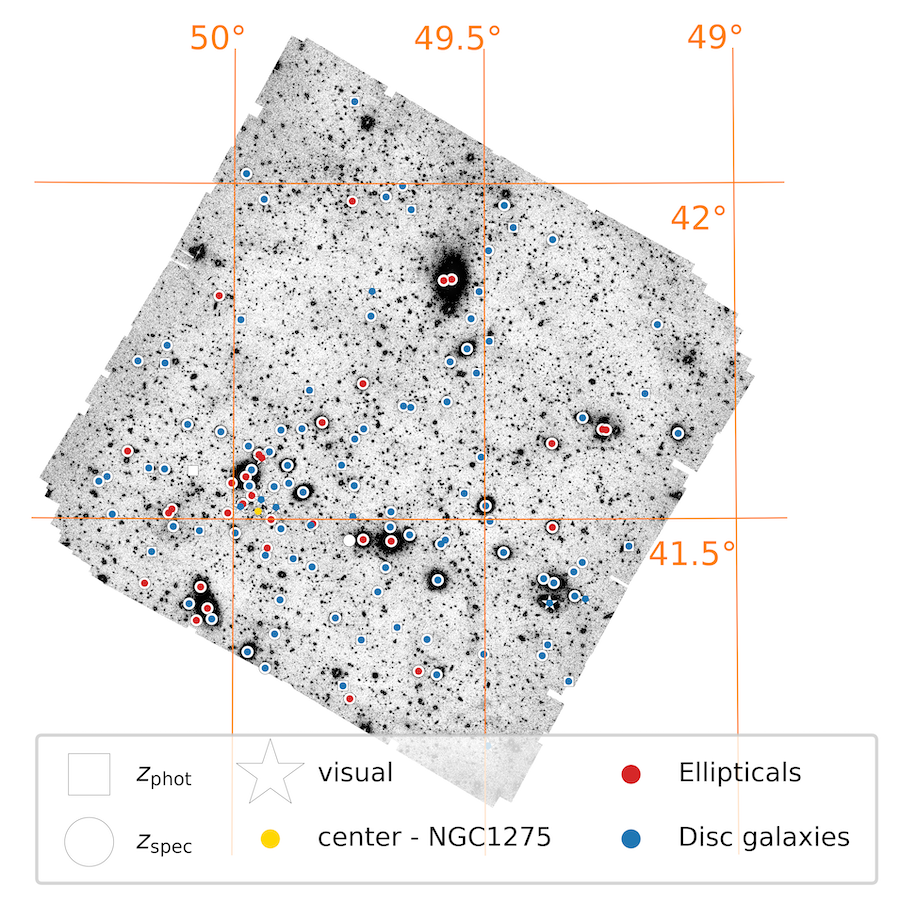}
 \caption{\IE image of the Perseus field of view after subtraction of the \ac{ICL}: blue dots indicate the position of each disc galaxy, red dots show the position of ellipticals. Orange lines shows the right ascension and the declination of \IE image. In the background, the method of identification for each galaxy is displayed, with visual markers distinguishing galaxies classified by photometric redshifts ($z_{\mathrm{phot}}$, square symbols) and spectroscopic redshifts ($z_{\mathrm{spec}}$, circular symbols). The yellow dot highlights the location of the cluster centre (NGC\,1275)} 
\label{fig:finalimage}
\end{figure}

\begin{figure*}[htbp!]
\includegraphics[width=1.\textwidth]{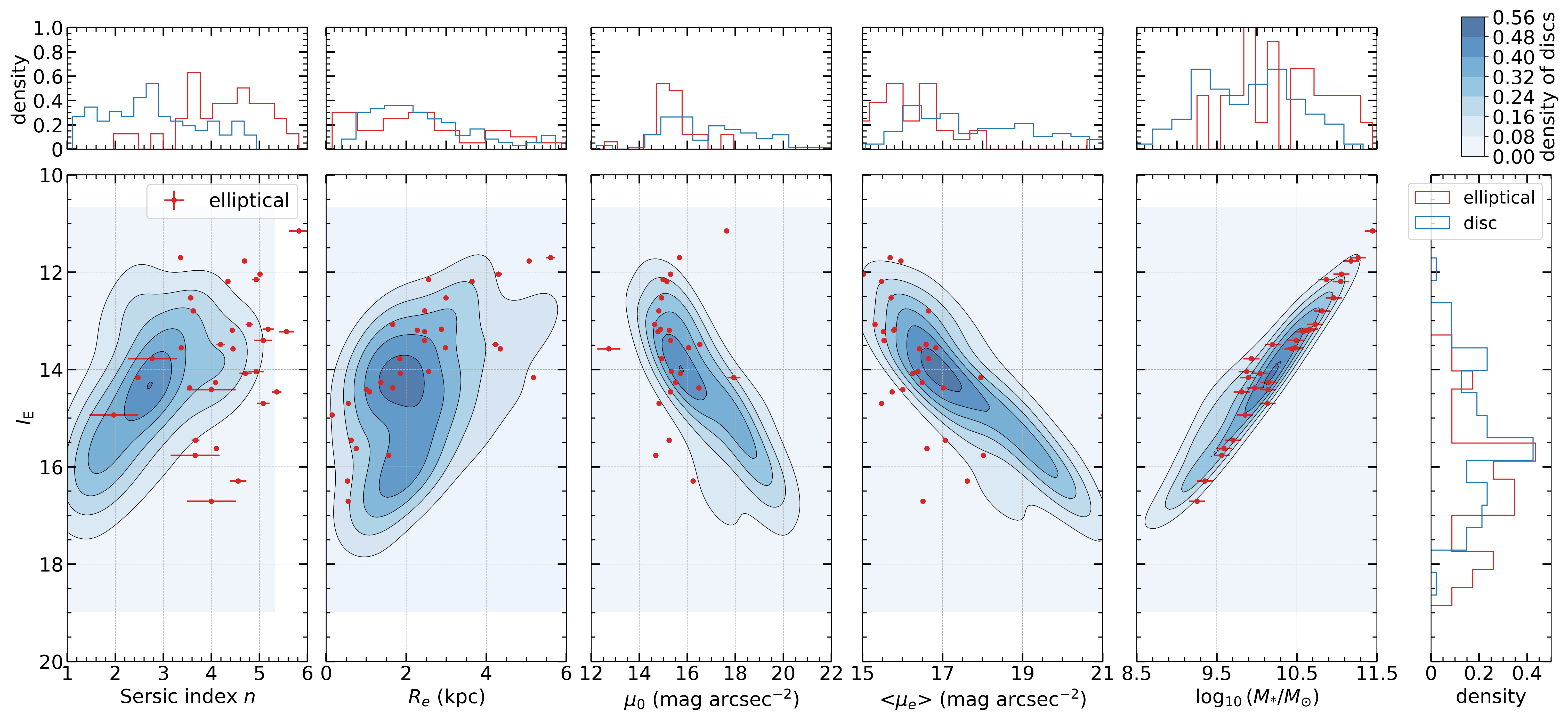}
\caption{Scaling relations between the S\'ersic index $n$, the effective radius $R_{\rm e}$, the central surface brightness $\mu_{0}$, mean effective surface brightness within $R_{\rm e}$, $\langle\mu_{\rm e}\rangle$, the mass $\log_{10}(M_{*}/M_{\odot})$ and the total magnitude \IE of galaxies measured using \texttt{AutoProf}/\texttt{AstroPhot}. The blue distribution shows the probability density function for the 102 cluster member bright disc galaxies while red dots are for cluster member bright ellipticals. The surface brightness is given in mag arcsec$^{-2}$. The panels on the top and side provide the normalised histograms of the parameters for discs (in blue) and for ellipticals (in red).} 
\label{fig:scalingrelations}
\end{figure*}

Galaxies in the \Euclid \ac{ERO} Perseus cluster were identified in \cite{LF} and \cite{Dwarfs}. The brightest galaxies were assigned to the cluster based primarily on available spectroscopic redshifts or alternatively photometric redshifts. For fainter or smaller galaxies, visual inspection of the images was also used to confirm their cluster membership. As described in \cite{LF}, 136 bright galaxies are members of the cluster within the 0.7 ${\rm deg}^{2}$ \ac{ERO} Perseus field. Figure \ref{fig:finalimage} shows the positions of the disc and elliptical galaxies overlaid on the \ac{LSB} image of the cluster in \IE. Note that isophotal models for NGC\,1275 (including the surrounding intracluster light) and NGC\,1272 were subtracted,, as well as the interstellar medium (using the Wise 12µm map), as detailed in \cite{ICL}. The influence of ICL on the classification and surface brightness profiles of these galaxies is explored in detail in \cref{appendix:ICL}. This appendix shows how important it is to ensure that down- or up-bending disc breaks are not overlooked. We show the scaling relations in Fig. \ref{fig:scalingrelations} -- which illustrates key structural parameters (Sérsic index $n$, effective radius $R_{\rm e}$, central surface brightness $\mu_{0}$, and mean effective surface brightness $\langle\mu_{\rm e}\rangle$) for disc and elliptical galaxies extracted from \cite{LF}. We identify 102 out of 136 galaxies as disc systems. We show the  distributions of disc galaxies in blue, while red points represent ellipticals, with normalised histograms offering a comparative view of each type’s parameter distributions. Details of this classification by morphology are given in \cref{sc:models}. 

\subsection{Gravitational interactions in the sample} \label{sc:interactions}

After selecting the large galaxies in the Perseus cluster, galaxies which are not considered as dwarfs in \cite{LF}, we inspect each one for signs of interaction in order to place them in the overall context of the cluster. The Perseus cluster, often perceived as a graveyard of evolved galaxies, is in fact a complex and dynamic environment where gravitational interactions, ram pressure and tidal forces play a crucial role in galaxy evolution. This framework enables us to better interpret the perturbations observed in the outer regions of these discy galaxies.
Multi-wavelength studies, including X--ray and radio observations \citep{vanWeeren2024}, have revealed numerous signs of past and present activity in this cluster. For example, the X-ray centre of the cluster \citep{Xstudy} is off-centred with respect to the gravitational centre of mass, an observation corroborated by the mis-centreing of the \ac{ICL} and globular clusters \citep{ICL}. This context highlights the diversity of environmental interactions affecting galaxies in clusters, particularly in the Perseus cluster.

Here, we observe at high resolution in \IE images that approximately 20\% exhibit significant signs of disturbances in their outer isophotes. When considering possible minor mergers/satellite galaxies, this rate rises to 50\%. It should be noted that these rates of disturbed galaxies are likely even higher as they are immersed in a sea of dwarf galaxies, the \ac{ICL} and the cluster's gravitational potential well.

\begin{figure*}[htbp!]
\centering
\includegraphics[width=1.\textwidth]{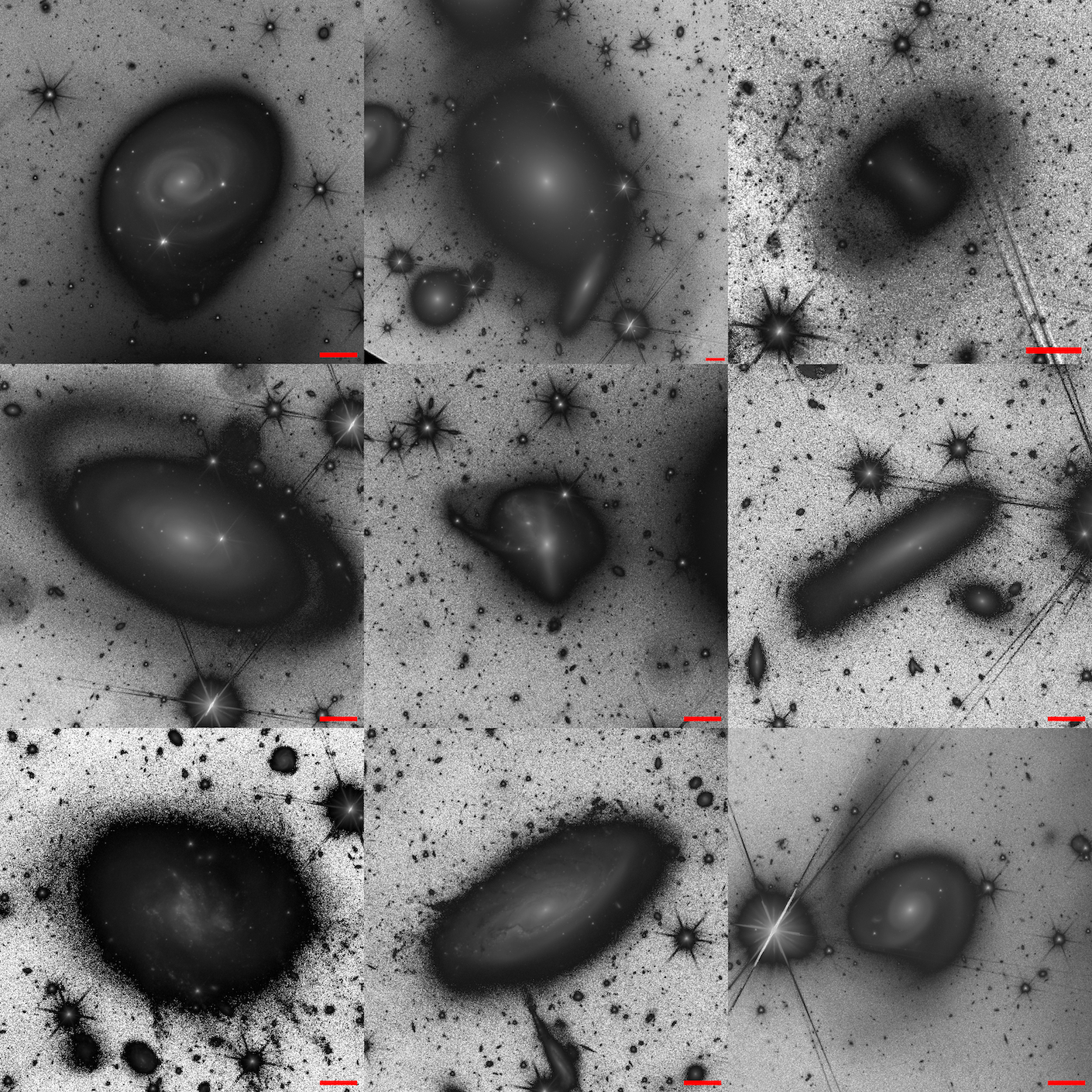}
\caption{Examples of different types of interactions observed in the Perseus cluster. Each panel shows a \ac{LSB} \IE image with high contrast of a galactic interaction within the cluster. A red line in the bottom-right corner representing a scale of 10\arcsec is provided at the bottom of each panel.
     {\it Top left}: NGC\,1268 -- a galaxy with a smooth, elongated shape, likely experiencing a close encounter with a neighbouring galaxy, causing mild distortion in its outer regions.
     {\it Top centre}: NGC\,1282 -- an interacting galaxy with a faint halo, possibly stripped due to gravitational forces from nearby massive galaxies.
     {\it Top right}: GALEXASC J031939.68+413105.6 -- a disrupted galaxy showing two tidal rings, suggesting a recent interaction or minor merger with another galaxy.
     {\it Middle left}: PGC\,012221 -- a galaxy with a clear spiral structure that appears distorted, possibly due to tidal forces.
     {\it Middle centre}: PGC\,012358 -- a major merger with an asymmetrical shape and tidal tails, showing evidence of material being pulled away.
     {\it Middle right}: PGC\,012520 -- an elongated galaxy with an asymmetric stretched halo, suggesting ongoing gravitational interactions or stripping by the cluster's dense environment.
     {\it Bottom left/centre}: MCG+07-07-070 and UGC\,02665 -- galaxies possibly affected by ram pressure stripping due to their motion through the intracluster medium. MCG+07-07-070 shows an asymmetric diffuse halo extending towards the lower right, while UGC\,02665 displays an umbrella-like morphology, both consistent with ram-pressure stripping (George et al. in prep).
     {\it Bottom right}: WISEA\,J032020.96+41225.4 -- a galaxy interacting with a larger galaxy, showing faint tidal features, which may indicate gravitational influence from a nearby massive galaxy.}
\label{fig:interactions}
\end{figure*}

This general framework lays the foundations for our photometric analysis, where we seek to characterise the perturbations of galaxy outer discs. Figure \ref{fig:interactions} illustrates the diversity of interactions revealed in \IE band \ac{LSB} images, with processes such as mergers and dynamic pressure clearly influencing morphologies. Note that in the \ac{RGB} image -- Fig. 1 in \cite{LF} -- streams and various outer isophotes marked by rather orange hues can also be observed. These features are highlighted using the \HE band image, which reveals diffuse and extended structures, often associated with stellar remnants from interacting galaxies. These red hues indicate older stellar populations, typical of the outer regions of post-interaction galaxies, where newly formed stars are absent, leaving a halo reddened by stellar ageing. 
Additionally, in galaxies undergoing ram pressure stripping, blue zones of star formation are visible in regions of the cluster within the gas stripped from the galaxy. The pockets of star-forming activity  at the faintest levels outside the galaxies are studied in George et al. (in prep.).

We also note that cosmological simulations, such as those in \cite{simu2012} and \cite{simu22012}, help improve our understanding of the detection of tidal debris and interaction remnants at different depths. For example, \cite{Mancillas2019} shows that at a limiting surface brightness of 29 $\mathrm{mag~arcsec^{-2}}$, only a portion of the tidal debris detectable at 33 $\mathrm{mag~arcsec^{-2}}$ becomes visible. Although these simulations often focus on a limited sample, they consistently reveal the persistence of tidal features and debris from different interaction epochs. This provides a realistic benchmark for our detection limits in ERO images, which reach down to 30 $\mathrm{mag~arcsec^{-2}}$ in the outer regions and around 27 $\mathrm{mag~arcsec^{-2}}$ in the cluster centre, where the \ac{ICL} strongly dominates the galaxy's flux \citep{ICL}.

With this big picture of the dynamical state of the Perseus cluster, we can now detail the method for photometrically extracting the luminosity profiles of the selected disc galaxies, focusing on quantifying the deformations and characteristics of the outer isophotes to better pinpoint the effects of the environment on these galactic structures.

\subsection{Extraction of surface brightness profiles}

In this study, we adopt the photometric extraction method detailed in \cite{LF} in order to extract the profiles. The performance of the VIS and \ac{NISP} cameras, as well the \ac{LSB} processing of the \ac{ERO} pipeline \citep{pipelineERO} have enabled unprecedented depth across this wide field. Additionally, the study of the \ac{ICL} in the Perseus cluster \citep{ICL} facilitates the subtraction of this diffuse flux from the original images, allowing for more precise photometry of the galaxies' outermost regions. The flux from the two central giant ellipticals, NGC\,1275 and NGC\,1272, is also subtracted from the \IE\ images to limit their contributions to neighbouring galaxies, based on the residual maps developed in \cite{ICL}. Since the surface brightness of the \ac{ICL} affects values around 27\,mag\,arcsec$^{-2}$, which is approximately 1 magnitude fainter than the range where profile breaks are typically observed (24–26\,mag\,arcsec$^{-2}$), we expected minimal impact on break detection. To confirm this, we compared galaxy profiles extracted from images with and without \ac{ICL} subtraction and verified that the presence of ICL does not significantly affect the detection of breaks.  We found that differences between the two cases are mainly notable for the most central galaxies. In these cases, the ICL subtraction step is crucial, as the galaxies are  located very close to the extended envelopes of the central ellipticals and deeply embedded in the diffuse intracluster light. At larger distances, beyond the median cluster-centric distance of our sample, discrepancies between profiles extracted with and without ICL subtraction appear only at very faint surface brightness levels, after 26\,mag\,arcsec$^{-2}$, and do not affect the identification of profile breaks, which occur at larger surface brightness levels. A detailed comparison illustrating this effect is provided in Appendix~\ref{appendix:ICL}.

From there, square tiles centred on each galaxy are then extracted. The size of each tile is chosen according to the angular size of the galaxy: 1\,{\rm k}\,$\times$1\,{\rm k} pixel (i.e. 1\arcminf7\,$\times$\,1\arcminf7), 2\,{\rm k}\,$\times$\,2\,{\rm k} pixel, (i.e 3\arcminf3\,$\,\times$\,3\arcminf3), or 4\,{\rm k}$\,\times$\,4\,{\rm k} pixel, (i.e. 6\arcminf7\,$\times$\,6\arcminf7) for the largest galaxies. After masking the bright stars and small galaxies near the main galaxies on each tile, the \texttt{AutoProf} tool \citep{ConnorAP} is used to extract the photometry of the individual galaxies. \texttt{AutoProf} is a pipeline that adjusts isophotes of varying semi-major axes around an isolated galaxy in an image to extract the galaxy's surface brightness, ellipticity, and PA as a function of the semi-major axis. Note that in our context, isolated refers to galaxies sufficiently separated from neighbours such that masking nearby objects is adequate for reliable profile extraction. The tool also provides estimates of different biases and noise levels. Due to the extended point-spread function (PSF) of high purity as detailed in \cite{pipelineERO}, the deconvolution of the profiles by the PSF is not performed here. Note, however, that the \cref{appendix:PSF} provides a rapid study of the influence of the extended PSF on model parameters. Instead, a simple point-source PSF model is used by \texttt{AutoProf} \citep{ConnorAP}. However, for some disc galaxies whose envelopes overlap with those of their neighbours: the \texttt{AstroPhot} tool \citep{ConnorAPh} is preferred for 15 out of the 102 disc galaxies in the sample, using 4\,{\rm k}$\,\times$ 4\,{\rm k} pixel tiles. \texttt{AstroPhot} is a tool designed to parameterise the surface brightness distribution of all galaxies and objects within an image. While \texttt{AstroPhot} could model all the galaxies within a 4\,{\rm k}$\,\times$ 4\,{\rm k} pixel tile, this approach requires significantly more computation time, compared to \texttt{AutoProf} -- approximately five times longer -- and the results are equivalent for isolated galaxies \citep{LF}.
We now detail in the two subsequent subsections how the profiles are obtained in practice from \texttt{AutoProf} and \texttt{AstroPhot}.

\subsubsection{\texttt{AutoProf} photometric profiles}

\begin{figure*}[htbp!]
\includegraphics[width=1.\textwidth]{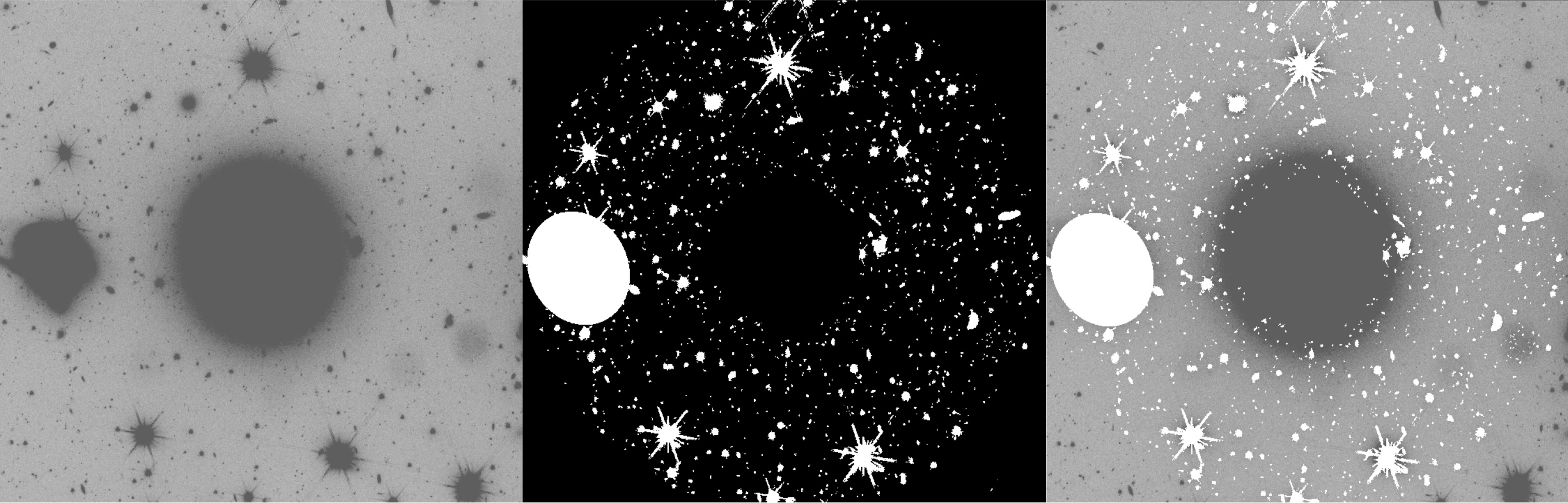}
\caption{Star masking process. {\it Left:} image of a 2\,{\rm k}$\,\times$ 2\,{\rm k} tile (i.e.  3\arcminf3$\,\times$\,3\arcminf3), centred on NGC\,1270. {\it Middle:} mask of stars and small background galaxies near NGC\,1270 in the image, shown in white. {\it Right:} original image overlaid with the mask in white, illustrating the regions excluded from subsequent profile extraction.}
\label{fig:mask}%
\end{figure*}

After extracting tiles corresponding to around 4 times the approximate apparent size of the galaxy, the next step is to mask the foreground stars that may contaminate the data. As detailed in \cite{LF}, the Perseus cluster is located close to the Galactic plane, and many bright stars overlap with the envelopes of the galaxies studied here. \texttt{SExtractor} \citep{Bertin}, as indicated in \cite{LF}, is used to detect the stars, particularly by utilising the star-galaxy separation parameter, supplemented with visual inspection to validate or add stars. The pixels corresponding to stars within a radius of 800 \IE pixels from the centre of the galaxy (80\arcsec), are masked using the separation parameter and a Fast Fourier Transform convolution of 10$\,\times$ 10 pixel in order to increase the mask. The star masks are sometimes adjusted based on visual inspection, especially when stars extend over multiple pixels due to saturation. Then, a mask for neighbouring main galaxies is created: the average ellipticity is obtained from the axis ratio previously provided in the catalogue from \cite{LF}, and a masking ellipse with a semi-major axis of five times the effective radius is applied on each neighbouring galaxies at the centre. Additionally, background galaxies identified by \texttt{SExtractor} detections are also masked, with visual validation ensuring that only relevant galaxies are obscured, thereby minimising data contamination. As an illustration, Fig. \ref{fig:mask} shows an example of masking for galaxy NGC\,1270.

\begin{figure*}[htbp!]
\centering
\includegraphics[width=1\textwidth]{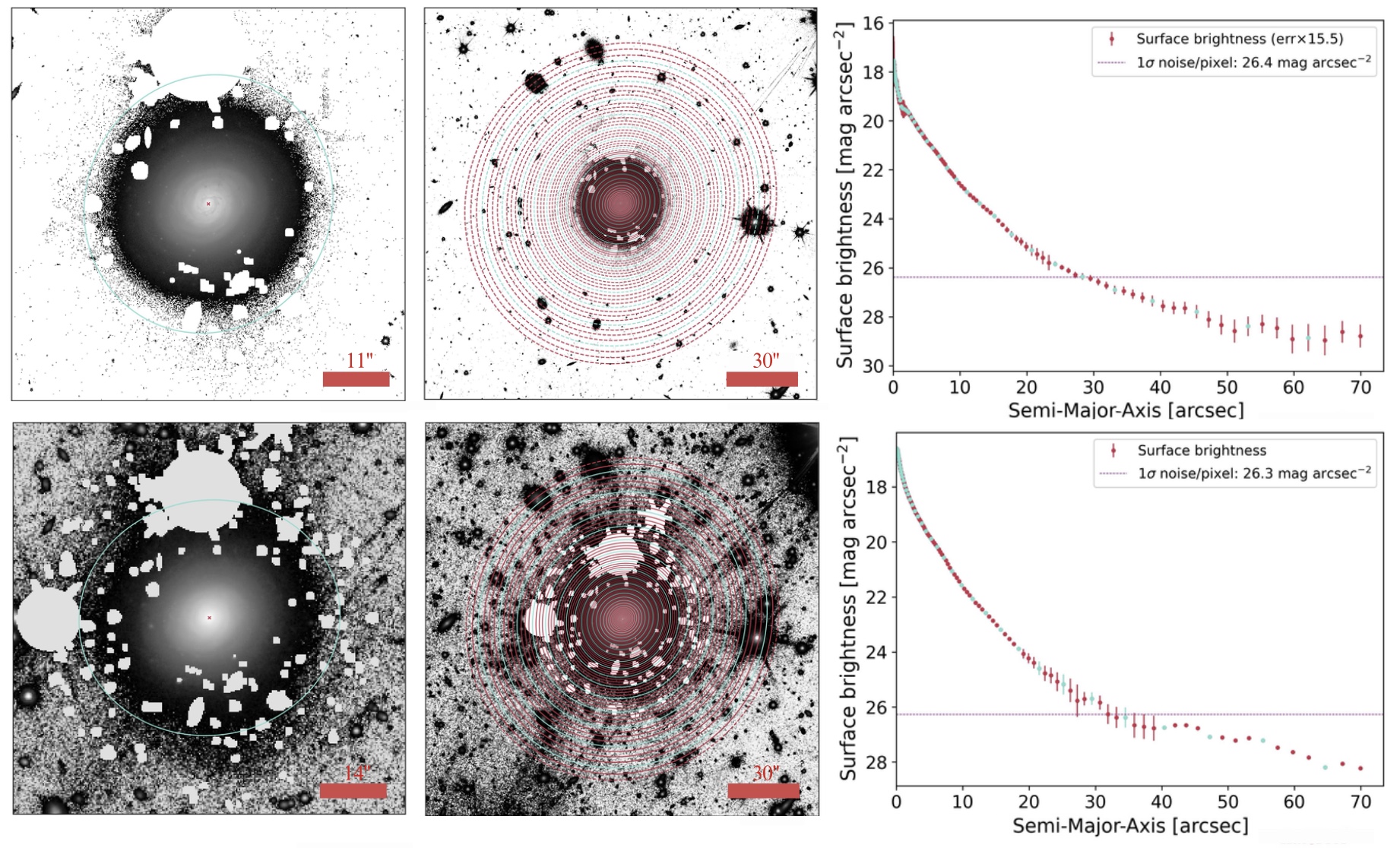}
\caption{Several steps of the \texttt{AutoProf} process for WISEA\,J031817.90+414031.0 for a \IE ({\it top}) and a \HE  ({\it bottom}) images. {\it Left:} Initialisation of the first ellipse in cyan around the central galaxy. {\it Middle:} Final isophote fitting. Using red and cyan colours enhances the visualisation and counting of isophotes \citep{ConnorAP} {\it Right:} Extraction of the radial surface brightness profile. Note that cyan points are drawn each four isophotes.} 
\label{fig:autoprof}
\end{figure*}

The masked image centred on a galaxy is then fed into the \texttt{AutoProf} run. Starting from a given position and using this image, a galaxy is fitted by the \texttt{AutoProf} pipeline. Several steps are involved in extracting the surface brightness profile of the galaxy. First, all bad pixels in the image are masked by the \texttt{AutoProf} procedure. A model of the background and then the PSF is derived to subtract a constant pedestal (the images being perfectly flat) starting from the given zero point, which is 30.132 for the ERO stacks. 
Note that the background subtraction is performed automatically, using the mode of the pixel flux distribution measured in the outermost regions of the image. This method, which applies a Gaussian-smoothing of the flux distribution to robustly estimate the sky background, is described in detail in \cite{ConnorAP} and ensures minimal sensitivity to contamination from bright sources, which is crucial for accurate surface brightness measurements at faint levels.
The centre of the image is also adjusted from an input centre. Using the masked image, elliptical isophotes are then initialised, fitted, and extracted to obtain the surface brightness profile. Radial isophotes are successively fitted until reaching the background level of the image. The pipeline outputs a 2D image showing the elliptical isophotes on the image, a 1D profile showing the surface brightness as a function of the semi-major axis of the isophotal ellipse, and a residual image. Figure \ref{fig:autoprof} illustrates some steps of the \texttt{AutoProf} process for extracting the surface brightness profile of our galaxies.

For the extraction of profiles in the \YE, \JE, and \HE bands, a method of forced photometry with \texttt{AutoProf} is applied, using the \IE profile as a reference. The principle remains the same as previously described, but it also leverages the solution from the \IE image to extract the isophotes \citep{ConnorAP}. Specifically, this approach fixes the centre, the position of the isophotes, as well as their orientation and ellipticity, based on the global isophote fit profile obtained from the \IE \texttt{AutoProf} process. Since the NISP resolution is three times lower than that of VIS (pixel scale of \IE band is equal to 0\arcsecf1 compared to pixel scale of \HE, \YE, \JE is 0\arcsecf3), this method benefits from the higher-resolution \IE photometry for consistent isophote geometry.

\subsubsection{\texttt{AstroPhot} photometric profiles}

In the context of these observations of the core of the Perseus cluster, some galaxies overlap with their neighbours along the line of sight. In such peculiar cases, \texttt{AutoProf} produces a combined profile for two galaxies. Therefore, \texttt{AstroPhot} -- which enables the comprehensive modelling of surface brightness profiles for all galaxies and objects within a large tile -- is clearly more suitable for the 15 galaxies we identified as problematic for \texttt{AutoProf}. In these cases, 4\,{\rm k}$\,\times$ 4\,{\rm k} tiles around them are extracted. A masking process, also using \texttt{SExtractor}, is performed to mask the foreground stars in these tiles. Each galaxy in the field is then initialised with a spline galaxy model up to a radius limit determined by visual inspection, corresponding to the point where the noise seems to blend with the galaxy's flux. The \ac{PA} and semi-major axis values are provided by the catalogue in \cite{LF}.

\begin{figure*}[htbp!]
\centering
\includegraphics[width=1.\textwidth]{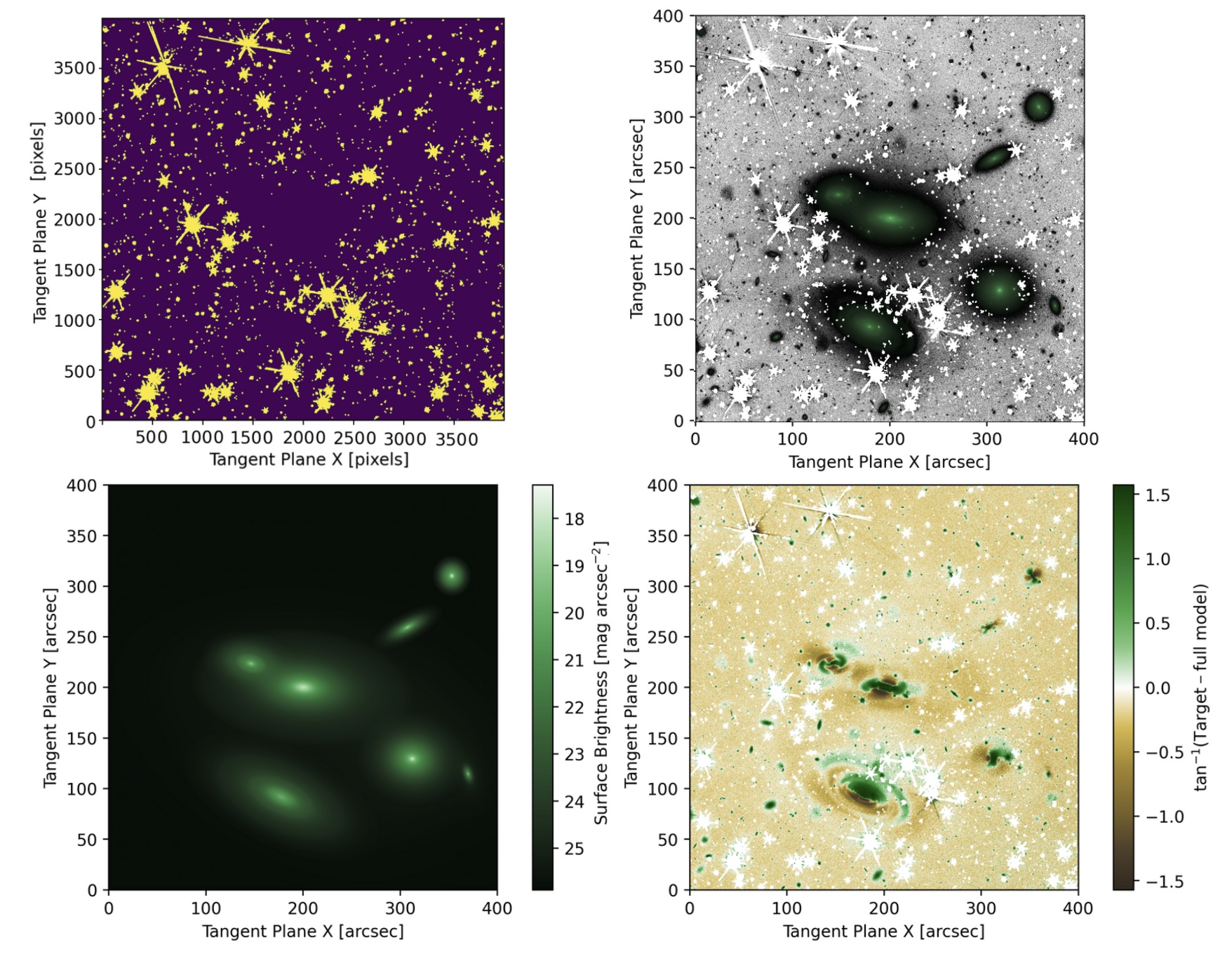}
\caption{Steps of the \texttt{AstroPhot} process for NGC\,1260 and its neighbouring galaxies for a \IE image.
{\it Top left:} Detection of stars (in yellow) from the \texttt{SExtractor} segmentation map. 
{\it Top right:} Masking of stars on the 4\,{\rm k}$\,\times
$ 4\,{\rm k} image centred on NGC\,1260. 
{\it Bottom left:} Final fitting of galaxies provided by \texttt{AstroPhot}, colour coded by the surface brightness value. 
{\it Bottom right:} Final residual map.
}

\label{fig:astrophot}
\end{figure*}

A spline radial light profile is interpolated for 50 points from the centre of the galaxy to the radius limit using a cubic spline interpolation of the stored brightness values. An initial model group, containing a model for each galaxy, is created. During the iterative fitting process, parameters such as the (\ac{PA}), ellipticity, and flux distribution are adjusted for each galaxy while keeping the radii of the isophotes fixed. 
Similar to the approach described in the \texttt{AutoProf} analysis, the sky background is explicitly modelled. In this case, a flat sky model is adopted, in which the background is assumed to be constant across the tile. The background level is treated as a free parameter in the fit, enabling a robust estimation that accounts for any residual large-scale background and ensures reliable surface brightness measurements in complex and crowded environments. This iterative approach ensures an optimal fit for each galaxy in the crowded field, resulting in a 2D profile and a residual image. Note that the 1D profile is also derived directly by \texttt{AstroPhot}. Figure \ref{fig:astrophot} illustrates the complete fitting process for these galaxies.

We note that the comparison of these photometric tools for isolated galaxies is discussed in \cite{LF} and shows a good consistency between the two methods, with the extracted profiles superimposed. Over a dozen galaxies tested, we observed a difference of less than 1\% on the radius measured at 25 $\mathrm{mag~arcsec^{-2}}$ (called $R_{\textrm{25}}$).

\subsection{Classification by profile type}

Each surface brightness profile extracted with \texttt{AutoProf} or \texttt{Astrophot} is then resampled with 300 points to obtain regularly spaced intervals in radius/semi-major axis along the surface brightness curve. This process involves applying a simple cubic interpolation and smoothing the curves. A careful visual inspection of the profiles was conducted to validate this step, ensuring the accuracy and consistency of the resampled data. Appendix \ref{appendix:meanprofiles} provides the complete set of profiles for both tools.

Moreover, before proceeding with the detailed analysis of the surface brightness profiles, the presence of a disc in each galaxy is first validated through a preliminary fitting process. According to \cite{LF}, 17 of the S0 galaxies from the initial catalogue have a S\'ersic index $n$ greater than 4, suggesting they are elliptical galaxies. However, after visual inspection of the galaxies and their profiles, this is attributed to two main phenomena: either the presence of a very bright and extended bulge or the presence of a bar (as suggested in \cite{Salo2015}, which similarly distorts the profile. We confirm the presence of a disc component in all cases by employing a more sophisticated modelling approach than the simple Sérsic profile. This advanced modelling extends to a surface brightness of 25 $\mathrm{mag~arcsec^{-2}}$ , surpassing the initial rough estimate provided by the Sérsic profile. As Quilley et al. (in prep.) point out, the simple Sérsic approach proves inadequate for galaxies where both bulge and disc significantly contribute to the overall structure, such as in lenticular and early-type spiral galaxies. The existence of a residual disc is clearly evident in the surface brightness profile beyond a certain radius. This is characterised by a distinctive slope of one in the profile, typically emerging at surface brightnesses between 22 and 24  $\mathrm{mag~arcsec^{-2}}$.

This involves fitting a simple exponential disc + bulge decomposition, using data up to magnitude 25,  as is traditionally done. The S\'ersic index \(n\) is fixed to four during the fitting to accurately model the bulge, while the disc is modelled with an exponential function. This initial fit allows us to confirm the presence of a disc structure by ensuring that the disc component dominates the light profile beyond the bulge region. The results from this preliminary fitting provide the basic parameters and ensure the reliability of the subsequent analysis. This validation is particularly important for the 17 galaxies that initially had a high value of S\'ersic indices for the fitting of a simple S\'ersic  model  up to \( R_{\textrm{25}} \).

From the radial surface brightness profiles of the galaxies, our goal is to extract the physical parameters of the discs and notably determine the presence of a break, indicative of a Type\,II or Type\,III profile. Note that the classification takes into account only breaks that occur between 22 and 27 $\mathrm{magarcsec^{-2}}$. The models adjust to account for the strongest break observed, although it's true that double breaks probably exist. For instance, in a preliminary analysis of two late Type,III galaxies, a small down-bending break around 21-22 $\mathrm{magarcsec^{-2}}$ was observed in their surface brightness profiles. However, these double breaks will not be discussed further in this paper. 

Different models are fitted to the profile using the method employed in several studies \citep{ErwinBeckmanPohlen2005, PohlenTrujillo2006, ErwinPohlenBeckman2008, MunozMateos2013, Laine2016}. A radial interval $[r_{\textrm{min}}, r_{\textrm{max}}]$ within which these models are applied is defined for each galaxy through visual inspection. The minimum radius roughly corresponds to the inner radius of the galaxy where the bulge is located, beyond which the disc flux begins to dominate. The profile is fitted by a de Vaucouleurs profile before $r_{\textrm{min}}$ and by a decreasing exponential in $[r_{\textrm{min}}, r_{\textrm{max}}]$. As suggested by several studies on bulge + disc decomposition of spiral galaxies \citep{Allen2006, Kim2016, Gao2019, Quilley2023}, allowing the S\'ersic parameter to vary freely between one and four enables a better characterisation of the bulge. However, our study focuses primarily on the characterisation of the discs. Taking an intermediate S\'ersic index like two risks blending the disc and the central bulge, which would contradict our intention to isolate the disc component. Therefore, a de Vaucouleurs profile seems sufficient as a first approximation, especially since it is particularly relevant for evolved galaxies with prominent bulges, such as those present in the Perseus cluster. 

Additionally, the maximum radius corresponds to the point where the flux reaches the background noise level estimated by the \texttt{AutoProf} pipeline. While the photometric zero point is at 30.132 for \IE and 30 for NISP \citep{pipelineERO}, this background level is estimated around  29 $\mathrm{mag~arcsec^{-2}}$, with punctual areas affected by structured galactic cirrus (well visible in this low galactic latitude field) limiting the detection at 28 $\mathrm{mag~arcsec^{-2}}$.

From the intensity profile defined as
\begin{equation}
\centering
   I(r) = 10^{\dfrac{ZP - \mu(r)}{2.5}},
\end{equation}
where \(ZP\) is the zero point and \(\mu(r)\) is the surface brightness at radius \(r\),
we apply different model functions namely i) a simple exponential profile defined as
\begin{equation}
I(r) = I_0 \exp\left(-\frac{r}{h_\textrm{d}}\right),
\end{equation}
where \(I(r)\) is the intensity at radius \(r\), \(I_0\) is the central intensity and \(h_\textrm{d}\) is the scale length of the disc;
and ii) a double exponential profile for Type\,II and Type\,III defined as
   \begin{equation}
   I(r) = S I_0 \exp\left({-\frac{r}{h_{\textrm{d1}}}}\right) \left\{1 + \exp[\alpha (r - R_{\textrm{break}})]\right\}^p,
   \label{eq:Ibreak}
   \end{equation}
   with $p = -\frac{1}{\alpha} \left(\frac{1}{h_{\textrm{d1}}} - \frac{1}{h_{\textrm{d2}}}\right)$ and $S^{-1} = 1 + \exp\left[-\alpha R_{\textrm{break}}\left(\frac{1}{h_{\textrm{d1}}} - \frac{1}{h_{\textrm{d2}}}\right)\right]$,\\
   where \(I_0\) is the central intensity, \(h_{\textrm{d1}}\) is the inner scale length, \(h_{\textrm{d2}}\) is the outer scale length, \(R_{\textrm{break}}\) is the break radius, and \(\alpha\) is a parameter that controls the sharpness of the transition between the two regions. This latter parameter is typically fixed to 0.5 \citep{ErwinPohlenBeckman2008, Laine2016}.

The values and uncertainties on the parameters are directly derived from the \texttt{curve\_fit} Python fitting function. These models can now help us to characterise the surface brightness profiles of the galaxies and identify the presence of breaks indicative of Type\,II or Type\,III profiles. 

For each model, the reduced chi-squared (\(\chi^2\)) value is calculated to determine the best fit. The model with the lowest value is considered to validate the visually observed down- or up-bending disc breaks. Note that around 10 galaxies have a break that is not really clear, and that the fitting procedure does not obtain a reduced chi-square that is very different from one model to another, so we decide to classify them as Type\,I. This analysis is performed for the \IE band, and the same procedure is applied to the NISP bands, using the forced photometry, as explained in \cite{LF}. This approach yields model parameters for each profile, which then allows us to characterise our sample. In Appendix \ref{appendix:exampletype}, Figs. \ref{fig:typeI} to \ref{fig:typeIII} present examples of the different profiles. We note a slight difference in the break position between the \IE and \HE profiles. However, this difference may not be significant as the r$_{\text{break}}$ values for each band fall within their respective error bars.

Additionally, to detect potential colour gradients, the radial colour (\IE$-$\,\HE) is calculated. In practice, \texttt{AutoProf} directly outputs the total magnitudes at each isophote radius. For \texttt{AstroPhot}, the surface brightness is integrated over each ellipse to obtain the total magnitude.

The distribution of disc galaxy types in our sample is summarised in Table \ref{tab:galaxy_types}. We clearly note the low proportion of Type\,II, which appears to be in agreement with \cite{Erwin2012}. However, unlike the study on the Virgo cluster, the proportion of down-bending break profile is not null if we consider all disc galaxies, not only the S0 galaxies. This result is therefore more consistent with those of the Coma cluster \citep{Head2015}. A more detailed discussion is provided in \cref{sc:Discussion}, where the proportion of Type\,III is also explored. Apart from the general classification of disc galaxies, another important structural feature to consider is the presence of bars. In our sample, the fraction of observed strong barred galaxies seems low (around 10\%) compared to \cite{Head2015} and \cite{Laine2016}, probably because we take into account only strong bars here in our identification. Indeed, these are the only ones confirmed by eye. These bars are often very long, of the order of the exponential scale of the disc, and remain visible as a stretched bump in the surface brightness profile up to $25\ \mathrm{mag}~\mathrm{arcsec}^{-2}$. In the remainder of this paper, we will explore the connection between the type of disc profiles and various indicators such as morphology, mass, and environment.

\begin{table}[htbp!]
\centering
\caption{Distribution of disc galaxy types in the sample}
\begin{tabular}{lccccc}
\hline
\noalign{\vskip 0.5pt}
Type & Number &  Percentage &  Bars & $\log_{10}(M_{*}/M_{\odot})$  \\
\hline
\noalign{\vskip 2pt}
I & 74 & (76$\pm$9) \% & 7 & 9.82$\pm$0.67 \\
II & 4 & (4$\pm$2)\% & 1 & 10.03$\pm$0.54\\
III & 24 & (20$\pm$4) \% & 0 & 10.03$\pm$0.57\\
Total & 102 & 100\% & 8 & 9.87$\pm$0.65\\
\hline
\end{tabular}
\label{tab:galaxy_types}
\end{table}

\subsection{Classification by morphology } \label{sc:models}

\begin{figure}
\includegraphics[width=\columnwidth]{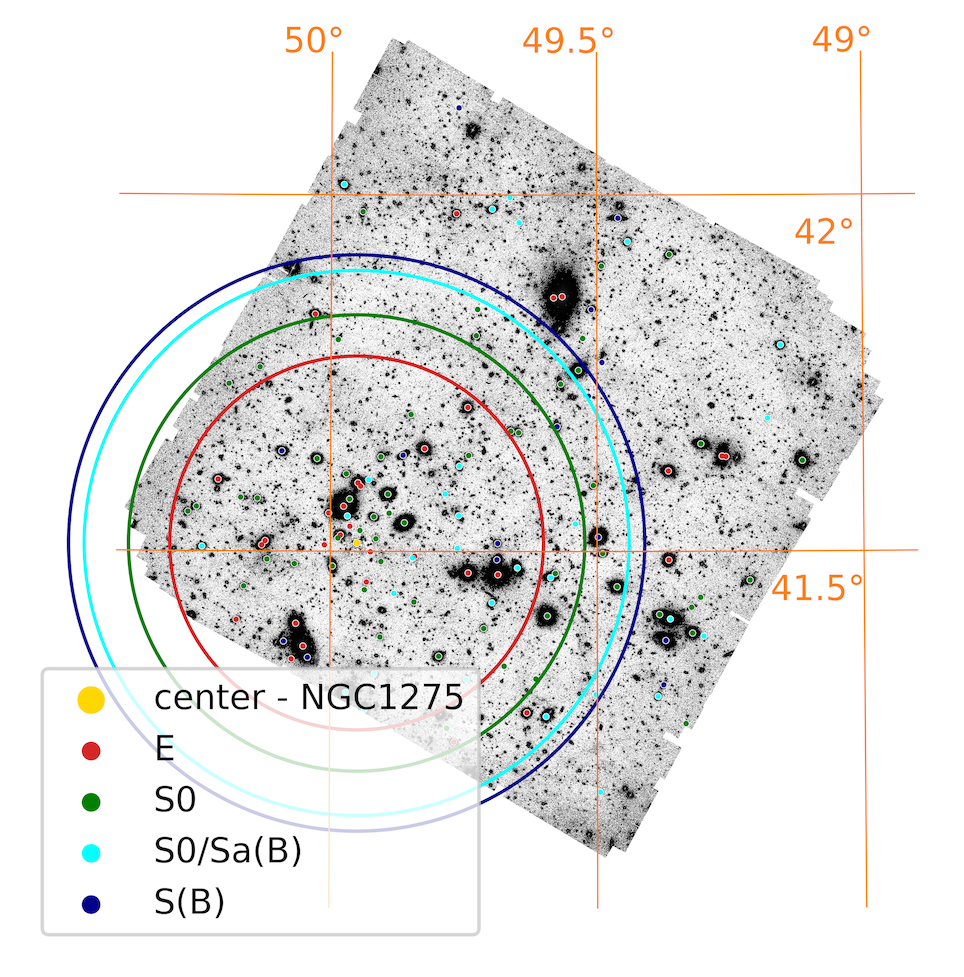}
\caption{Residual image in the \IE field of the Perseus cluster (after ICL subtracting): The centres of elliptical (E) galaxies are marked with red dots, S0 galaxies with green dots, intermediate types between S0 and spirals with cyan dots, and spirals with dark blue dots. Coloured circles, centred on the central elliptical galaxy NGC\,1275, indicate the average projected distance for each morphological type.
} 
\label{fig:VISmorpho}
\end{figure}

In order to study the influence of galaxy morphology on profile type, a classification into three subgroups among disc galaxies is performed. The first group consists of spiral galaxies, primarily corresponding to Sa/barred Sa galaxies according to the Hubble-de Vaucouleurs sequence, identified by the presence of observable spiral arms. Some less evolved galaxies are included in this category for our statistical study. The second group corresponds to S0 galaxies, distinguished by the absence of spiral structure. These galaxies typically exhibit less prominent discs and significant bulges, as indicated by their visual aspect and surface brightness profiles. This visual classification is validated by the NED reference catalogue\footnote{The NASA/IPAC Extragalactic Database (NED) is funded by the National Aeronautics and Space Administration and operated by the California Institute of Technology.} for most galaxies in these two groups. Finally, a third group includes galaxies whose classification between the two main groups is uncertain and for which no relevant information was found in the NED reference catalogue. Figure \ref{fig:VISmorpho} illustrates the spatial distribution of galaxies by morphological type within the Perseus cluster. Each galaxy is represented by a circle whose size reflects its morphological classification: larger circles correspond to earlier-type galaxies, while smaller circles represent later-type systems. A clear trend emerges, with early-type galaxies predominantly concentrated near the cluster centre around NGC\,1275, and late-type galaxies becoming more common at larger projected distances. This spatial segregation of morphologies is consistent with the well-established morphology–density relation observed in galaxy clusters \citep{Dressler1980, Postman1984, Whitmore1993, Fasano2015}.

In addition, Table \ref{tab:galaxy_morpho} shows the number of galaxies and the mean S\'ersic index for a single-S\'ersic model used in the catalogue from \cite{LF} for each morphological group. We note that the average S\'ersic index increases slightly with the morphological group, which is consistent with our classification. \cref{appendix:morphologicalclass} provides some plots validating this morphological classification.

\begin{table}[htbp!]
\centering
\caption{Distribution of disc galaxy morphology in our sample}
\begin{tabular}{lccc}
\hline
\noalign{\vskip 1pt}
 Parameter &  S/SB &  S0 -- S/SB &  S0/S0B \\
\hline
\noalign{\vskip 1pt}
Code morphology & 1 & 0 & $-$1 \\
Number of galaxies & 16 & 30 & 56 \\
Mean single-S\'ersic & 2.61$\pm$1.16 & 2.67$\pm$1.16 & 2.87$\pm$0.84 \\
Number of bars & 4 & 3 & 1 \\
\hline
\end{tabular}
\label{tab:galaxy_morpho}
\end{table}

\subsection{Influence of other parameters}

The sample was also characterised by its mass distribution, determined in \cite{LF}. The disc galaxies have masses in the range $\log_{10}({M_{*}}/{M_\odot})\in$  [7.7, 11.3] with a median value of $\log_{10}({M_{*}}/{M_\odot})$ = 9.92.

The link between profile type and galaxy mass is now studied through classification into three bins, each containing 34 disc galaxies. This arbitrary choice delivers meaningful statistics for each bin given the small number of objects.

Considering the fact that the role of the environment is highly debated in the literature, as indicated previously, we adopt in our study four distinct regions within the 0.7 deg² field of view of Perseus. They are defined in the RA/Dec plane using a kernel density estimation (KDE) plot, i.e. the \texttt{gaussian\_kde} function. This approach allows us to visualise the global distribution of cluster galaxies, considering bright and faint dwarf galaxies. Each galaxy counts as one in this KDE plot and is not weighted by mass in order to only take into account spatial distributions. For more details on the distribution of dwarfs, we refer the readers to \cite{Dwarfs}.  Analysing the density contours generated by the KDE plot enables a more precise characterisation of the spatial distribution of galaxies within the defined regions, as depicted in \cref{sc:generaltrend}.

\section{Characterisation of the disc breaks }\label{sc:charectisation}

Our global sample is thus characterised in terms of morphologies and profile types. This section now describes down- and up-bending break discs in more detail. Specifically, the parameters of the double exponential fitting are first characterised. Finally, the influence of the environment will be studied through other parameters such as mass and morphology.

\subsection{Parameters of the break}

\begin{figure*}[htbp!]
\centering
\includegraphics[width=1\textwidth]{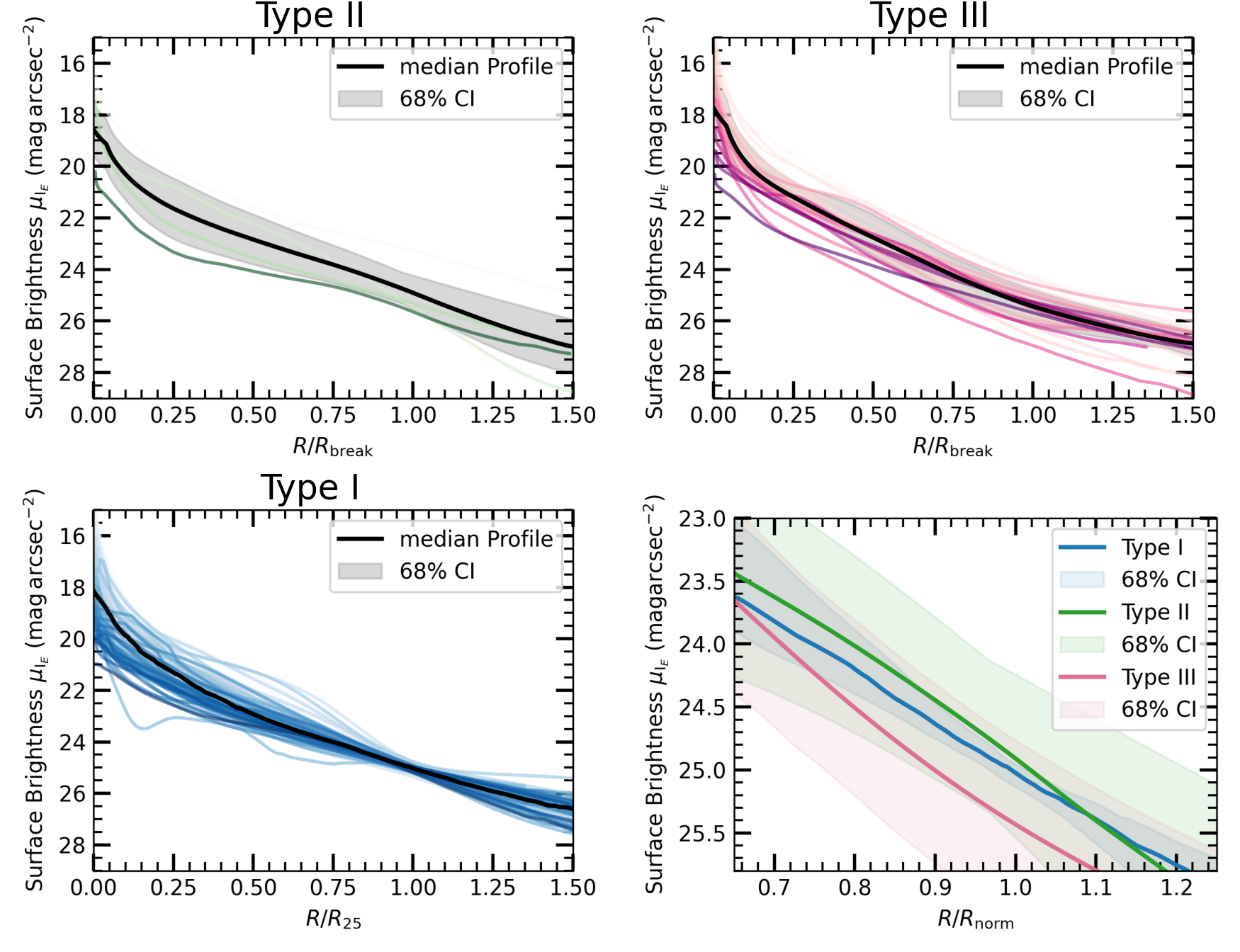}
\caption{Distribution of normalised surface brightness profiles for different types of galaxies. The subplots show the profiles for Type\,II (top left), Type\,III (top right), Type\,I (bottom left), and the combined median profiles for the three types around $R = R_{\textrm{norm}}$ (bottom right). Individual profiles are displayed with a colour gradient, while the median profile is represented in black. The shaded region around the median indicates the 68\% confidence interval, reflecting the variability among individual profiles. Note that $R_{\textrm{norm}}$ corresponds to $R_{\textrm{25}}$ for Type\,I or $R_\textrm{break}$ for Type\,II and \,III. 
}
\label{fig:meanprofiles}
\end{figure*}

Tables \ref{tab:parametersII} and \ref{tab:parametersIII} provide the mean and median parameters of \cref{eq:Ibreak} in our sample, both for the down-bending double, meaning Type\,II, and the up-bending, meaning Type\,III, double exponential profiles respectively. The uncertainties associated with the mean and median values correspond to the standard deviation within our sample. This characterisation allows for the identification of the break occurring for Type\,II and Type\,III profiles around magnitude 25, as hinted in \cite{Laine2016}. Type\,III profiles show a slightly higher average and median surface brightness at the break compared to Type\,II profiles. However, when considering the scatter, this difference is not statistically significant. In addition, the average and median break radius for both Type\,II and Type\,III profiles is approximately equal to the radius at magnitude 25 in the visible, which is consistent with previous results \citep{Laine2016}. 
Figure \ref{fig:meanprofiles} illustrates the normalised surface brightness profiles for the different profile types in our sample. To construct the median profiles, each individual galaxy profile was first normalised in radius: by the break radius $R_{\text{break}}$ for Type\,II and Type\,III galaxies, and by the isophotal radius $R_{25}$ for Type\,I galaxies. Each normalised profile was then interpolated onto a common radial grid spanning from 0 to 1.5 in units of the normalisation radius. It highlights the distinct characteristics of Type\,II, Type\,III, and Type\,I profiles, alongside a combined view of their median profiles. For Type\,II profiles (top left panel), the break is marked by a down-bending transition, consistent with break disc observed in previous studies. In contrast, Type\,III profiles (top right panel) display an up-bending transition, indicative of an extended outer disc structure. The bottom left panel shows Type\,I profiles, where no significant break is evident. The combined median profiles (bottom right panel) reinforce the typical trends for each profile type, with shaded regions indicating the variability among individual profiles within each category.

\begin{table}[htbp!]
\centering
\caption{Parameters of the double exponential model for the four identified down-bending disc breaks (Type\,II) in our sample
}
\begin{tabular}{lcccc}
\hline
\noalign{\vskip 1pt}
 Parameter &  Units &  mean &  median \\
\hline
\noalign{\vskip 1pt}
$\mu_0$ & $\mathrm{mag~arcsec^{-2}}$ & 21.1$\pm$1.3 & 21.2$\pm$1.7\\
$\mu_{\textrm{break}}$ & $\mathrm{mag~arcsec^{-2}}$ & 24.5$\pm$0.9 & 24.7$\pm$1.1 \\
$h_{\textrm{d1}}$ & $\mathrm{arcsec}$ & 8.8$\pm$1.5& 8.5$\pm$1.9 \\
$h_{\textrm{d2}}$ & $\mathrm{arcsec}$ & 5.9$\pm$1.8& 5.7$\pm$2.3 \\
$R_\textrm{break}$ & $\mathrm{arcsec}$ & 25.2$\pm$4.6& 24.2$\pm$5.8 \\
\hline
\end{tabular}
\label{tab:parametersII}
\end{table}

\begin{table}[htbp!]
\centering
\caption{Parameters of the double exponential model for the 24 identified up-bending disc breaks (Type\,III) in our sample}
\begin{tabular}{lcccc}
\hline
\noalign{\vskip 1pt}
 Parameter &  Units &  mean  &  median\\
\hline
\noalign{\vskip 1pt}
$\mu_0$ & $\mathrm{mag~arcsec^{-2}}$ & 19.8$\pm$0.9 & 19.7$\pm$1.1 \\
$\mu_\textrm{break}$ &  $\mathrm{mag~arcsec^{-2}}$ & 25.3$\pm$0.7 & 25.2$\pm$0.9 \\
$h_{\textrm{d1}}$ & $\mathrm{arcsec}$ & 5.4$\pm$2.4 & 5.2$\pm$3.0 \\
$h_{\textrm{d2}}$ & $\mathrm{arcsec}$ & 13.8$\pm$8.8 & 10.5$\pm$10.0 \\
$R_\textrm{break}$ & $\mathrm{arcsec}$ & 27.8$\pm$12.2 & 26.3$\pm$15.2\\
\hline
\end{tabular}
\label{tab:parametersIII}
\end{table}

\begin{figure}[htbp!]
\includegraphics[width=0.5\textwidth]{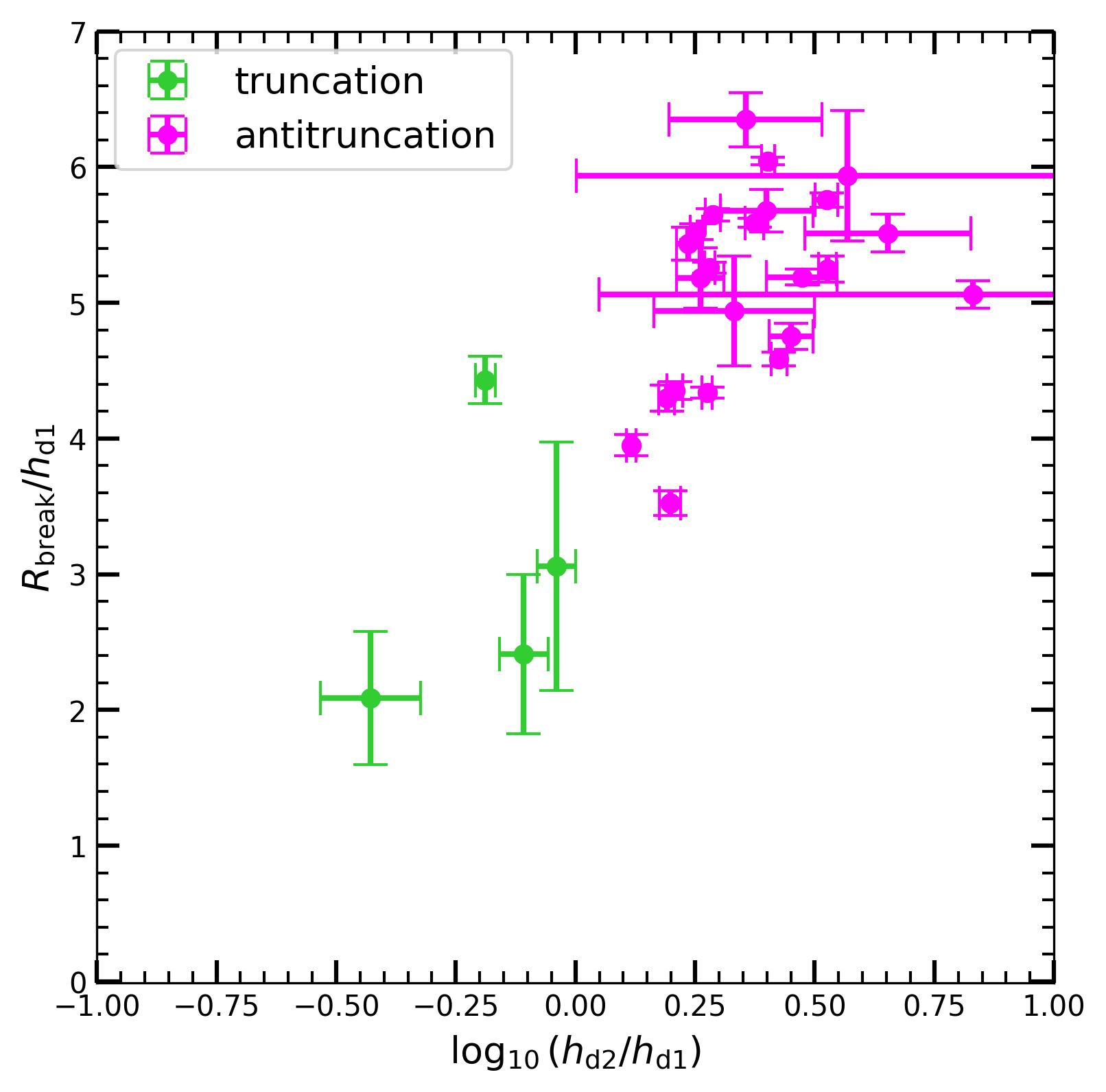}
\caption{Ratio of the truncation radius to the first disc scalelength ($R_\textrm{break}/h_{\textrm{d1}}$) as a function of the logarithm of the ratio between the disc scalelengths ($h_{\textrm{d1}}/h_{\textrm{d2}}$). This plot visually separates Type\,II galaxies (green dots) from Type\,III galaxies (magenta dots). The error bars are directly extracted from the uncertainties on the parameters obtained during the surface brightness profile fitting process.
} 
\label{fig:R_ratios}
\end{figure}

As indicated in previous studies \citep{vanderKruit1987, PohlenTrujillo2006, Comeron2012, Laine2016}, the plane displaying the ratio of the break radius to the first disc scalelength ($R_\textrm{break}/h_{\textrm{d1}}$) as a function of the ratio between the disc scalelengths ($h_{\textrm{d1}}/h_{\textrm{d2}}$) is widely used to distinguish between down- and up-bending disc breaks. Figure \ref{fig:R_ratios} shows the distribution of down- (in green) and up-bending break (in pink) profile parameters in this space, clearly identifying the Type\,II and Type\,III groups. In this diagram, as expected, for up-bending break profiles (in magenta), we observe $h_{\textrm{d2}} > h_{\textrm{d1}}$, meaning the slope decreases beyond the break radius, while for down-bending break profiles (in green), we see $h_{\textrm{d2}}<h_{\textrm{d1}}$, indicating that the slope increases beyond the break.

An interesting observation is that, on average, $R_\textrm{break}/h_{\textrm{d1}}$ is smaller for down-bending disc breaks compared to up-bending ones. This is consistent with the findings of \cite{Comeron2012}, who noted that thick discs tend to truncate at lower relative radii than thin discs, likely due to the longer inner scalelength of the thick disc. \cite{PohlenTrujillo2006} reported that the break typically occurs at about 2.5 times the inner scalelength ($h_{\textrm{d1}}$), with a surface brightness of $\mu_\textrm{break} \sim 23.5 \ \mathrm{mag}~\mathrm{arcsec}^{-2}$ for Type\,II galaxies. In contrast, Type\,III galaxies exhibit breaks further out, at about 4.9 times the inner scalelength, with a lower surface brightness of $\mu_\textrm{break} \sim 24.7\ \mathrm{mag}~\mathrm{arcsec}^{-2}$.

Our results are broadly consistent with these trends, although we find that for Type\,II galaxies, the break occurs at a slightly larger radius, approximately $2.9 h_{\textrm{d1}}$ on average, and at a fainter surface brightness of $\mu_\textrm{break} \sim 24.5\ \mathrm{mag}~\mathrm{arcsec}^{-2}$. For Type\,III galaxies, we observe a break radius of approximately $5.2h_{\textrm{d1}}$ on average, with an even lower surface brightness of $\mu_\textrm{break} \sim 25.3\ \mathrm{mag}~\mathrm{arcsec}^{-2}$. These differences may reflect variations in sample selection or environmental factors but confirm the general trend that down-bending disc breaks occur at smaller radii with brighter surface brightness, while up-bending disc breaks are found further out in regions with fainter surface brightness.

Note that we provide in \cref{appendix:histoparam} the distributions of all fitting parameters for Type\,II and Type\,III profiles.

\subsection{Role of the cluster environment}

Early studies, such as those by \cite{PohlenTrujillo2006}, revealed no significant differences between field and cluster galaxies, potentially due to variations in sample selection. However, more recent research presents a contrasting view, suggesting that galaxy types are significantly shaped by their environments. For instance, \cite{Laine2016} conducted analyses on similar galaxy populations across varied settings, including both field and cluster environments like the Virgo cluster. This research highlighted a statistically significant correlation between the inner and outer disc scale lengths and the Dahari parameter \footnote{The Dahari parameter quantifies the strength of gravitational interactions between a galaxy and its neighbouring companions, providing a measure of the local interaction environment.}, which measures the strength of gravitational interactions between a galaxy and its nearby companions. This correlation is particularly notable in Type\,III and Type\,I profiles, indicating that interactions within the cluster might play a crucial role in shaping these profiles. Further supporting this perspective, \cite{Pranger2017} demonstrated that the prevalence of each galaxy type varies considerably with environmental factors.

To investigate the role of the Perseus cluster environment more precisely, we aim to cover the nuanced impact of the cluster on galaxy evolution. We explore the spatial position of down- and up-bending break galaxies in \cref{sc:generaltrend}, as well as the effects of morphological types in \cref{sc:morpho} and mass in \cref{sc:massinfluence}.

\subsubsection{Spatial distribution of profile type} \label{sc:generaltrend}

\begin{figure}[htbp!]
\includegraphics[width=1.\columnwidth]{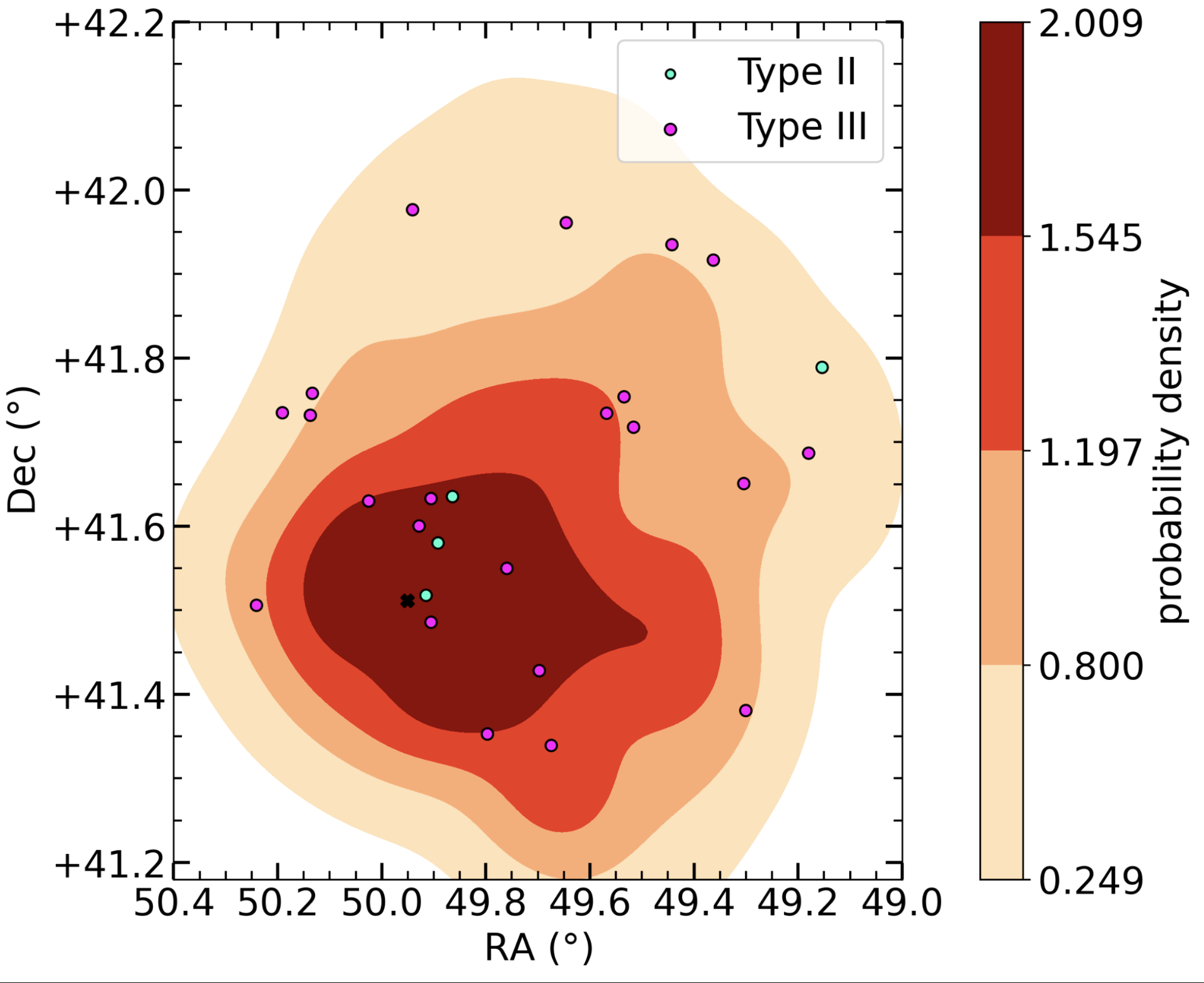}
\caption{Kernel density estimation plot of the distribution of the complete catalogue (dwarfs + bright galaxies) in the right ascension versus declination plane: four probability density bins are indicated in colour, showing higher probability density in the centre in red and lower density in the outskirts in pale yellow. Dots are overplotted on this distribution to indicate the positions of Type\,II galaxies in green and Type\,III galaxies in magenta. The black cross indicates the centre of NGC\,1275.}
\label{fig:densitytot}
\end{figure}

We first study in this section the spatial distribution of the different profile types so as to provide insights into how environmental conditions within the cluster influence their shape. To illustrate this, a gaussian KDE plot was created to visualise the distribution of the complete catalogue, which includes both dwarf and bright galaxies, in the right ascension versus declination plane, as shown in Fig. \ref{fig:densitytot}. The limits of the colourbar, and consequently the bins, are dynamically calculated based on the probability densities estimated over the entire set of massive galaxies. It is important to note that only four bins are chosen to ensure statistically meaningful results. The four probability density bins are depicted with colours ranging from intense red, marking the core of the cluster, to pale yellow, describing the outskirts. On top of this density visualisation, we overlay the specific positions of Type\,II (green dots) and Type\,III (magenta dots) galaxies. This method shows us with a detailed analysis of how the spatial distribution correlates with different profile types. Figure \ref{fig:densitytot} shows indeed that both down-bending break Type\,II and Type\,III profiles appear to be present throughout the cluster. Type\,I galaxies (profiles without breaks) initially dominate, representing approximately 80\% of the population in the innermost bin. This fraction slightly increases in the second bin but then decreases towards the cluster outskirts, where Type\,I galaxies make up only about 50\% of the population. This shift suggests that Type\,I profiles, initially predominant in the cluster’s central regions, gradually become less common in the outer parts.
Type\,III galaxies appear to be more uniformly distributed throughout the cluster. We note a small trend that Type\,III galaxies  show a tendency to avoid the cluster core, as suggested in Fig. \ref{fig:densitytot}. Their population increases in the outskirts, supporting the idea that environmental effects in the cluster centre may inhibit the processes leading to the formation or survival of up-bending break profiles. In contrast, the spatial distribution of Type\,II galaxies shows a more concentrated pattern, with three out of four Type\,II galaxies located in the core of the cluster and only one in the periphery. This peripheral Type\,II galaxy might suggest distinct characteristics compared to its counterparts in the core, potentially exhibiting additional features such as a bar structure or other morphological peculiarities. These differences in spatial distribution may hint at varying formation and evolutionary processes for Type\,II and Type\,III galaxies within the cluster. It is important to note, however, that the number of galaxies is small, especially for Type\,II. 

\begin{figure}[htbp!]
\includegraphics[width=1.\columnwidth]{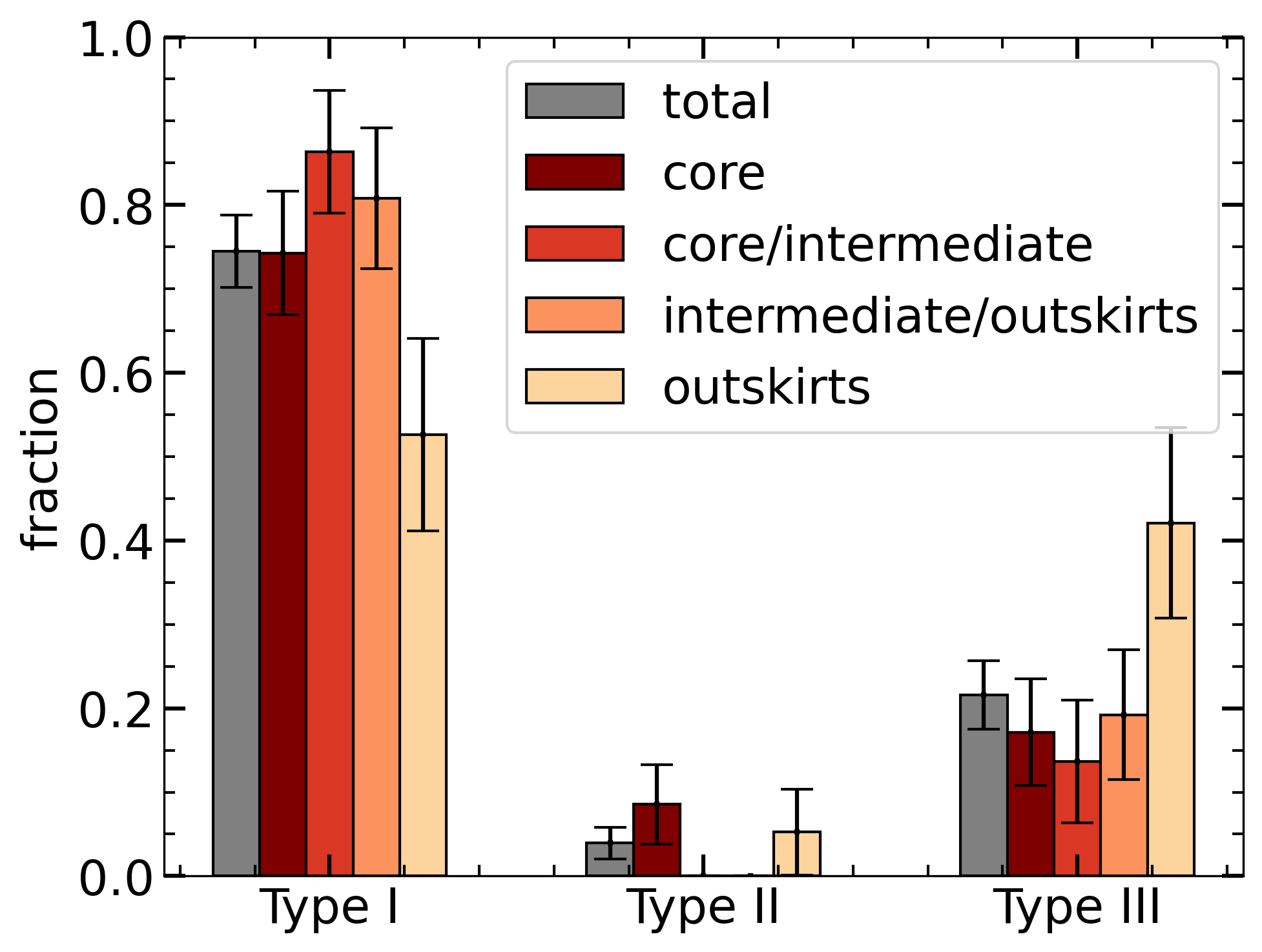}
\caption{Fraction of each type within the total population of spiral to S0 galaxies: Type\,I, Type\,II and Type\,III. The grey bars indicate the total fraction of each type. The coloured bars, ranging from dark red to pale yellow, show the fraction of each type within each region from the core to the outer regions. Error bars represent the $1\sigma$ binomial uncertainties calculated as $\sigma = \sqrt{f(1-f)/N}$, where $f$ is the measured fraction and $N$ is the total number of galaxies in the corresponding bin.}
\label{fig:percentage_distance_typetot}
\end{figure}

One step further in this analysis, Fig. \ref{fig:percentage_distance_typetot} shows the fraction of each profile type within the total population of spiral to S0 galaxies, highlighting the variations from the core (dark red) to the outer regions (yellow). This detailed breakdown allows for an examination of how the prevalence of different profile types varies with their location within the cluster, further quantifying the impact of environmental conditions on galaxy evolution. 
Overall, Type\,I galaxies are prevalent (75\%), while roughly between 20 and 25\% are up-bending break profiles, and a few percent down-bending break profiles. This distribution indicates that while non-down-bending break profiles are the most numerous, up-bending break profiles also constitute a significant proportion of the population, and down-bending break profiles are less common but still present across the cluster. More precisely, when examining these same fractions across different probability density bins, it is observed that the fraction of up-bending break profiles, meaning Type\,III, slightly increases as we move away from the central bin. In the outermost bin, Type\,III profiles account for about 40\% of disc galaxies. Conversely, for Type\,II profiles, there are three down-bending break profiles in the core and one down-bending break profile in the outer region. However, despite the small statistics, down-bending break profiles (Type\,II) are present both in the core and the outer regions, which contrasts with earlier conclusions suggesting the absence of Type\,II profiles in clusters like Virgo \citep{Erwin2012}. The presence of Type\,II profiles in our study aligns more closely with the findings of \cite{Laine2016} and \cite{Head2015}, who showed that down-bending disc breaks can persist even in dense environments such as the core of the Coma cluster. Finally, it is important to note that our density bins are defined in projection, meaning that some objects classified in the core bin could actually be located at significant distances from the centre in 3D space. This distinction is crucial when interpreting the spatial distribution of profile types within the cluster.

\subsubsection{Morphological influence}\label{sc:morpho}

\begin{figure}[htbp!]
\includegraphics[width=1.\columnwidth ]{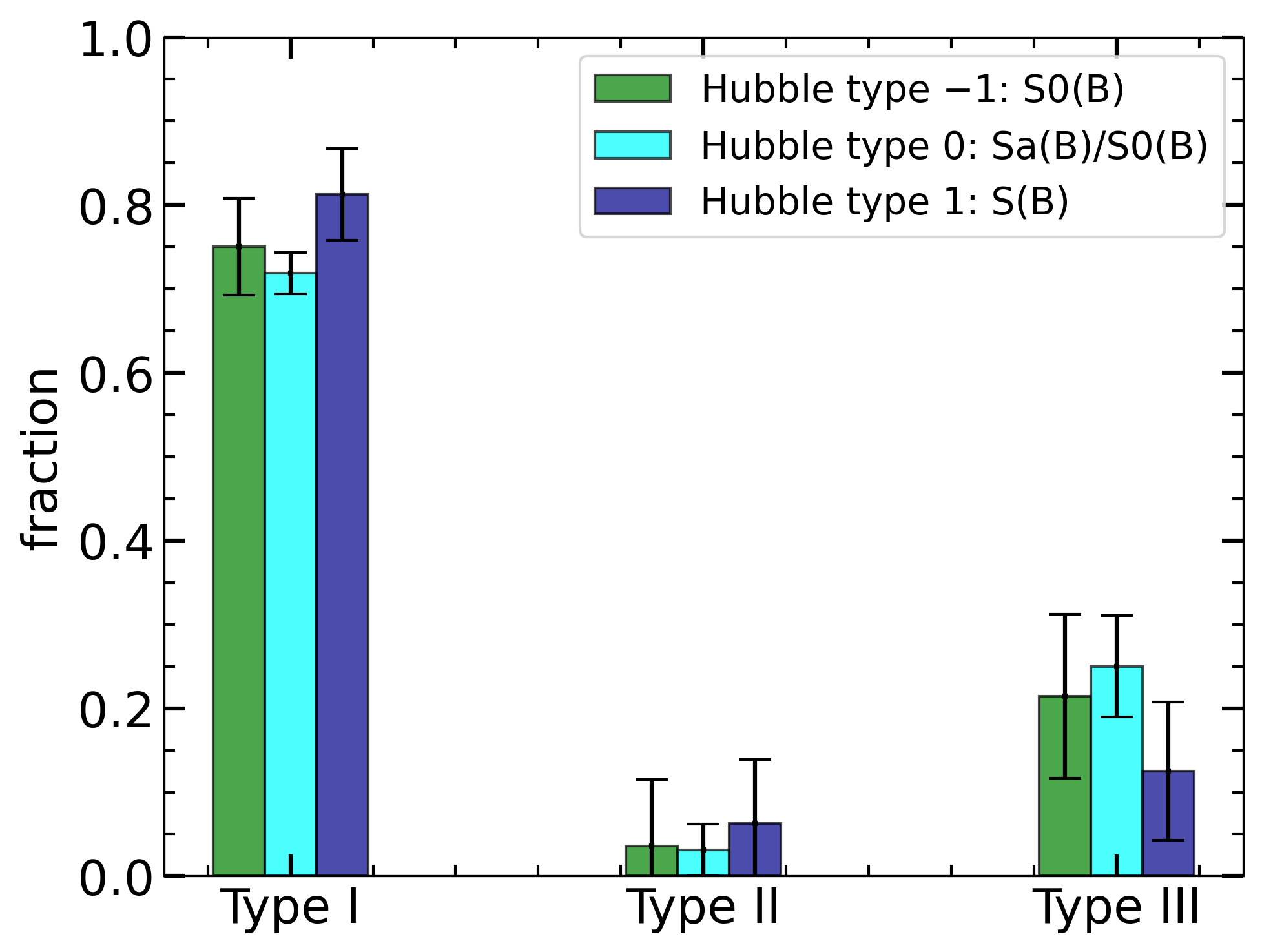}
\caption{Fraction of galaxies in each density bin for each Type profiles for the different Hubble morphological type: one corresponds to spirals (darkblue), zero to S0/spiral intermediates (cyan), and minus one to confirmed S0 galaxies (green). Error bars represent the $1\sigma$ binomial uncertainties calculated as $\sigma = \sqrt{f(1-f)/N}$, where $f$ is the measured fraction and $N$ is the total number of galaxies in the corresponding bin.}
\label{fig:fraction_Hubbletype}%
\end{figure}

Let us now turn to the role of morphological types. As indicated in \cref{sc:method}, we define three classes of discs, ranging from S0 galaxies (type $=$ $-$1) to spirals (t $=$ 1). Figure \ref{fig:fraction_Hubbletype} shows the evolution of the total fraction of each profile type with the associated morphological type, regardless of the cluster position. It is observed that Type\,I profiles dominate in fraction across all galactic morphologies. However, this fraction is slightly higher for spirals compared to the other morphological types. On the other hand, the fraction of Type\,III profiles is slightly lower for spirals. The proportion of Type\,II profiles appears relatively consistent across different morphological types, though the small sample size warrants caution in drawing definitive conclusions.

\begin{figure*}[htbp!]
\includegraphics[width=2.\columnwidth]{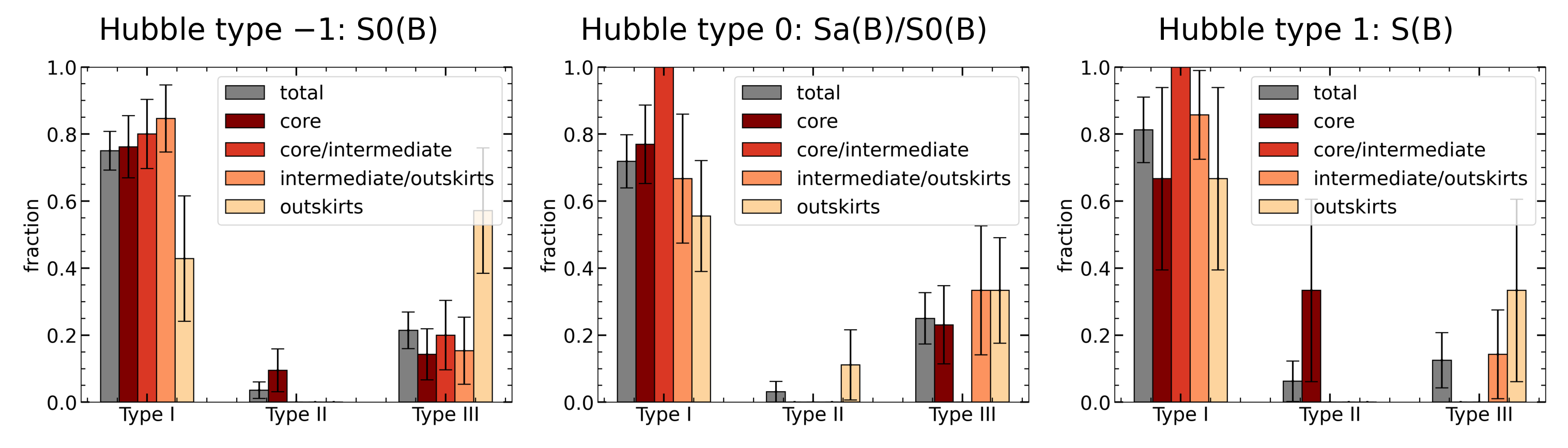}
\caption{Fraction of each profile type (I, II, III) according to the angular projection within the cluster (from red to yellow) similar to Fig. \ref{fig:percentage_distance_typetot} but for different galaxy morphology from Hubble type $-$1 on the left-hand panel, to one on the right-hand panel. Error bars represent the $1\sigma$ binomial uncertainties calculated as $\sigma = \sqrt{f(1-f)/N}$, where $f$ is the measured fraction and $N$ is the total number of galaxies in the corresponding bin.
} 
\label{fig:percentage_distance_morpho}
\end{figure*}

Aiming at studying the interplay between profile types and both morphology and environment, we display in 
Fig. \ref{fig:percentage_distance_morpho}, the abundance of Type\,I, II, III 
across the different probability density bins for each galaxy morphology. In the left panel, the fraction of Type\,III profiles for S0 galaxies is shown to dominate in the outermost region of the cluster compared to Type\,I profiles. In this region, Type\,II profiles are not present for S0 galaxies. For spiral galaxies, Type\,II profiles are found in the cluster core, while Type\,I and Type\,III profiles are more common in the lower density bins. 

With respect to both spatial distribution and morphological characteristics within the cluster, differences are observed between different morphologies but no significant general trend emerges.

\subsubsection{Mass influence}\label{sc:massinfluence}

\begin{figure*}[htbp!]
\centering
\includegraphics[width=1.\textwidth]{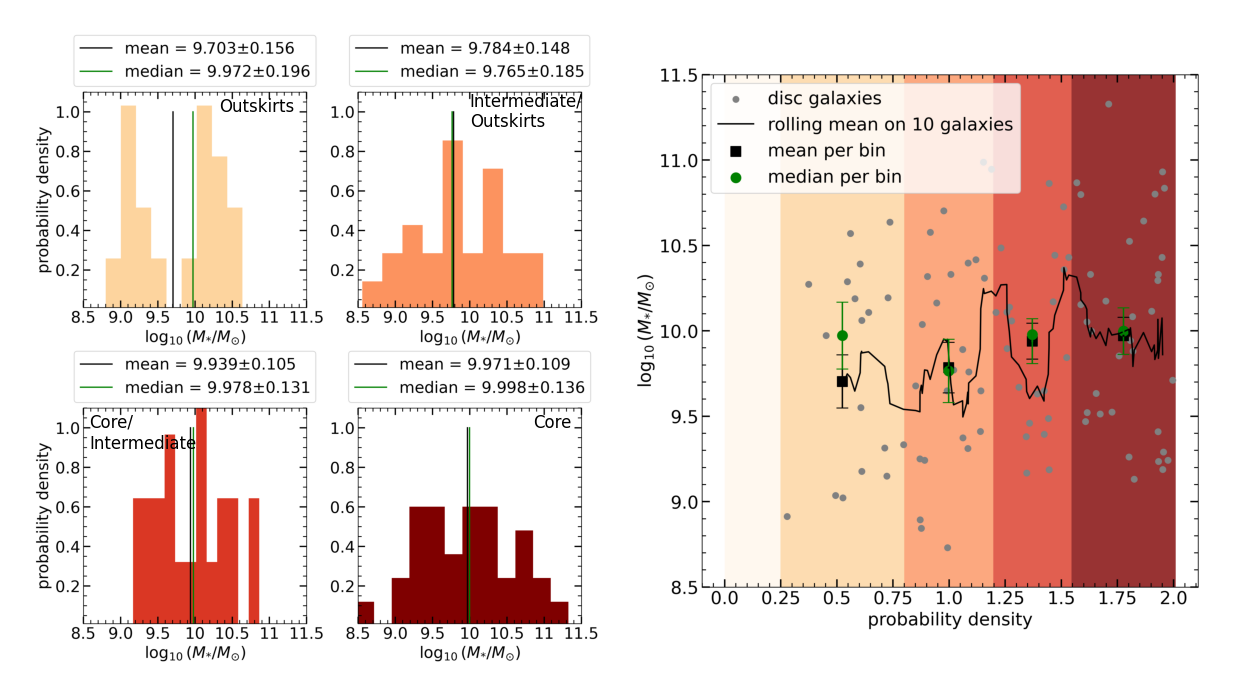}
\caption{Stellar mass distribution of galaxies according to their position in the cluster. Left-hand panel: stellar mass distribution of galaxies across the different environments (from outskirts -- top left panel -- to core -- bottom right panel), each represented by a different colour as labelled. Right-hand panel: galaxy mass as a function of probability density, with individual galaxies represented by grey dots. The black curve indicates the rolling average over 10 galaxies, black squares show the mean, and green dots represent the median for each density bin. The error bars corresponds to the standard error on the mean and median.}
\label{fig:distance_mass_study}
\end{figure*}

We now turn to the potential correlations between the mass of disc galaxies and the proportion of each profile type. Figure \ref{fig:distance_mass_study} presents the mass distribution of disc galaxies across galaxy density bins within the cluster. Although asymmetry appears in the distributions, the mean and median disc masses remain fairly consistent across bins. Examining a rolling mean over 10 galaxies reveals a slight decrease in mass from denser regions to the outskirts, though this trend in mean/median mass is weak. Notably, the distribution tails indicate that the most massive galaxies are preferentially located closer to the cluster centre, consistent with a general trend of mass segregation. \citet{vanderBurg2018} discuss how massive galaxies are commonly found near the core, while lower-mass galaxies dominate the outer regions.

This mass segregation may result from various gravitational and dynamical processes: dynamical friction, for example, might cause massive galaxies to lose orbital energy over time, drawing them towards the cluster centre \citep{Chandrasekhar1943}. Additionally, while high velocities inhibit mergers in dense cores, slower encounters at the outskirts could lead to accretion and central galaxy growth \citep{Merritt1985}. Recent studies support this mass-dependent distribution; for instance, \citet{Barsanti2016} observed lower velocity dispersions for massive galaxies, suggesting that dynamical interactions influence their central distribution. However, within the innermost regions, particularly within $0.25R_{\textrm{200}}$, mass variations are modest, with only slightly more massive galaxies at the centre. This suggests a relatively homogeneous stellar mass population in dense cluster cores, consistent with a gradual gradient in mass segregation at small scales \citep{Haines2015}.

In our sample within 0.25$R_{\textrm{200}}$, within the \Euclid field of view \citep{LF}, the mass distribution follows this pattern, showing a weak but noticeable gradient in stellar mass outwards, as described Fig. \ref{fig:distance_mass_study}. This aligns with expectations that gravitational and environmental processes, such as dynamical friction or tidal stripping, tend to homogenise the mass distribution in the densest cluster regions, resulting in a modest overall mass variation within 0.25$R_{\textrm{200}}$.

\begin{figure*}[htbp!]
\includegraphics[width=1.\textwidth]{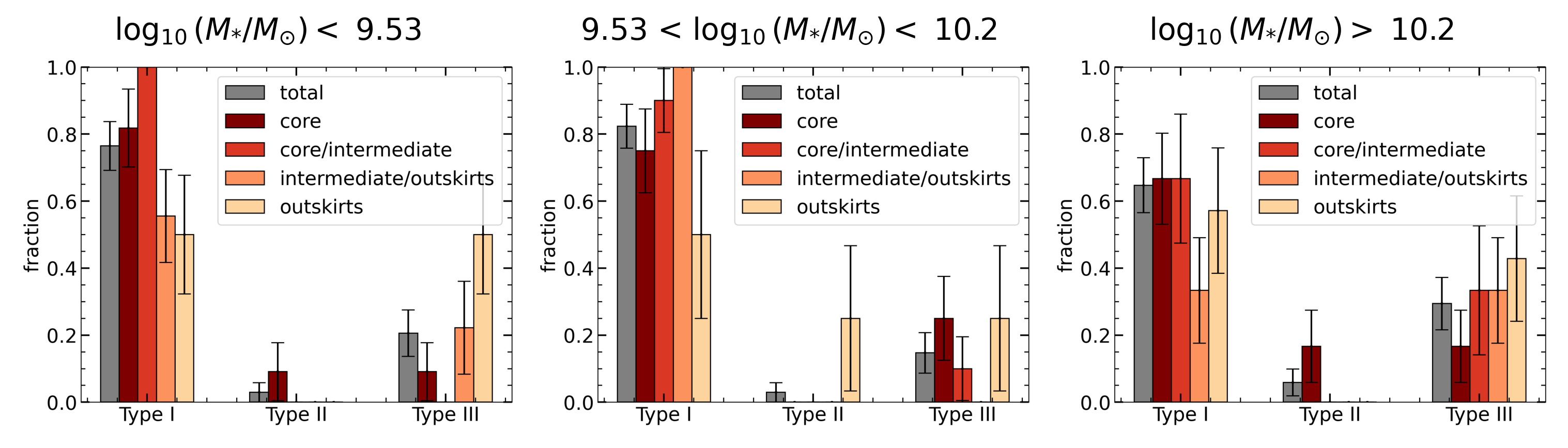}
\caption{The fraction of each profile type (I, II, III) according to the angular projection within the cluster (from red to yellow) similar to Fig. \ref{fig:percentage_distance_typetot} but for three different stellar mass bins from left/lower masses to right/larger masses as labelled. Error bars represent the $1\sigma$ binomial uncertainties calculated as $\sigma = \sqrt{f(1-f)/N}$, where $f$ is the measured fraction and $N$ is the total number of galaxies in the corresponding bin.}
\label{fig:percentage_distance_binmass}
\end{figure*}

Subsequently, three mass bins were defined, each containing an equal number of disc galaxies. The fractions of each profile type within each mass bin were then analysed, as shown in Fig. \ref{fig:percentage_distance_binmass}. The overall trend shows minimal variation in the fractions across different probability density bins, specifically different environments. For low-mass galaxies, the general trend observed globally is similar to our results of \cref{sc:generaltrend}, with a slightly decreasing number of Type\,I profiles and an increasing proportion of Type\,III profiles in the less dense, outer regions of the cluster. Type\,II profiles, however, are confined to the cluster core and constitute a small overall fraction. For high-mass galaxies, the same trend is observed but is less pronounced: in the outer regions, Type\,III profiles represent 40\% of the profiles, a value that rises to 50\% in the first mass bin. For intermediate-mass galaxies, the correlation is less clear. In the first three denser bins near the centre, the fractions remain relatively constant. Nevertheless, it is notable that down- and up-bending break profiles collectively constitute nearly 50\% of all profiles in the outermost probability density bin. These findings are consistent with recent studies \citep{Callum2024} showing that stellar masses tend to be higher in denser environments, such as the cluster core, although the differences are relatively modest compared to other environments, such as groups or the cluster outskirts. In summary, the most massive galaxies, as well as the more evolved morphological types, are concentrated near the cluster centre, as expected.

However, we must highlight that the statistical significance of these present results is limited, and the differences observed here should be interpreted with caution, as they may not be very significant due to the small sample size.  A clearer result could be expected if the 2D distribution could be deprojected into 3D. 

\subsubsection{Colour gradients as a function of radius}

Our observations in the NIR in the \HE-band allow us to map these regions of outer discs in more detail than previous optical studies, such as those by \cite{Zheng2015}, which were often limited in depth. NIR data provide a unique view of LSB regions dominated by older stars. At these wavelengths, the effects of dust extinction are minimised, and the mass-to-light ratio remains more stable, offering a clearer indication of stellar mass. Unlike optical studies, which are heavily influenced by younger stars with high luminosity/mass ratios, our NIR data more accurately reflect the mass distribution of the galaxy. This red gradient, therefore, indicates older and redder stellar populations in the peripheries of galaxies \citep{Bakos2008, Zheng2015, Watkins2016}. This approach offers two main advantages. First, focusing on older stellar populations in the outer disc creates a smoother profile, facilitating the identification of structures such as upturns and downturns in disc breaks. The U-shaped profiles observed in some galaxies underscore this effect, revealing an age gradient across the disc, with older and redder populations concentrated in the peripheries. Secondly, by working in the NIR, we avoid some of the complex corrections required in optical data for estimating stellar mass, where dust extinction and age-related luminosity variations can complicate direct mass mapping. Additionally, \cite{Zheng2015} shows that, across different Hubble types, galaxies tend to have simple exponential mass profiles, indicating that these variations reflect differences in stellar populations rather than changes in mass distribution.

\begin{figure*}[htbp!]
\includegraphics[width=1\textwidth]{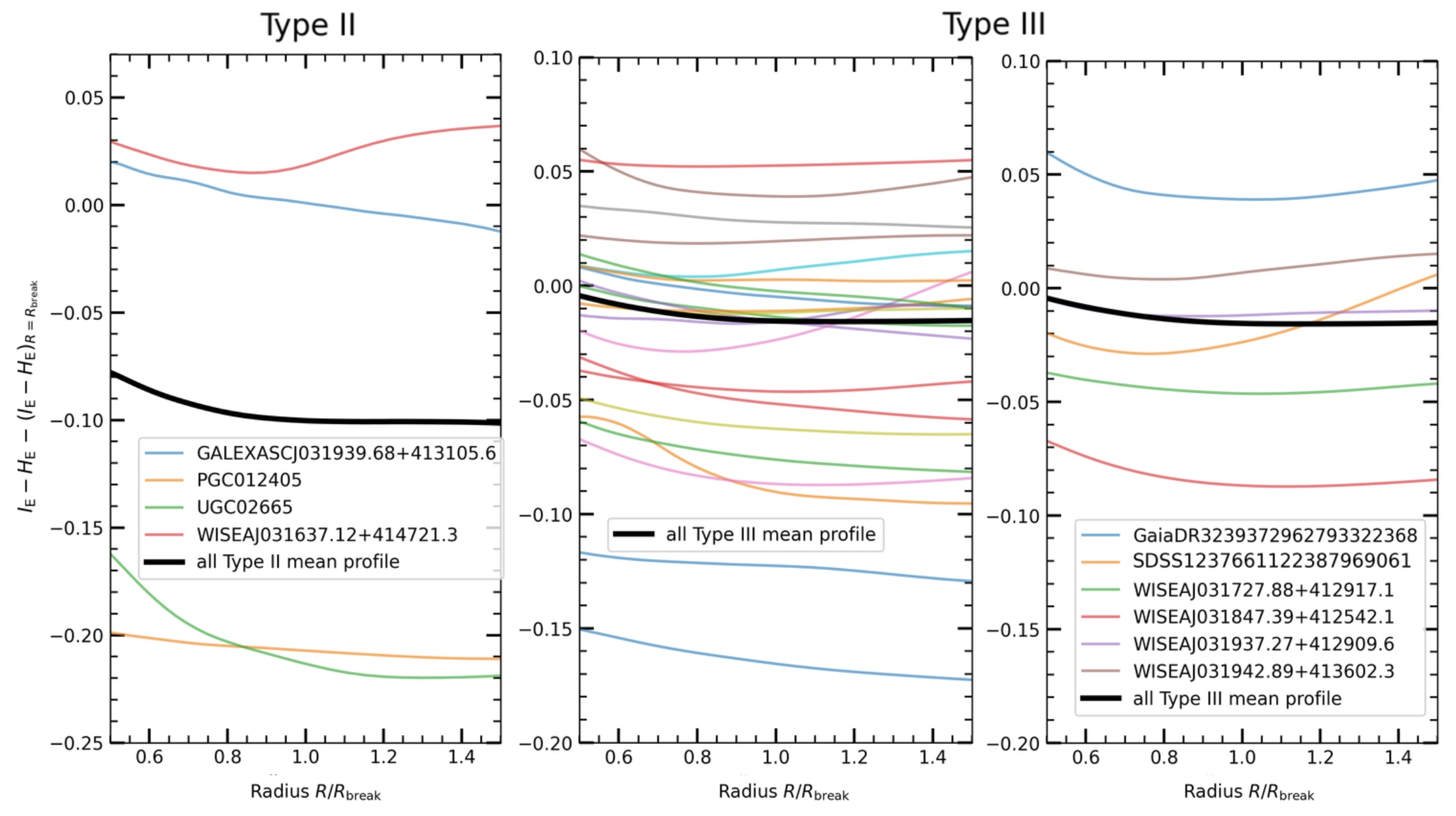}
\caption{
Normalised magnitude difference \IE$-$\,\HE as a function of radius $R / R_{\textrm{break}}$ for galaxies Type II, Type III and Type III with a detected \textit{U}-shaped profile. The profiles are normalised around $R = R_{\textrm{break}} $. The minimum value within the range $0.5 \leq R / R_{\textrm{break}} \leq 1.5$ is identified.}
\label{fig:typeUall}%
\end{figure*}

\begin{figure*}[htbp!]
\centering
\includegraphics[width=1.\textwidth]{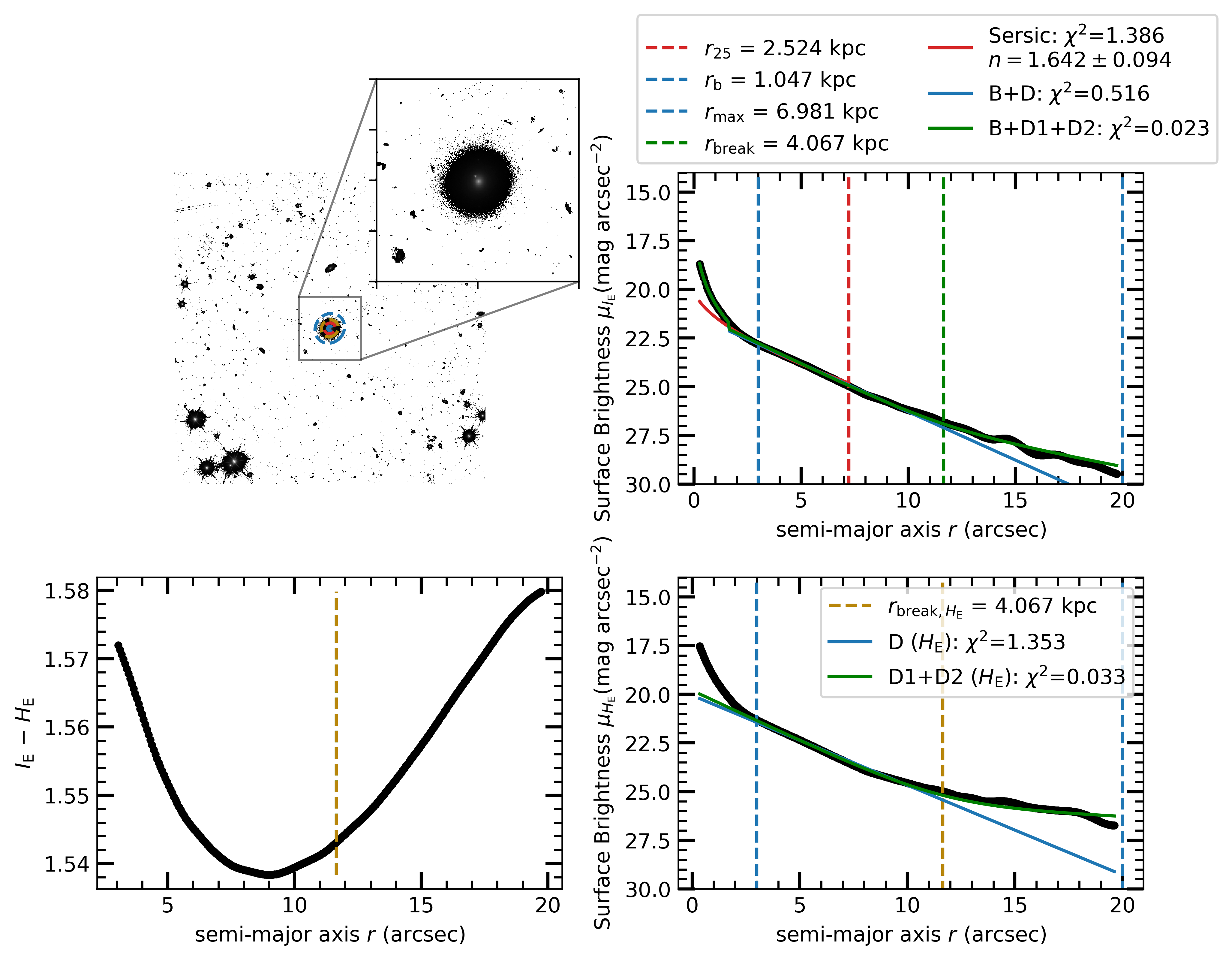}
\caption{
Profile of the galaxy SDSS\,1237661122387969061.
\textit{Top Left Panel:} Image of the galaxy with associated radii indicated by ellipses of different colours: green/yellow for the break positions in the surface brightness profiles of \IE\ ($\mu_{\IE}$) and \HE\ ($\mu_{\HE}$) bands, red for the radius at $\mu_{\IE} = 25\,\mathrm{mag}\,\mathrm{arcsec}^{-2}$, and blue for the boundary regions of interest in semi-major axis $r$ between [$r_{\textrm{b}}$, $r_{\textrm{max}}$].
\textit{Top Right Panel:} Surface brightness profile ($\mu_{\IE}$) versus semi-major axis ($r$) of the fitted ellipses in the \IE\ band. Solid lines indicate the fitted models: Single S\'ersic model in red, bulge/disc decomposition model in blue, bulge/disc1/disc2 decomposition model in green. The vertical dashed lines indicate in blue the boundary of the region of interest in $r$ where the break may potentially be found, in green the break position, and in red the radius corresponding to $\mu_{\IE} = 25\,\mathrm{mag}\,\mathrm{arcsec}^{-2}$.
\textit{Bottom Left Panel:} \IE$-$\HE\ colour as a function of the semi-major axis $r$ between [$r_{\textrm{b}}$, $r_{\textrm{max}}$].
\textit{Bottom Right Panel:} Surface brightness profile $\mu_{\HE}$ versus semi-major axis ($r$) of the fitted ellipses. The solid lines indicate the fitted models after the radius of the bulge: disc model in blue, disc1/disc2 model in green. The dashed lines indicate in blue the boundary of the region of interest in $r$ where the break may potentially be found and in yellow the break position.
}
\label{fig:typeU}%
\end{figure*}

Figure \ref{fig:typeUall} presents the normalised \IE-\HE colour profiles of Type II and Type III galaxies. As the left panel shows, among the Type II galaxies studied, only one exhibits a U-shaped colour profile, characterised by a marked colour gradient. This finding is significant as it indicates a relatively low occurrence of this phenomenon in Type II galaxies, which is lower than the frequency generally reported in cluster studies, such as that by \cite{Roediger2012}. The consensus is that this U-shape comes from the stellar radial migration \citep{Roskar2008}. Moreover, Fig. \ref{fig:typeUall} reveals that among Type III galaxies, six of them, or one quarter, exhibit U-shaped colour profiles. This observation is particularly interesting as it suggests additional complexity in the evolutionary processes of Type III galaxies. \cite{Roediger2012} demonstrates that U-shaped galaxies can be present for all three types, although it is primarily Type II galaxies that are predominantly of this shape. \cite{Si-YueYu2024}, who have recently been exploring discs at higher redshifts, confirm that U-shaped profiles are not exclusive to Type II galaxies but also exist for Type III. Figure \ref{fig:typeU} provides an example of a galaxy with an upward curvature break profile and its strong U-shape. We observe that the break in the profile \ref{fig:typeU} occurs before the expected break, as indicated by \cite{Bakos2008}. We observe that the break in the profile, marked by the green dotted line in the top left panel, is around 0.7$R_\textrm{break}$, indicated by the golden dotted line in the reddest NISP band, the \HE band. This is also the case for two other galaxies in purple and brown on the right panel of Fig. \ref{fig:typeUall}. However, most of these U-shapes, four out of six, are not very strong and thus only show a small colour gradient, tending towards a plateau, a known feature highlighted by \cite{Bakos2008}. Thus, the few Type III discs observed could be the result of recent star formation in the outer regions of the disc, as suggested by \cite{Si-YueYu2024}. The others have rather a plateau profile after a minimum or a slight decrease, as shown in the middle panel of Fig. \ref{fig:typeUall} and expected by \cite{Bakos2008}, \cite{Roediger2012}, and \cite{Zheng2015}.

These observations in \ac{NIR} represent a step forwards in probing \ac{LSB} regions in nearby galaxies, making it possible to reach the faint outermost regions of galactic discs with greater precision. By mapping the distribution and mass of these older stellar populations, this dataset provides a more refined perspective on the outer structures of galaxies and the evolutionary mechanisms shaping them, such as radial migration and satellite accretion.

\section{Discussion}\label{sc:Discussion}

In this section, we discuss our findings on galaxy disc profile types within the Perseus cluster and compare them with results from both simulations and other clusters. Our observations reveal possible trends in the presence of Type\,II and Type\,III profiles in this cluster environment, potentially shaped by both internal dynamics and the external cluster environment. These findings highlight the role of galaxy-cluster interactions in shaping galaxy morphology, supporting or challenging various theoretical interpretations and previous observational results. 

\subsection{Comparison to simulations of galaxy-cluster interaction}\label{sc:Simu}

To understand the rarity of Type\,II profiles within the Perseus cluster, contrasted with their frequency in field surveys, and the persistence of Type\,III profiles in this environment, we use results from a series of simulations (Mondelin et al. in prep, hereafter Paper~II). These simulations start with disc galaxies displaying a Type\, I, \,II, or III profile, formed through internal instabilities at high redshift ($z$ between one and three). A Type\,II disc galaxy, for example, remains stable in isolation, with its radial break intact over a few billion years. However, when subjected to the tidal forces of a Perseus-like cluster, this truncation weakens, leading the galaxy to adopt a Type\,I profile within one or two cluster crossing times (i.e. at most 1 or 2\,Gyr).

\begin{figure}[htbp!]
\centering
\includegraphics[width=0.5\textwidth]{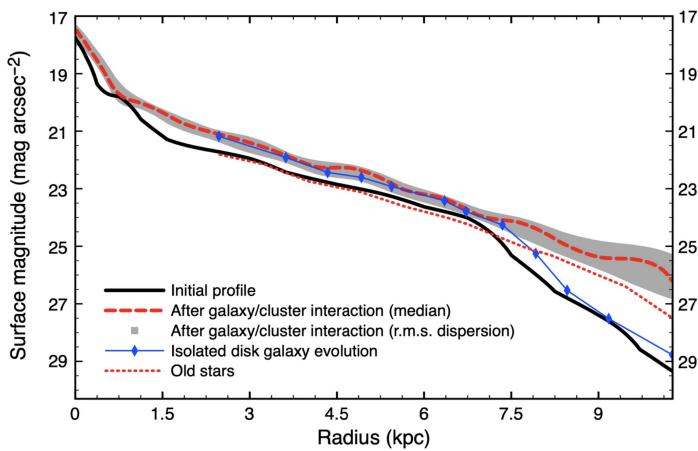}
\caption{Hydrodynamical simulations of the interaction of a Type\,II galaxy with a Perseus-like cluster. The initial disc galaxy (solid black curve) has a double exponential, Type\,II radial profile, and a central bulge, formed by disc instabilities at high redshift, with a break at 6.5\,kpc and a gas fraction of 17\% in the disc. These initial conditions are taken from simulations of high-redshift disc instabilities and Type\,II profile formation in \citep{Bournaud2007}. Simulations (Paper~II) evolve this disc galaxy in the tidal field of the Perseus-like cluster for about one cluster crossing time. The median profile over six simulated orbits (red dashed curve) and the RMS dispersion (shaded area) are displayed, showing that the galaxy evolves most of the time towards a Type\,I disc, while the same galaxy in isolation during the same timescale would remain Type\,II (thin blue line with diamonds). Stars pre-existing to the galaxy/cluster interaction (red dotted line) are scattered radially over increasingly eccentric orbit, accounting for about two thirds of the replenishment of the outer disc beyond the initial break radius, the other third coming from the triggering of turbulence, shocks, and star formation in the outer disc by the tidal field of the cluster (see Paper~II for individual orbits and detailed interpretation).} 
\label{fig:simus}%
\end{figure}

The results of these simulations, summarised in Fig. \ref{fig:simus}, show that the transition from Type\,II to Type\,I profiles results primarily from the disturbance of stellar orbits, which become more elongated, and from the triggering of star formation in the galaxy's outer disc by the cluster's tidal field. This suggests that the cluster environment can indeed inhibit Type\,II profiles, aligning well with our observations in the Perseus cluster. Thus, our findings provide observational support for the idea that Type\,II discs may be transformed or erased by cluster interactions, while Type\,III discs can survive due to different physical mechanisms. 

\subsection{Comparison to other works}

We now turn to how our findings on disc profiles in the ERO-Perseus field compare with previous observational studies and theoretical models, especially regarding the environmental factors that influence profile transformations.

\subsubsection{Down-bending disc break profile}

Our study identified a few Type\,II galaxies (4), a number which is relatively low compared to field surveys, aligning with other cluster studies, such as \citet{Erwin2012}, which found no Type\,II profiles within the Virgo cluster core. As we have also verified using hydrodynamical simulations in \cref{sc:Simu}, this scarcity in dense environments suggests that Type\,II profiles may struggle to survive cluster conditions. 

Moreover, approximately 25\% of the galaxies observed in the 2D inner bins actually originate from the outskirts in 3D, assuming an isotropic distribution. This finding emphasises the need for incorporating precise 3D measurements, such as spectroscopic redshifts and velocity measurements, corrected for projection effects like the finger-of-god effect, in order to confirm whether Type\,II profiles genuinely survive the harsh conditions of dense cluster environments.

\begin{figure}[htbp!]
\centering
\includegraphics[width=0.5\textwidth]{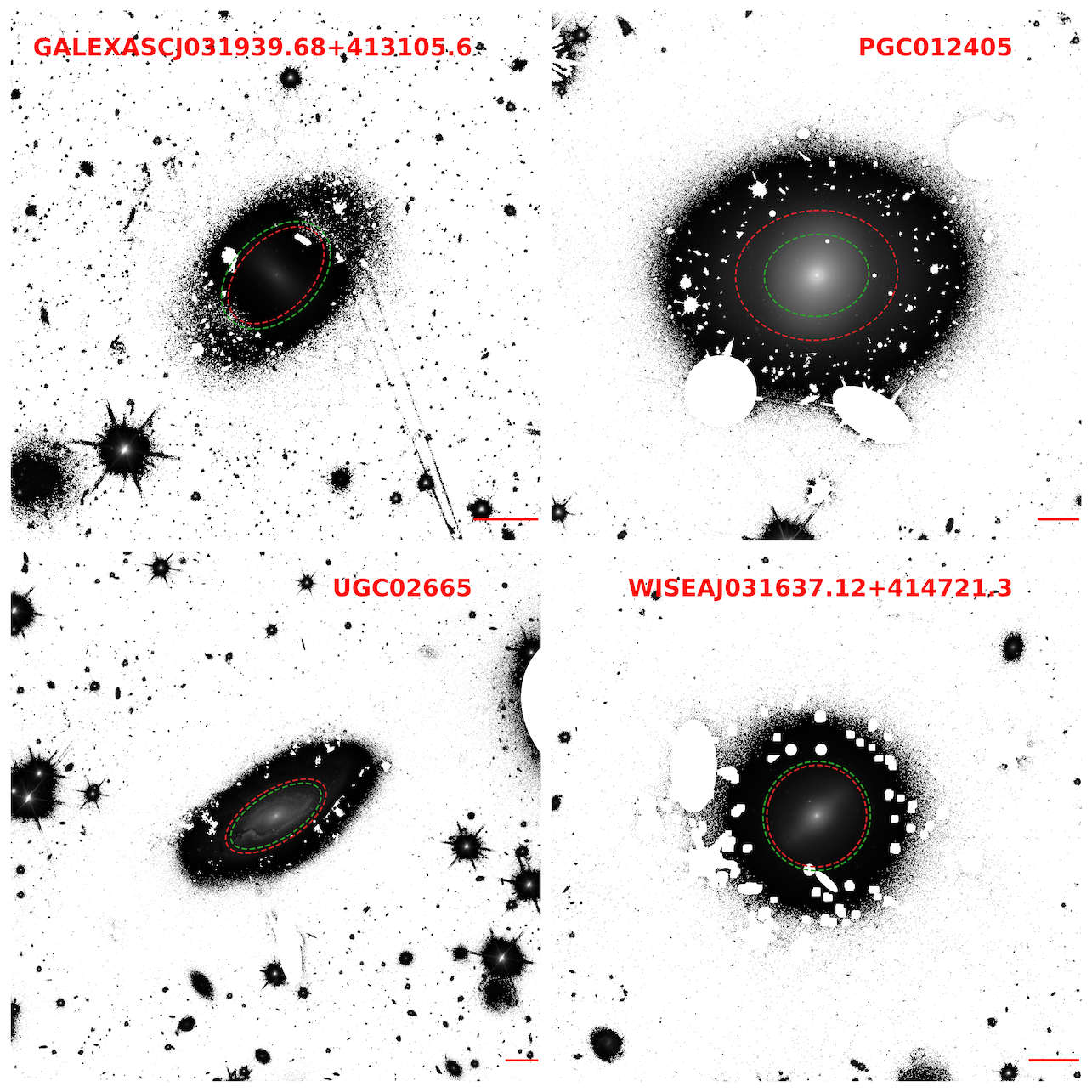}
\caption{\IE LSB images of four down-bending break galaxies. The truncation is indicated by the green dashed ellipse, i.e the isophote with a semi-major axis equal to that of the break. The outer region lies beyond this isophote. The red ellipse denotes the radius at $\mu_{\IE} = 25~\mathrm{mag~arcsec^{-2}}$. We note that the red circle is close to the green circle. The line at the bottom right of each image gives the scale, which corresponds to 10\arcsec.} 
\label{fig:typeIIimages}%
\end{figure}

However, if we consider that some Type\,II galaxies do survive within Perseus’ inner regions, it is likely due to certain stabilising effects. Structural features, such as bars, may have a significant role in maintaining their persistence \citep{Elmegreen2014}. Figure \ref{fig:typeIIimages} presents \IE images of four galaxies with down-bending disc breaks, where structures like bars and outer Lindblad resonances (OLR) appear to contribute to their stability for at least two of them. For example, the galaxy WISEA\,J031637\_12+414721\_3 shows evidence of a recent bar formation, suggesting that such features might help retain a down-bending break profile even in the cluster's core. Notably, UGC\,02665 displays clear signs of truncation likely due to ram pressure stripping, where it is losing gas and stars as it moves through the cluster. The different truncation mechanisms observed hint at a variety of structural and environmental interactions contributing to the persistence or suppression of Type\,II profiles in the Perseus cluster.

These observations in Perseus are consistent with trends in the Coma cluster, where bars have also been linked to the formation and maintenance of broken discs \citep{Head2015}. However, comparisons with Virgo and Coma reveal that the proportions of Type\,II and Type\,III profiles vary between clusters, suggesting that local environmental conditions and assembly histories could influence profile distribution. Moreover, \citet{Pranger2017} compared spiral galaxies in clusters and those in the field and shows that the fraction of Type\,I galaxies is higher in clusters, while the fraction of Type\,II galaxies is lower compared to the field. Specifically, Type\,I galaxies are approximately 2.5 times more frequent in clusters than in the field, which is in agreement with our observations. 

\subsubsection{Up-bending disc break profile}

We observe that about a third of the Type\,III galaxies in Perseus display U-shaped colour gradients, indicating older stellar populations in the outer regions, which may be due to radial migration or accretion. This finding aligns with those in other clusters, such as Virgo \citep{Erwin2012}, where a mix of Type\,I and Type\,III profiles was found. Interestingly, we observe a location-dependent variation in Perseus, with Type\,III profiles representing about 20\% in the cluster core but up to 40\% in the outskirt, as shown in Fig. \ref{fig:percentage_distance_typetot}. This gradient could be due to environmental effects, though detecting up-bending disc breaks near the cluster centre remains challenging due to the high density of galaxies. 

\citet{Pranger2017} found that the fraction of Type\,III galaxies remains consistent between cluster and field environments. Interestingly, they observed that Type\,III cluster galaxies tend to reside significantly closer to the cluster centre compared to other break types. This finding contrasts with our observations in Perseus, where we see a higher proportion of Type\,III profiles in the cluster outskirts compared to the core. However, it is important to note that our observations are limited to within 0.25 R200 of Perseus, and we lack statistical data for regions beyond this radius. Moreover, \cite{Sanchez2023} shows that in more isolated environments, there are mainly Type\,II and I profiles, and that Type\,III profiles are primarily due to interactions, such as major mergers. This is consistent with our results where the number of Type\,II profiles is low for cluster galaxies and Type\,III profiles are numerous, potentially due to their merger history and frequent interactions in a dense environment such as Perseus. 

\begin{figure*}[htbp!]
\centering
\includegraphics[width=1\textwidth]{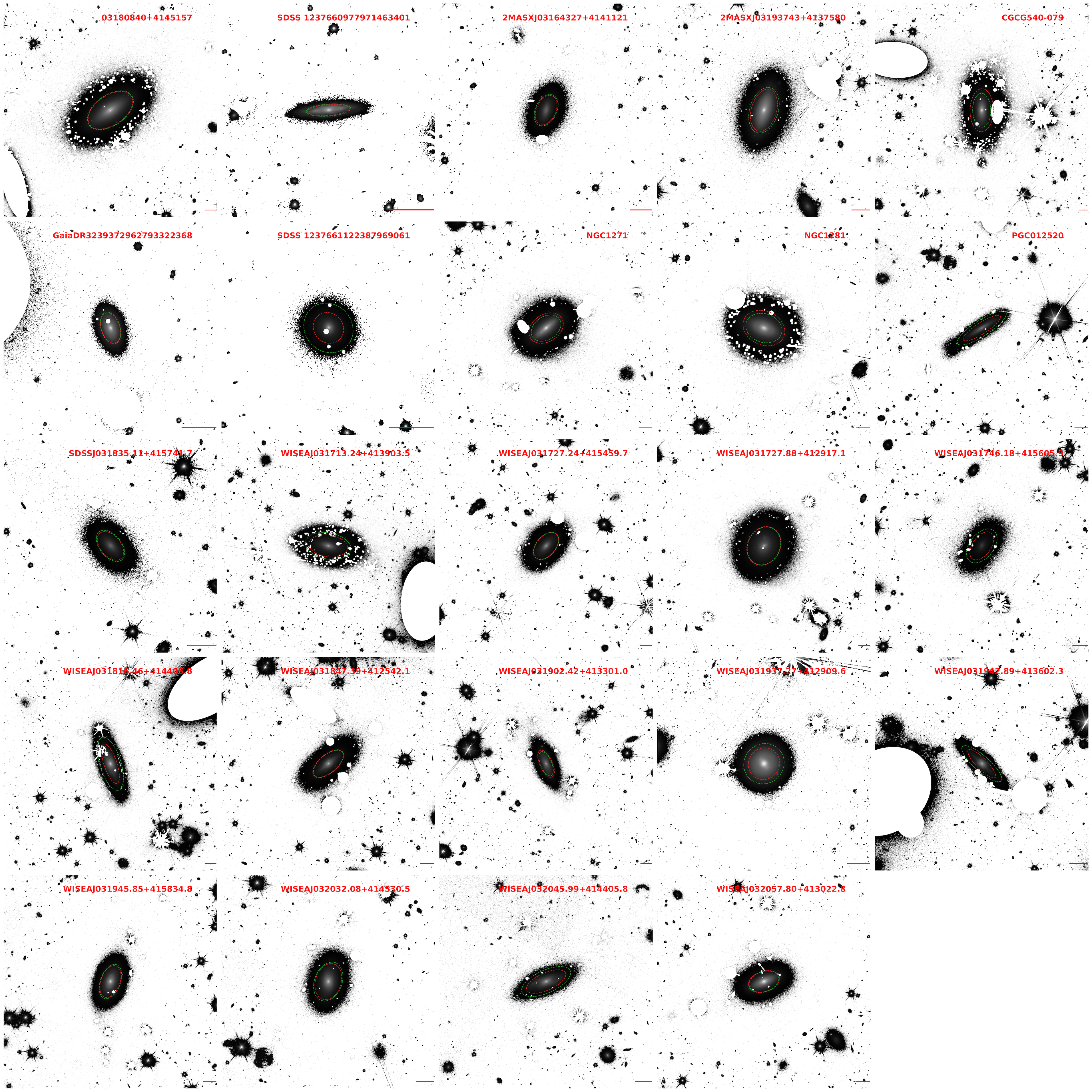}
\caption{\IE LSB images of up-bending break galaxies. The break is indicated by the green dashed ellipse, i.e the isophote with a semi-major axis equal to that of the break. The red ellipse denotes the radius at $\mu_{\IE} = 25 $  \text{mag}~\text{arcsec}$^{-2}$. We note that the red circle is close to the green circle, in most of the cases. The line at the bottom right of each image gives the scale, which corresponds to 10\arcsec.} 
\label{fig:typeIIIimages}
\end{figure*}

In the Coma cluster, Type\,III profiles are frequently associated with bars, with nearly 71\% of these galaxies displaying this feature. By contrast, barred Type\,III galaxies in Perseus are almost absent, as shown in Fig. \ref{fig:typeIIIimages}. This discrepancy suggests that while bars are essential for broken discs in Coma, other mechanisms, such as minor mergers or accretion, might be more relevant in Perseus. Notably, \citet{ChambaHayes2024} find that Type\,III profiles often correlate with extended \ion{H}{i} reservoirs and smoother \ion{H}{i} gradients, highlighting the role of gas dynamics in shaping these profiles. Extended \ion{H}{i} reservoirs provide material for star formation in the outer disc, while smoother gradients facilitate stellar redistribution via radial migration, contributing to the observed U-shaped colour gradients. External processes, such as minor mergers or satellite interactions, can also drive the formation of up-bending break profiles. Future spatially resolved \ion{H}{i} observations of Perseus could further illuminate the connection between gas reservoirs and the persistence of Type\,III profiles in cluster environments.

The Coma study also links up-bending break discs to bulge growth, likely driven by mergers or starbursts, supporting the view that Type\,III profiles can result from significant internal restructuring. Our findings in Perseus, particularly the prevalence of older stellar populations, align with this hypothesis. The comparison to previous studies further reinforces that Type\,II profiles are typically suppressed in dense environments, while Type\,III profiles remain more stable, likely sustained by mechanisms such as radial migration or mergers. The differing proportions of Type\,III profiles across Perseus, Coma, and Virgo may be linked to the clusters’ dynamical states and formation histories. For instance, the lack of barred Type\,III galaxies in Perseus compared to Coma might reflect differences in the role of secular processes in more virialised clusters. Meanwhile, Virgo’s ongoing accretion of substructures could create an environment where gas accretion and minor mergers dominate the formation of up-bending break profiles. Future studies combining detailed stellar population analyses and \ion{H}{i} maps could provide further insights into how these histories influence the evolution of galaxy profiles.

Our methodological approach, employing \texttt{AutoProf}/\texttt{AstroPhot} and high-resolution, wide-field \Euclid images, allows for improved profile detection, even in dense regions like the Perseus core. This approach contrasts with methods used in previous studies, such as the \texttt{IRAF} task ellipse and \texttt{GALFIT}, which are effective but, as \texttt{AutoProf}, limited in crowded cluster environments. By achieving greater precision in profile identification, especially at faint magnitudes, we expand our understanding of galaxy morphology across cluster environments.

\section{\label{sc:Conclusions}Conclusions}

In this work, we have studied the distribution and characteristics of galaxy profiles within the Perseus cluster, focusing on the roles of mass, bars, and morphology. The main findings can be summarised as follows.

\begin{itemize}
    \item {\it Type\,II profiles}:
    Among the 102 massive disc galaxies identified in the Perseus cluster, we classified 74 as Type\,I, four as Type\,II, and 24 as Type\,III. Of the Type\,II galaxies, three are projected close to the cluster core, within a few hundred kiloparsecs of the central galaxy. Bars and specific resonance effects, such as the OLR, were found in two out of the four Type\,II galaxies, suggesting that they may play a role in stabilising these down-bending break profiles, even in the dense cluster environment.
    
    However, it is important to consider that older processes, such as previous interactions or internal dynamics, might have caused breaks that could have been erased upon entering the cluster. The harsh conditions of the cluster, such as tidal forces or ram pressure stripping, may lead to a smoothing of the surface brightness profile, masking older breaks. Additionally, if the truncation is related to a threshold effect in star formation, or to ram pressure stripping of the outer gas, this would more prominently affect the colours of the galaxy. Specifically, the region beyond the break would appear old and red due to the cessation of star formation. These aspects suggest that both the age of the stellar populations and the colour profiles are critical in understanding the origin and persistence of Type\,II profiles in cluster environments. Moreover, as mentioned in \cref{sc:Simu}, simulations of Type\,II profiles in a cluster context reveal the role of the cluster's gravitational potential, explaining the low fraction of this Type\,In the Perseus cluster. In Paper~II, we aim to describe and further develop these simulations, with the goal of demonstrating how initial conditions -- such as the star formation threshold or the entry orbit into the cluster -- can significantly explain the observed variations in profile types. 
    
    \item {\it Type\,III profiles}: Approximately one quarter of the Type\,III galaxies in the Perseus cluster show a small U-shaped colour gradient with a minimum around 0.7$R_\textrm{break}$, indicative of older star populations in the outer regions. Unlike the Coma cluster, where bars are commonly associated with broken discs, no visibly barred Type\,III galaxies were confirmed in Perseus. This suggests that other mechanisms, such as mergers and radial migration, may be more significant in the formation of up-bending break profiles in this cluster.
    
    \item {\it Role of mass and morphology}: The study also underscored the influence of galaxy mass and morphology on profile types. The spatial distribution of these profiles varies, with certain morphological types and mass ranges more prevalent in specific regions of the cluster. In particular, it was observed that more massive galaxies tend to have more complex structures and profile types. This is consistent with the notion that massive galaxies are more likely to have undergone significant interactions and mergers, which can lead to the formation of diverse structural components such as bulges and extended discs. The distribution of mass within the cluster also appears to influence the types of profiles observed, with less massive galaxies being more susceptible to environmental effects like ram pressure stripping.
    
    \item {\it Spatial distribution}: The findings suggest that while the overall frequency of Type\,II galaxies may decrease in clusters, certain stabilising factors, such as the presence of bars and specific resonance effects, can maintain these profiles even in harsh conditions. The spatial distribution of profiles within the cluster reflects the complex interplay between internal dynamics and environmental influences.
\end{itemize}

These findings highlight the diversity of mechanisms that can lead to the formation and stabilisation of Type\,II profiles, even in dense cluster environments. While the overall frequency of Type\,II galaxies may decrease in clusters, certain stabilising factors, such as the presence of bars and specific resonance effects, can maintain these profiles even in harsh conditions. Further studies, including 3D spatial analysis, would be necessary to confirm the true position and resistance of these galaxies within the cluster. Extending the study to a larger field around the cluster would provide more statistical power and help clarify the broader environmental influences on galaxy evolution. Additionally, acquiring 3D velocity maps would allow for a non-projected representation of galaxy positions, offering more precise insights into their true local environment. Although this is currently challenging due to the dominant effect of galaxies’ proper motions within the cluster, the upcoming \Euclid DR1 data release will allow us to expand the field of view, enabling analysis of galaxies in less dense environments as well. This will provide a valuable comparison to cluster environments and deepen our understanding of the environmental factors shaping galaxy evolution.

\begin{acknowledgements}
\AckERO  
\AckEC
Fernando Buitrago acknowledges support from the GEELSBE2 project with reference PID2023-150393NB-I00 funded by MCIU/AEI/10.13039/501100011033 and the ESF+.
The authors thank Mireia Montes for their useful comments and discussions.
\end{acknowledgements}

\bibliography{Perseus}

\begin{appendix}
\section{\label{appendix:ICL} Influence of Intracluster Light on surface brightness profiles}

\begin{figure}[htbp!]
\includegraphics[width=0.5\textwidth]{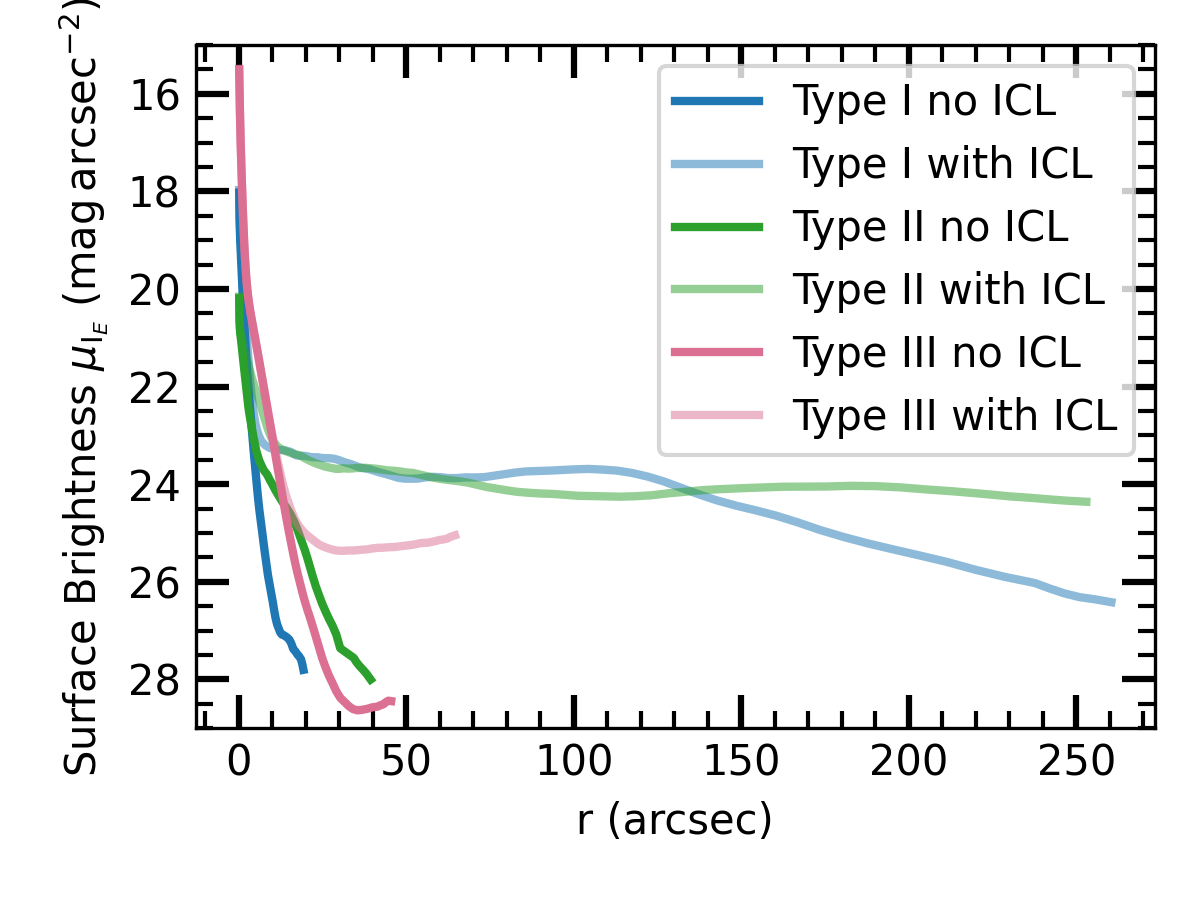}
\caption{
Surface brightness profiles $\mu_{\IE}$ (in mag~arcsec$^{-2}$) for galaxies located at the minimum distance from the centre. The curves represent different types (Type\,I, Type\,II, and Type\,III) with or without the ICL component.
} 
\label{fig:ICLmin}
\end{figure}

\begin{figure}[htbp!]
\includegraphics[width=0.5\textwidth]{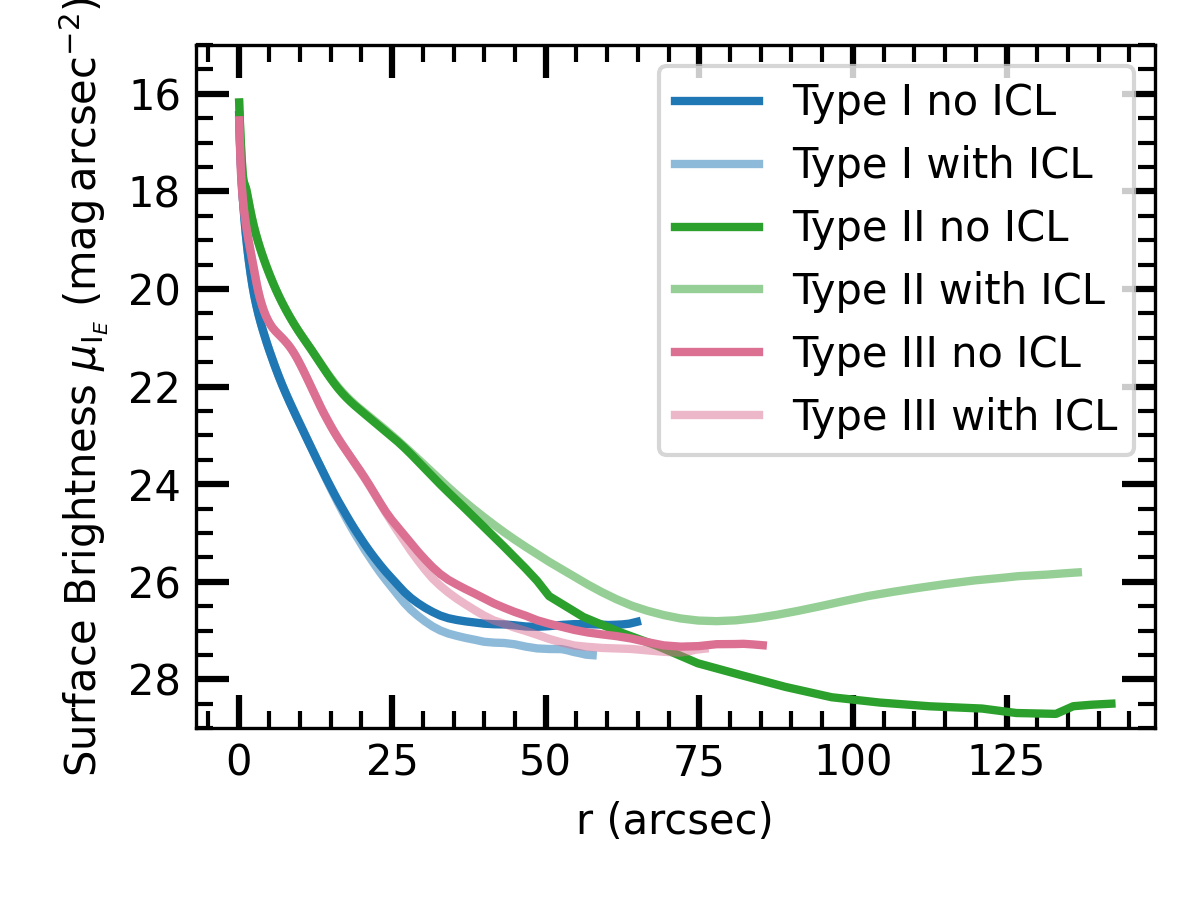}
\caption{
Surface brightness profiles $\mu_{\IE}$ (in mag~arcsec$^{-2}$) for galaxies located at a median distance from the centre. The curves represent different types (Type\,I, Type\,II, and Type\,III) with or without the ICL component.
} 
\label{fig:ICLmed}
\end{figure}

\begin{figure}[htbp!]
\includegraphics[width=0.5\textwidth]{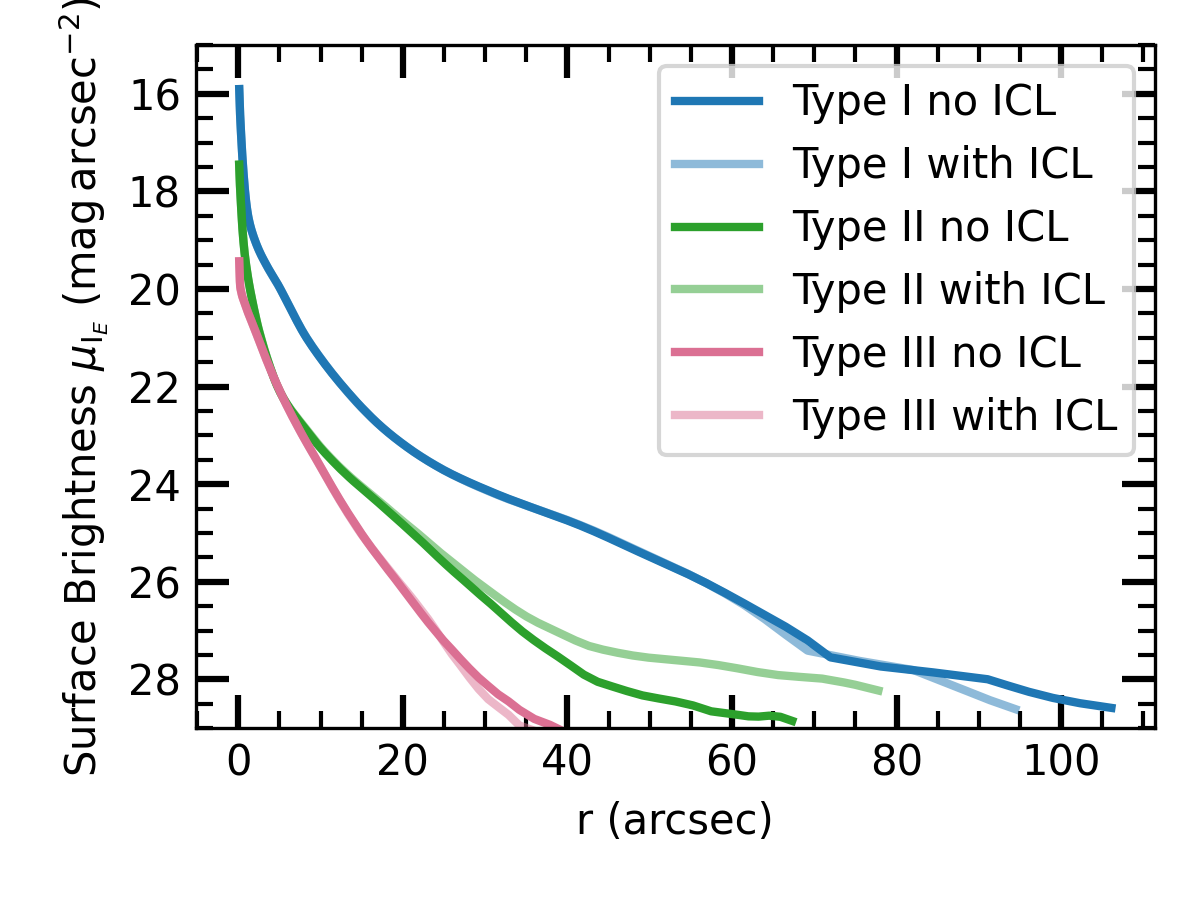}
\caption{
Surface brightness profiles $\mu_{\IE}$ (in mag~arcsec$^{-2}$) for galaxies located at the maximum distance from the centre. The curves represent different types (Type\,I, Type\,II, and Type\,III) with or without the ICL component.
} 
\label{fig:ICLmax}
\end{figure}

In this appendix, we aims at characterising the influence of ICL on the surface brightness profiles of galaxies across different morphological types (Type\,I, Type\,II, and Type\,III). As shown in \cite{ICL}, the effect of the ICL decreases with increasing distance from the cluster centre. While the ICL does not significantly impact the detection of profile types, it was crucial for verifying that no down- or up-bending disc break was missed in Type\,I profiles. Figures \ref{fig:ICLmin} to \ref{fig:ICLmax} illustrate how the ICL impacts surface brightness profiles, with the strongest effect near the cluster centre and a decreasing influence at larger distances. Despite the added ICL, the detection of different profile types remains robust.

\section{\label{appendix:PSF}Influence of the extended PSF on type profiles}

We propose to investigate the impact of the extended PSF of Euclid's cameras on the shapes of surface brightness profiles. As indicated in \cite{pipelineERO}, the fluxes are minimally affected by the extended PSF because it is very pure and extremely sharp. However, some recent studies, as \cite{Trujillo2016} and \cite{Borlaff2017}, suggest that the shape of the disc profiles, specifically the positions of the breaks, can be significantly impacted by a poorly behaved extended PSF. Here, we aim to quantify this effect in the case of Euclid's pristine extended PSF.

\begin{figure*}[htbp!]
\includegraphics[width=1\textwidth]{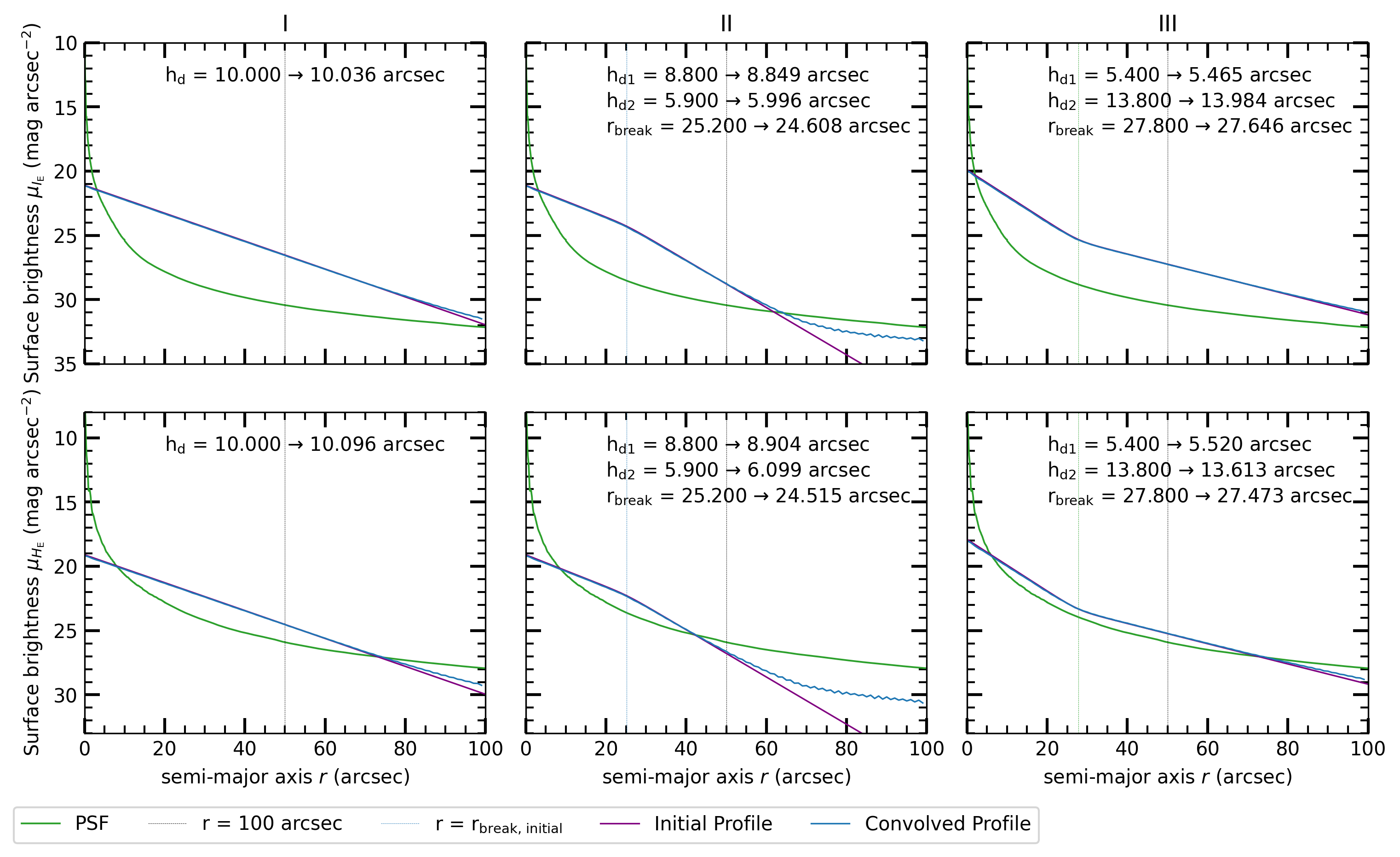}
\caption{Influence of the extended PSF (\IE on {\it top} and \HE on {\it bottom}) on surface brightness profiles of Type\,I ({\it right}), Type\,II ({\it middle}), and Type\,III ({\it left}): the green line is the \IE PSF \citep{pipelineERO}; the purple lines give the initial profile which is pure simple exponential for Type\,I or double exponential for Type\,II; the blue lines correspond to the profile from the convolution of the initial profile and the PSF.} 
\label{fig:PSF}
\end{figure*}

We begin our analysis with models of Type\,I, II, and III galaxies, using parameters corresponding to the median values measured in Table \ref{tab:parametersII} and \ref{tab:parametersIII}. We then convolve this model with the extended PSF model provided in \cite{pipelineERO}. The results are presented in Fig. \ref{fig:PSF}.

We focus on the outer regions of the galaxies, i.e. beyond the sharp region of the PSF. In these cases, the PSF becomes influential only at a radius exceeding 40\arcsec for the three profiles. For Type\,II and III profiles, the corresponding area is beyond the down-bending disc break radius, which in median around 25\arcsec in the case of disc galaxies in the field. Therefore, the fitting parameters of the models vary by less than 4\% between the original and convolved profiles. This justifies our decision not to account for the PSF in the profile fitting adjustments; specifically, we can say that the U-shaped colour profile is not simply due to the PSF influence. 

\section{\label{appendix:meanprofiles}Mean profile from \texttt{AutoProf} and \texttt{AstroPhot}}

We present the distributions of normalised surface brightness profiles obtained using two extraction tools: \texttt{AutoProf} and \texttt{AstroPhot}. These profiles allow us to examine the variations in galaxy surface brightness as a function of the radius normalised by $R_{\textrm{25}}$, while highlighting general trend. Each profile, originally consisting of approximately 20 points, has been resampled to 300 points using interpolation to ensure consistent representation and facilitate comparison across all profiles.

\begin{figure}[htbp!]
\includegraphics[width=0.5\textwidth]{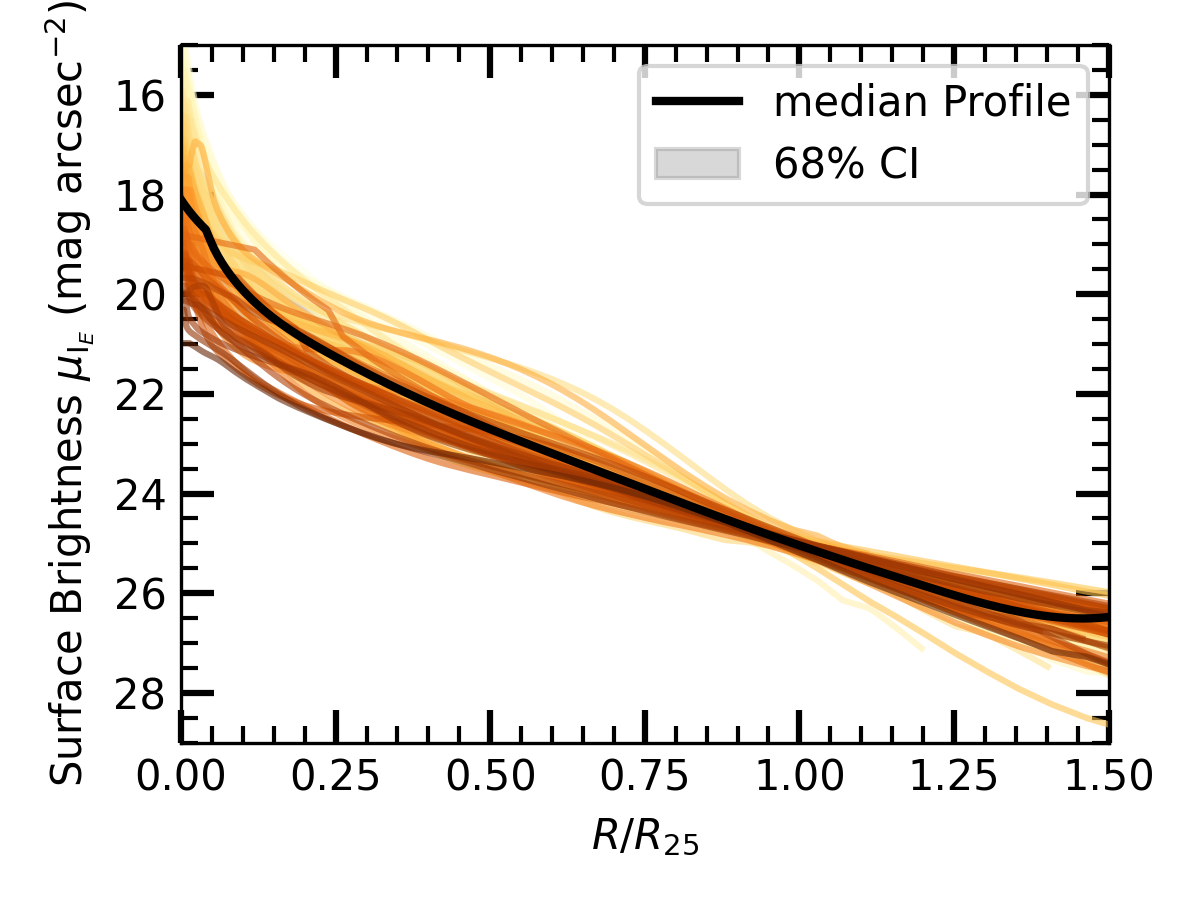}
\caption{Distribution of normalised surface brightness profiles extracted with \texttt{AutoProf}. Individual profiles are displayed with a colour gradient, while the median profile is represented in black. The shaded region around the median indicates the 68\% confidence interval, reflecting the variability among individual profiles.
} 
\label{fig:AutoProfprofiles}
\end{figure}

\begin{figure}[htbp!]
\includegraphics[width=0.5\textwidth]{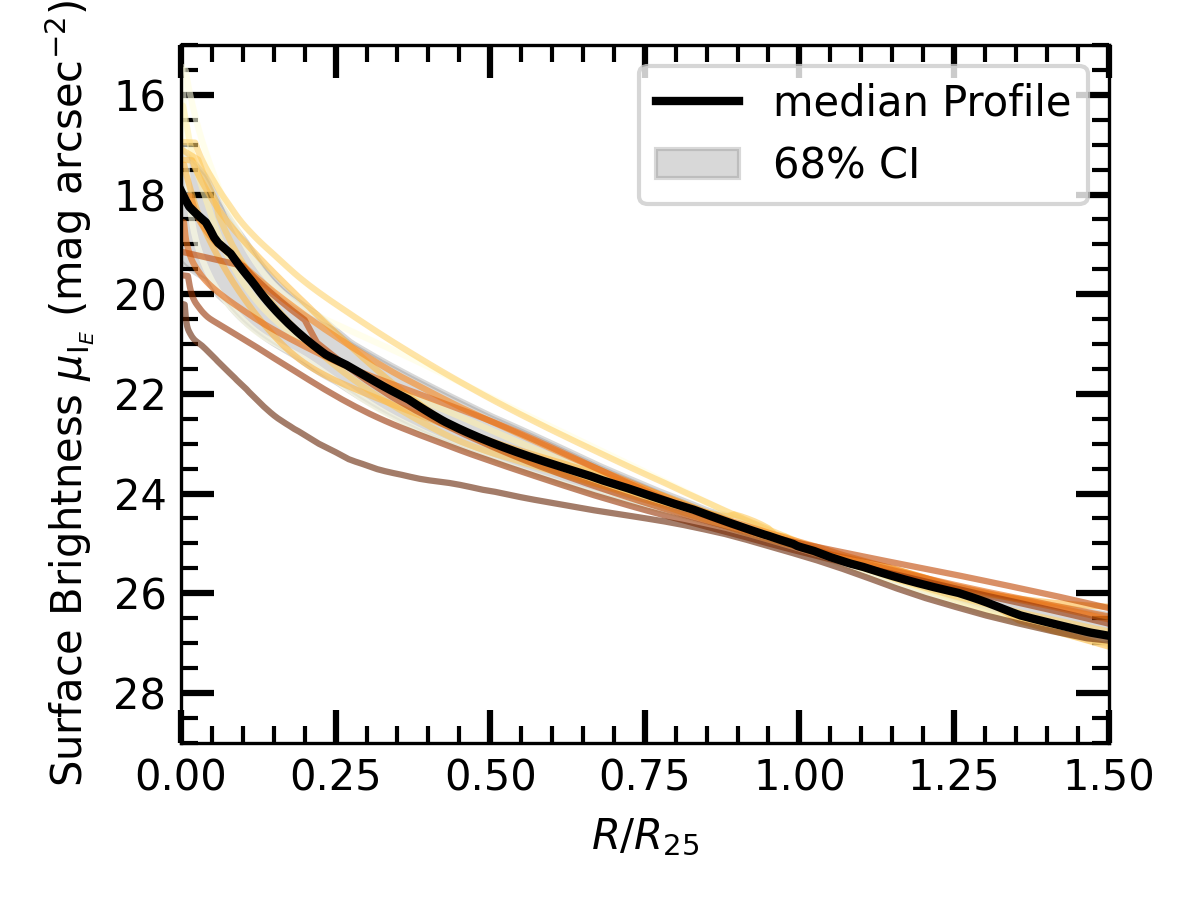}
\caption{Distribution of normalised surface brightness profiles extracted with \texttt{AstroPhot}. Individual profiles are displayed with a colour gradient, while the median profile is represented in black. The shaded region around the median indicates the 68\% confidence interval, reflecting the variability among individual profiles.
} 
\label{fig:AstroPhotprofiles}
\end{figure}

\section{\label{appendix:exampletype}Type profile}

\begin{figure*}[htbp!]
\centering
\includegraphics[width=0.7\textwidth]{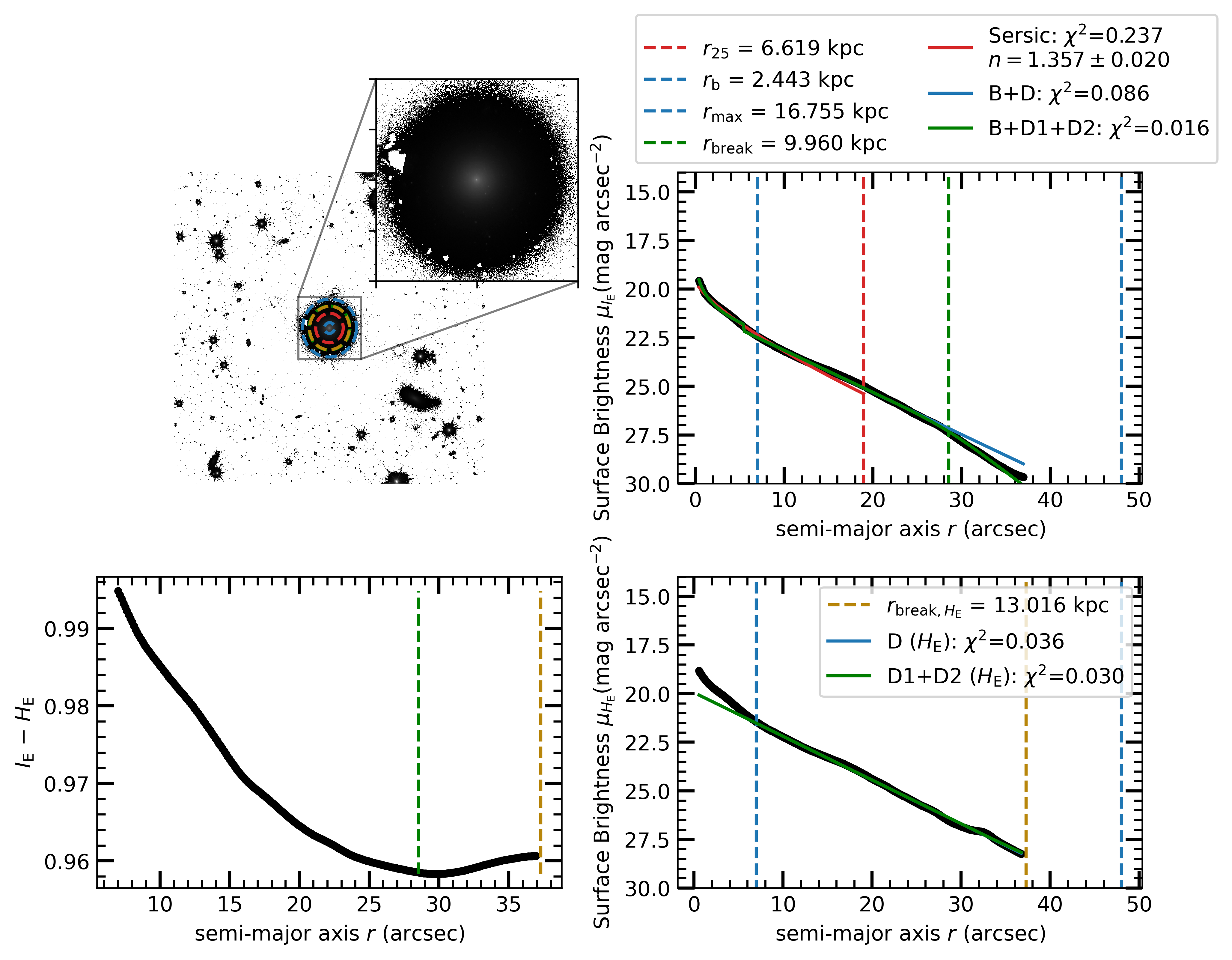}
\caption{Profile of the galaxy WISEA\,J031851\_10+412332\_5 -- Type\,I} 
\label{fig:typeI}%
\end{figure*}

\begin{figure*}[htbp!]
\centering
\includegraphics[width=0.7\textwidth]{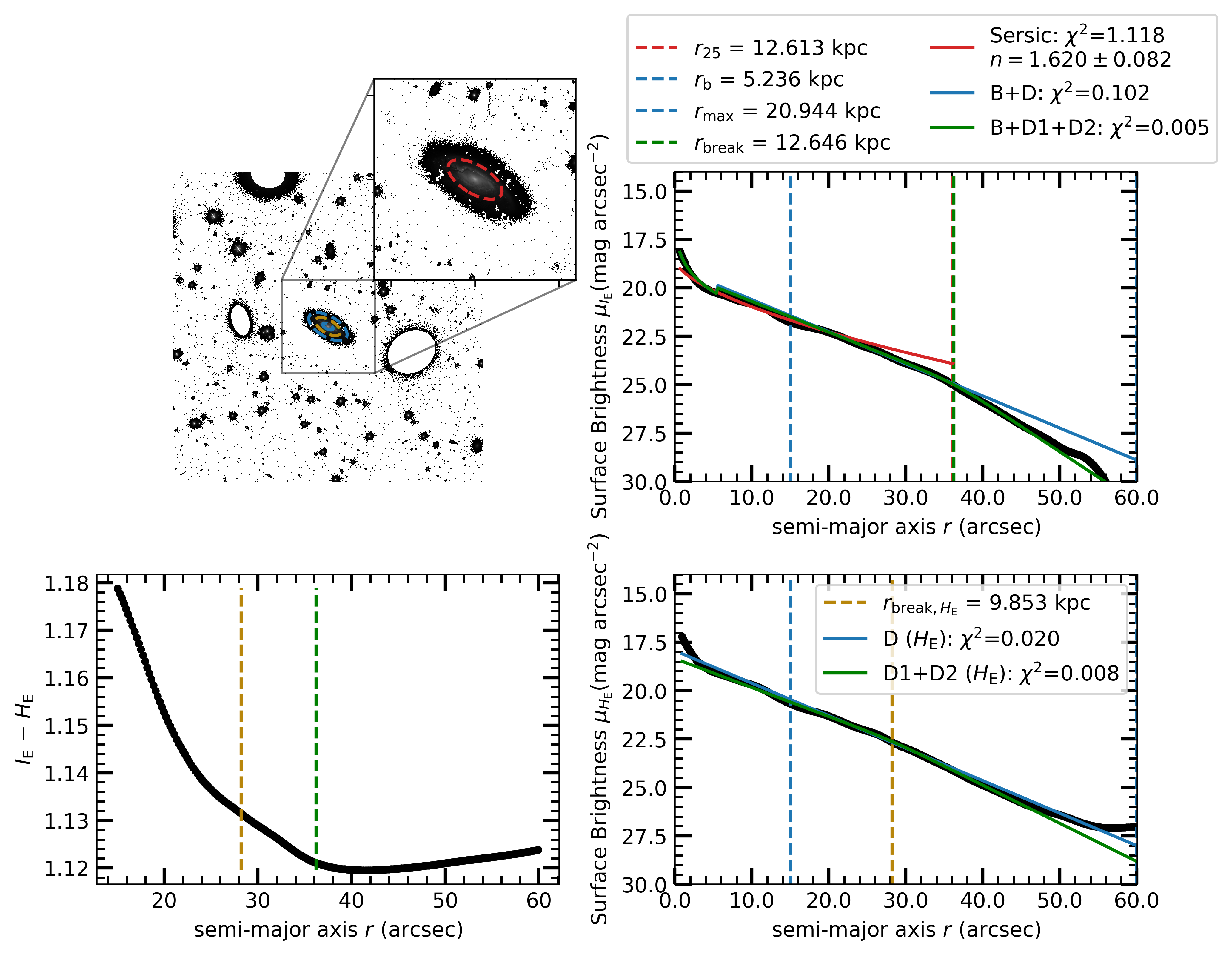}
\caption{Profile of the galaxy UGC\,02665 -- Type\,II}
\label{fig:typeII}%
\end{figure*}

\begin{figure*}[htbp!]
\centering
\includegraphics[width=0.7\textwidth]{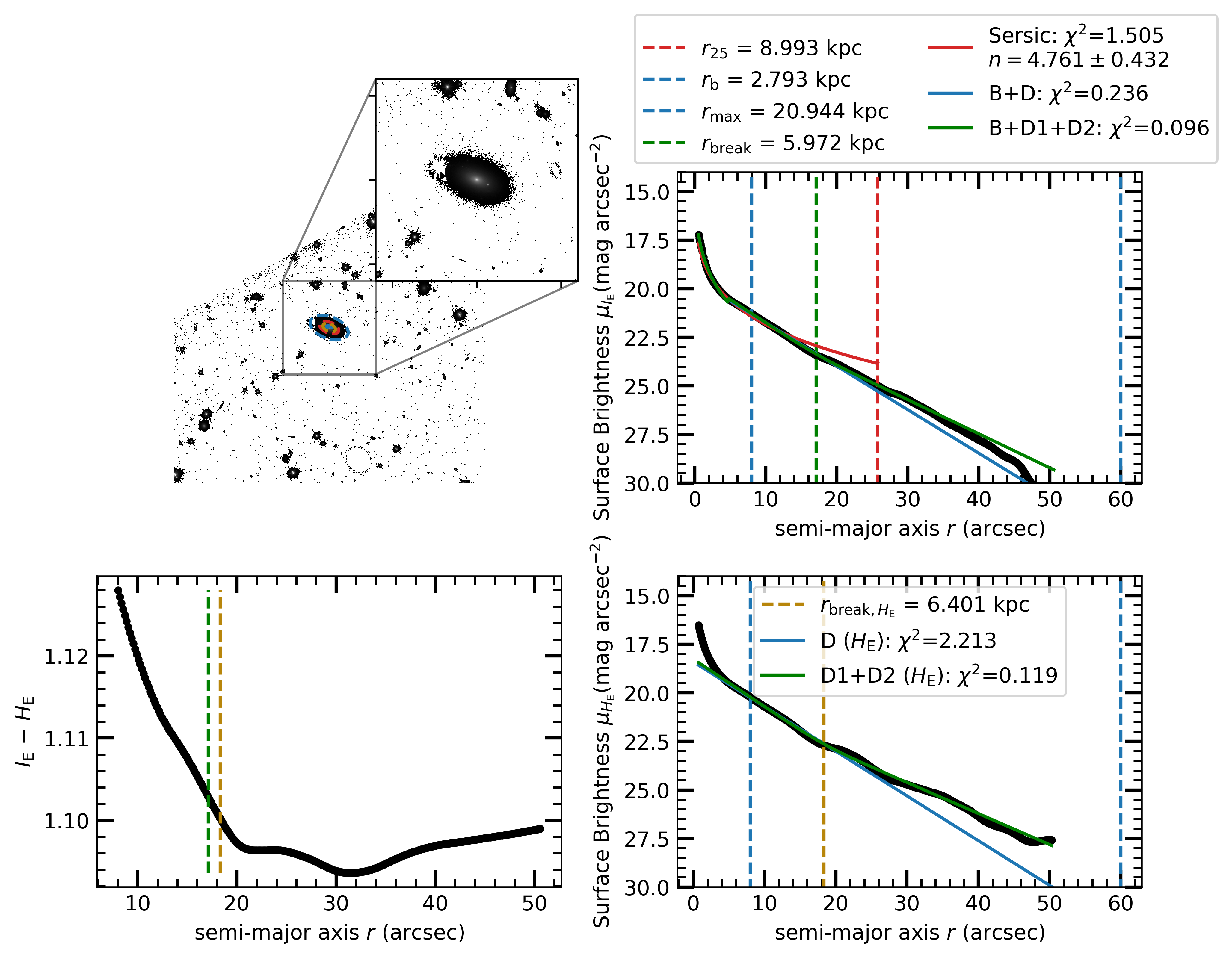}
\caption{Profile of the galaxy WISEA\,J032042\_17+412414\_2 -- Type\,III
} 
\label{fig:typeIII}%
\end{figure*}

In this section, examples of the three types of surface brightness profiles are provided in Figs. \ref{fig:typeI} to \ref{fig:typeIII}. We display four panels for each type as follows:

{\it Top Left Panel:} Image of the galaxy with associated radii indicated by ellipses of different colours: green/yellow for the break positions in the surface brightness profiles of \IE($\mu_{\IE}$) and \HE ($\mu_{\HE}$) bands, red for the radius at $\mu_{\IE} = 25 \, \textrm{mag}\,\textrm{arcsec}^{-2}$, and blue for the boundary regions of interest in semi-major axis $r$ where the break may potentially be found.

{\it Top Right Panel:} Surface brightness profile ($\mu_{\IE}$) versus axis ratio ($r$) of the fitted ellipses in the \IE band. Solid lines indicate the fitted models: single S\'ersic model in red, bulge/disc decomposition model in blue, bulge/disc1/disc2 decomposition model in green. Dashed lines indicate: blue for the boundary regions of interest in $r$ where the break may potentially be found, green for the break position, red for the radius at $\mu_{\IE} = 25 \, \textrm{mag}~\textrm{arcsec}^{-2}$.

{\it Bottom Right Panel:} Surface brightness profile ($\mu_{\HE}$) versus axis ratio ($r$) of the fitted ellipses in the \HE band. Solid lines indicate the fitted models after the radius of the bulge: disc model in blue, disc1/disc2 model in green. Dashed lines indicate the boundary regions of interest in $r$ where the break may potentially be found in blue and the break position in yellow.

{\it Bottom Left Panel:} Magnitude difference (\IE$-$\,\HE) as a function of $r$ in the blue region of interest.

In Fig. \ref{fig:typeI}, the top right panel shows that the single S\'ersic model -- red line -- gives a low S\'ersic index and does not fit the data well beyond a certain radius. Its chi-square value is higher compared to the other two models. It is observed that the disc models provide a better fit, with the double disc model indicating a very late break around a surface brightness of 28, which does not correspond to a typical Type\,II or Type\,III profile.

In contrast, Fig. \ref{fig:typeII} clearly shows a down-bending disc break near a magnitude of 25. The double disc model in green gives the best chi-square value. 

Finally, in Fig. \ref{fig:typeIII}, a Type\,III profile is clearly identified by the up-bending exponential profile model. 

It can also be noted that the bottom left panels, showing the magnitude variation between the \IE and \HE bands, indicate no gradient for any of these examples, particularly after the break for Type\,II and III profiles.


\section{\label{appendix:morphologicalclass}Characterisation of the morphological classification}

This appendix is dedicated to the validation of the morphological classification by verifying several aspects. In addition to comparisons for a large portion of the galaxies, approximately 60, to the literature, we also perform measurements of the bulge luminosity and the bulge-to-total light ratio for the morphological types S0 (type $-$1), intermediate between S0 and pure spirals (type 0), and spiral galaxies (type 1). These metrics are widely used in the literature to classify the morphologies of disc galaxies through the characterisation of their bulge \citep{Quilley2023}.

\begin{figure*}[htbp!]
\centering
\includegraphics[width=0.8\textwidth]{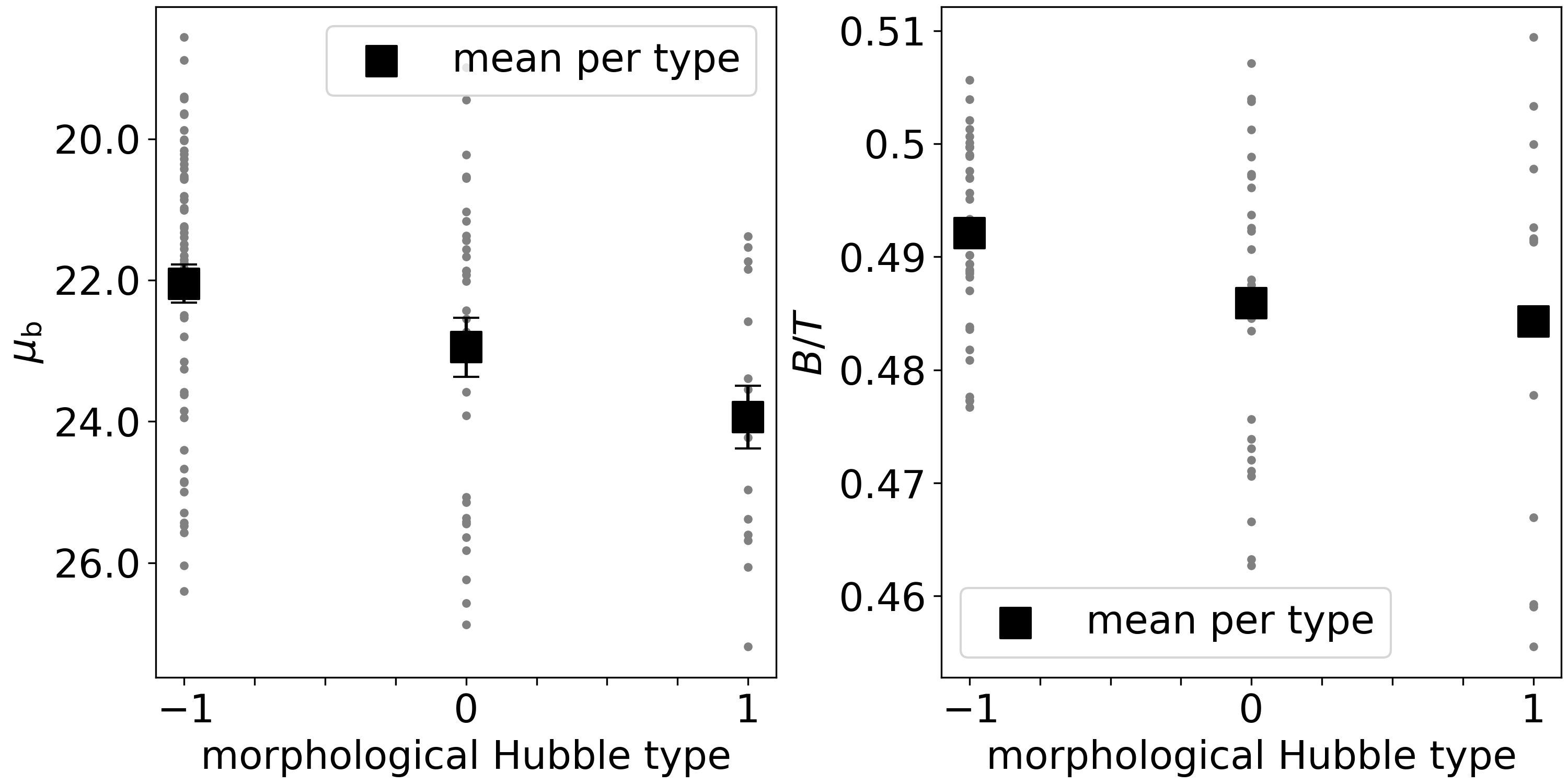}
\caption{Validation of morphological classifications. {\it Left Panel:} Bulge surface brightness measurements for disc galaxies according to morphological classes (S0, intermediate, spirals). Grey points represent individual measurements, while the dark squares give the mean and standard deviation of each distribution. {\it Right Panel:} Bulge-to-total light ratio as a function of morphological type. } 
\label{fig:mBmorphotype}
\end{figure*}

In Fig. \ref{fig:mBmorphotype}, the grey points represent these measurements for each disc galaxy according to its morphological class. The mean and standard deviation for each value are also provided. Clearly, as expected, the average surface brightness of the bulges increases with the evolutionary stage of the disc galaxies. Similarly, the right panel shows a slight decrease with the bulge flux fraction from S0 to spiral galaxies. Overall, our classification is validated by these results.

\section{\label{appendix:histoparam}Fitted parameters}

\begin{figure*}[htbp!]
\centering
\includegraphics[width=1\textwidth]{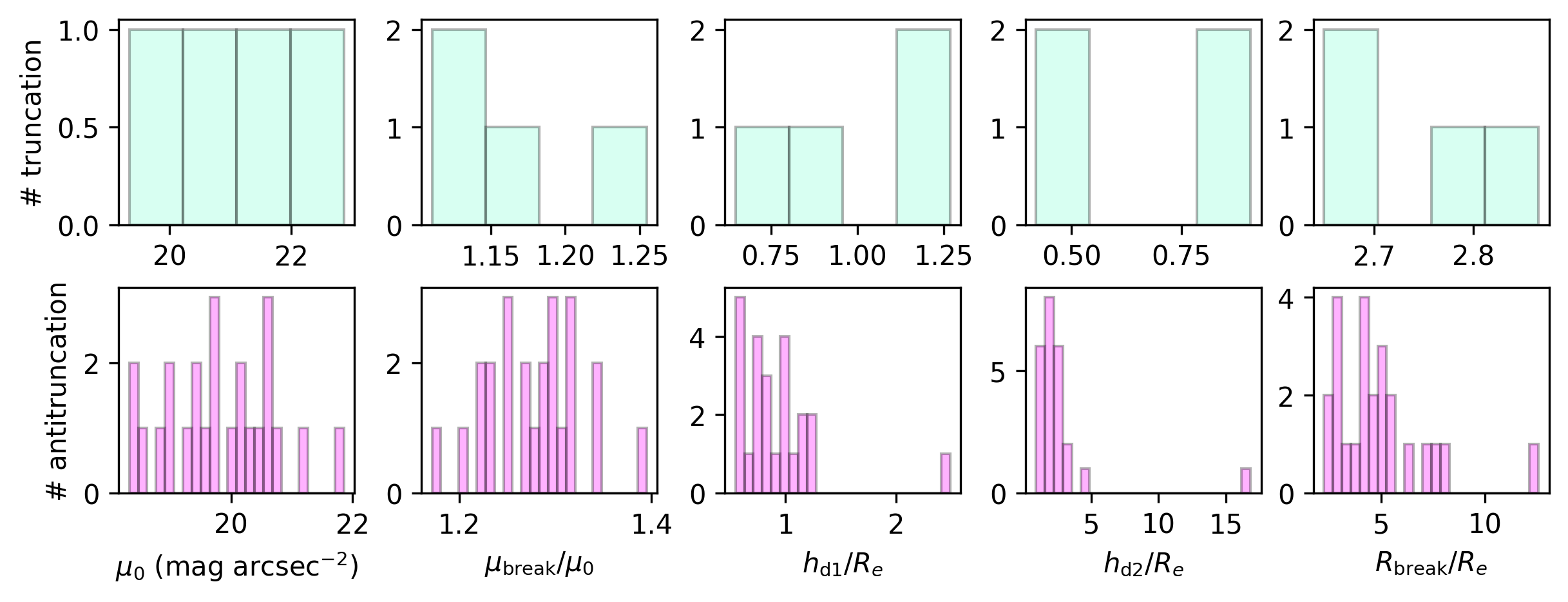}
\caption{Distribution of fitting parameters using the double exponential profile in the \IE band for Type\,II (down-bending break in higher panels) and Type\,III (up-bending break in lower panels) profiles. The scalelengths of the discs and down-bending break radii are compared to the effective radii from the main Perseus catalogue \cite{LF}. {\it First panel:} Distribution of the central surface brightness of the discs. {\it Second panel:} Distribution of the break surface brightness compared to the central surface brightness of the discs. {\it Third panel:} Distribution of first disc scalelengths. {\it Fourth panel:} Distribution of second disc scalelengths. {\it Fifth panel:} Break radii compared to the effective radius of the galaxy.} 
\label{fig:histoparameters}%
\end{figure*}

Figure \ref{fig:histoparameters} presents the distribution of fitted parameters obtained using the double exponential profile in the \IE band for Type\,II (down-bending break) and Type\,III (up-bending break) profiles. The disc scalelengths and down-bending break radii are compared to the effective radii provided in the catalogue from \cite{LF}. It is observed that the break occurs well beyond the effective radius of the galaxy, as shown in the rightmost panel. The scalelengths of the first disc are of the order of the galaxy's effective radius. One of the up-bending break galaxies has a value significantly higher than this effective radius, but no significant issue in the fit was found. For the scalelength of the second disc, as expected, values significantly lower than the effective radius are found for Type\,II profiles, while higher values are found for Type\,III profiles.

\begin{figure*}[htbp!]
\includegraphics[width=1.\textwidth]{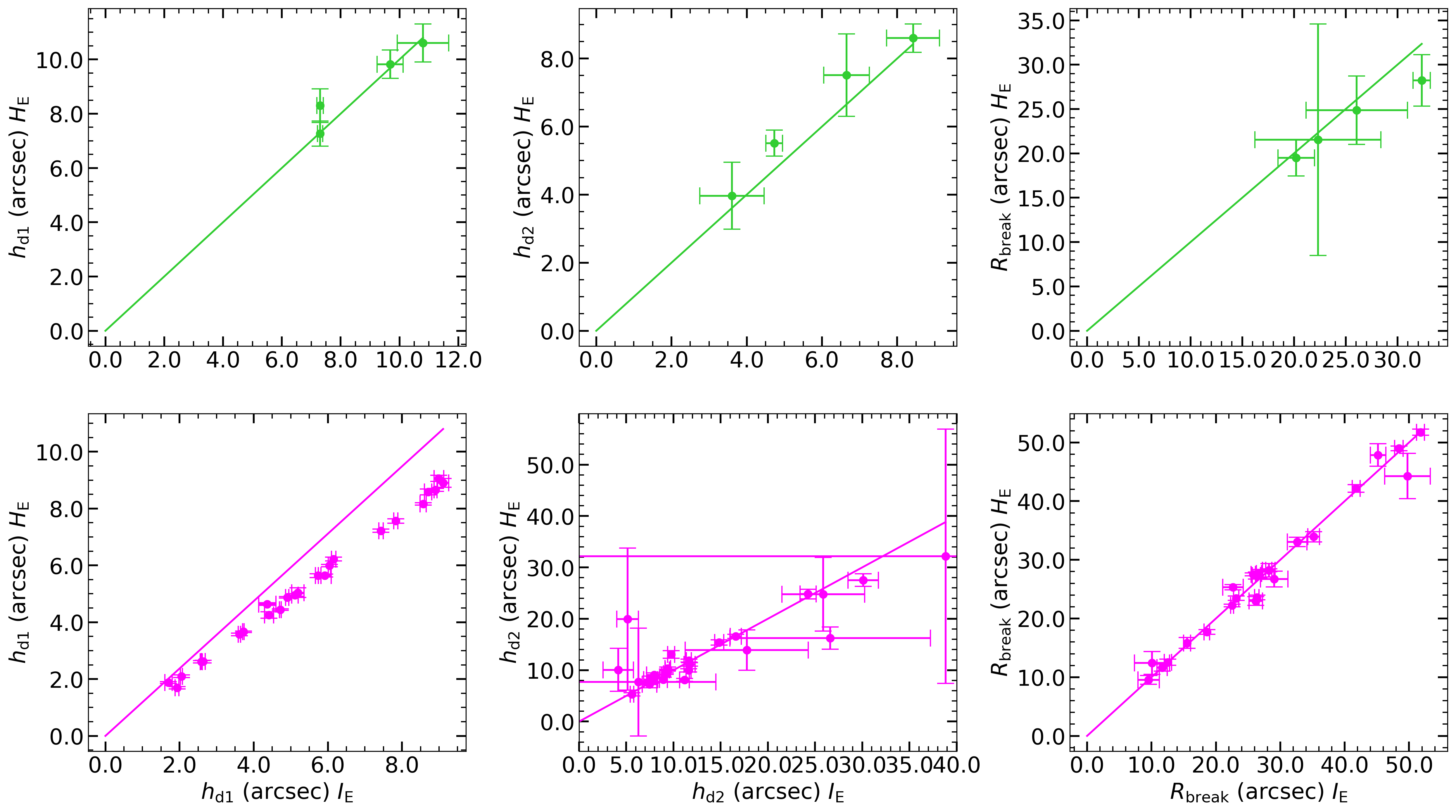}
\caption{Distribution of disc scalelengths and break radii for Type\,II (higher panels) and for Type\,III (lower panels) in the \HE band as a function of those in the \IE band. The solid line represents the identity line for comparison between the two bands.} 
\label{fig:RRH}%
\end{figure*}

In Fig. \ref{fig:RRH}, the distribution of disc scalelengths and break radii from fits in the \HE band are compared with those in the \IE band. The solid line represents the identity line to facilitate comparison between the two bands. The parameters are consistent across both bands, with a slight tendency for the scalelength of the first disc to be shorter in the infrared than in the visible, a trend noted in \cite{Laine2016}.

\end{appendix}

\label{LastPage}

\end{document}